\documentclass{iopart}
\newcommand{\araa}{{\it Ann. Rev. Astron. Astrophys. }}
\newcommand{\areps}{{\it Ann. Rev. Earth Planet. Sci.}}
\newcommand{\arnps}{{\it Ann. Rev. Nucl. Part. Sci. }}
\newcommand{\astrobio}{{\it Astrobio. }}
\newcommand{\aanda}{{\it Astron. Astrophys. }}
\newcommand{\aarev}{{\it Astron. Astrophys. Rev. }}
\newcommand{\aj}{{\it Astron. J. }}
\newcommand{\app}{{\it Astropart. Phys. }}
\newcommand{\apj}{{\it Astrophys. J. }}
\newcommand{\apjs}{{\it Astrophys. J. Suppl. Ser. }}
\newcommand{\apss}{{\it Astrophys. Space Sci. }}
\newcommand{\cjp}{{\it Can. J. Phys. }}
\newcommand{\car}{{\it Carbon }}
\newcommand{\cp}{{\it Chem. Phys. }}
\newcommand{\cpl}{{\it Chem. Phys. Lett. }}
\newcommand{\chinpl}{{\it Chin. Phys. Lett. }}
\newcommand{\epja}{{\it Eur. Phys. J. A }}
\newcommand{\epjc}{{\it Eur. Phys. J. C }}

\newcommand{\esr}{{\it Earth-Sci. Rev. }}
\newcommand{\far}{{\it Faraday Discuss. }} 
\newcommand{\gca}{{\it Geochim. Cosmochim. Acta }} 
\newcommand{\grl}{{\it Geophys. Res. Lett. }}
\newcommand{\hedp}{{\it High Energy Den. Phys. }}
\newcommand{\hyperf}{{\it Hyperf. Int. }}
\newcommand{\ica}{{\it Icarus }}
\newcommand{\iramp}{{\it Int. Rev. At. Mol. Phys. }}
\newcommand{\ijms}{{\it Int. J. Mass Spectrom. }}
\newcommand{\jcp}{{\it J. Chem.  Phys. }}
\newcommand{\jcap}{{\it J. Cosmo. Astropart. Phys. }}
\newcommand{\jcppcb}{{\it J. Chim. Phys. Phys.-Chim. Biol. }}
\newcommand{\jgr}{{\it J. Geophys. Res. }}
\newcommand{\jhep}{{\it J. High Energy Phys. }}
\newcommand{\jmr}{{\it J. Mater. Res. }} 
\newcommand{\jmsp}{{\it J. Mol. Spec. }}
\newcommand{\jmst}{{\it J. Mol. Struct. }} 
\newcommand{\josab}{{\it J. Opt. Soc. Am. B }}
\newcommand{\jpd}{{\it J. Phys. D }}
\newcommand{\jpca}{{\it J. Phys. Chem. A }}
\newcommand{\jpcs}{{\it J. Phys. Conf. Ser. }}
\newcommand{\jqsrt}{{\it J. Quant. Spectrosc. Radiat. Transfer }}
\newcommand{\JETP}{{\it Sov. Phys. JETP }}
\newcommand{\JETPL}{{\it JETP Lett. }}
\newcommand{\mps}{{\it Metorit. Planet. Sci. }}
\newcommand{\mnras}{{\it Mon. Not. R. Astron. Soc. }}
\newcommand{\molphys}{{\it Mol. Phys.  }}
\newcommand{\nat}{{\it Nature }}
\newcommand{\natphy}{{\it Nature Phys. }}
\newcommand{\njp}{{\it New J. Phys. }}
\newcommand{\nonlin}{{\it Nonlinearity }}
\newcommand{\nim}{{\it Nucl. Instrum. Meth. }}

\newcommand{\npa}{{\it Nucl. Phys. A }}
\newcommand{\npb}{{\it Nucl. Phys. B }}

\newcommand{\oleb}{{\it Orig. Life Evol. Biosph. }}
\newcommand{\pasp}{{\it Publ. Astron. Soc. Pacific }}
\newcommand{\pccp}{{\it Phys. Chem. Chem. Phys. }}
\newcommand{\physcr}{{\it Phys. Scr. }}
\newcommand{\pfa}{{\it Phys. Fluids A }}

\newcommand{\plb}{{\it Phys. Lett. B }}
\newcommand{\pr}{{\it Phys. Rep. }}
\newcommand{\pra}{{\it Phys. Rev. A }}
\newcommand{\prb}{{\it Phys. Rev. B }}
\newcommand{\prc}{{\it Phys. Rev. C }}
\newcommand{\prd}{{\it Phys. Rev. D }}
\newcommand{\pre}{{\it Phys. Rev. E }}
\newcommand{\prl}{{\it Phys. Rev. Lett. }}
\newcommand{\physplas}{{\it Phys. Plasmas }}
\newcommand{\phystoday}{{\it Phys. Today }}
\newcommand{\pss}{{\it Planet. Space Sci. }}
\newcommand{\ppcf}{{\it Plasma Phys. Control. Fusion }}
\newcommand{\pnasusa}{{\it Proc. Nat. Acad. Sci. U.S.A. }} 
\newcommand{\ppslsa}{{\it Proc. Phys. Soc. London Sect. A }}
\newcommand{\pSPIE}{{\it Proc. SPIE }}
\newcommand{\ptrsa}{{\it Phil. Trans. R. Soc. A }}
\newcommand{\pos}{{\it Proc. Sci. }}
\newcommand{\pasj}{{\it Publ. Astron. Soc. Japan }} 
\newcommand{\rpp}{{\it Rep. Prog. Phys. }}
\newcommand{\rsi}{{\it Rev. Sci. Instrum. }}
\newcommand{\rmp}{{\it Rev. Mod. Phys. }}
\newcommand{\sci}{{\it Science }}
\newcommand{\solphys}{{\it Sol. Phys. }}
\newcommand{\ssr}{{\it Space Sci. Rev. }}

\newcommand{\gax}{\hbox{${_{\displaystyle>}\atop^{\displaystyle\sim}}$}}

\usepackage{graphicx}
\begin{document}

\title[Impact of Laboratory Astrophysics]
{The Impact of Recent Advances in Laboratory Astrophysics on our Understanding of the Cosmos}

\author{
D.~W.~Savin$^1$, 
N.~S.~Brickhouse$^2$, 
J.~J.~Cowan$^3$, 
R.~P.~Drake$^4$, 
S.~R.~Federman$^5$, 
G.~J.~Ferland$^6$, 
A.~Frank$^7$, 
M.~S.~Gudipati$^8$,
W.~C.~Haxton$^9$, 
E.~Herbst$^{10}$, 
S.~Profumo$^{11}$, 
F.~Salama$^{12}$, 
L.~M.~Ziurys$^{13}$ and
E.~G.~Zweibel$^{14}$
}

\address{$^1$Columbia Astrophysics Laboratory, Columbia University,
New York, NY 10027, USA} 
\address{$^2$Harvard-Smithsonian Center for Astrophysics, 60 Garden 
Street, Cambridge, MA 02138, USA} 
\address{$^3$Homer L. Dodge Department of Physics and Astronomy, 
University of Oklahoma, Norman, OK 73019, USA} 
\address{$^4$Department of Atmospheric, Oceanic and Space Sciences,
University of Michigan, Ann Arbor, MI 48109, USA} 
\address{$^5$Department of Physics and Astronomy, University of 
Toledo, Toledo, OH 43606, USA} 
\address{$^6$Department of Physics, University of Kentucky, 
Lexington, KY 40506, USA} 
\address{$^7$Department of Physics and Astronomy, University of 
Rochester, Rochester, NY 14627, USA} 
\address{$^8$Science Division, Jet Propulsion Laboratory, California 
Institute of Technology, Pasadena, CA 91109, USA}
\address{$^9$ Department of Physics, University of California,
Berkeley, and Lawrence Berkeley National Laboratory, Berkeley, CA 97420}
\address{$^{10}$Departments of Chemistry, Astronomy and Physics,
University of Virginia, Charlottesville, VA 22904, USA}
\address{$^{11}$Department of Physics, ISB 325, University of 
California, 1156 High Street, Santa Cruz, CA 95064, USA}
\address{$^{12}$Space Science Division, NASA Ames Research Center, 
Moffett Field, CA 94035, USA} 
\address{$^{13}$Departments of Chemistry and Astronomy, Arizona 
Radio Observatory and Steward Observatory, University of Arizona, 
Tucson, AZ 85721, USA}
\address{$^{14}$Departments of Astronomy and Physics, University of 
Wisconsin, 6281 Chamberlain Hall, 475 North Charter Street, 
Madison, WI 53706, USA}

\begin{abstract}
An emerging theme in modern astrophysics is the connection between
astronomical observations and the underlying physical phenomena that
drive our cosmos.  Both the mechanisms responsible for the observed
astrophysical phenomena and the tools used to probe such phenomena --
the radiation and particle spectra we observe -- have their roots in
atomic, molecular, condensed matter, plasma, nuclear and particle
physics.  Chemistry is implicitly included in both molecular and
condensed matter physics.  This connection is the theme of the present
report, which provides a broad, though non-exhaustive, overview of
progress in our understanding of the cosmos resulting from recent
theoretical and experimental advances in what is commonly called
laboratory astrophysics.  This work, carried out by a diverse
community of laboratory astrophysicists, is increasingly important as
astrophysics transitions into an era of precise measurement and high
fidelity modeling.
\end{abstract}

\maketitle

\section{Introduction}

Laboratory astrophysics and complementary theoretical calculations are
the foundations of astronomy and astrophysics and will remain so into
the foreseeable future.  The impact of laboratory astrophysics ranges
from the scientific conception for ground-based, airborne and
space-based observatories, all the way through to the scientific
return of these projects and missions.  It is our understanding of the
underlying physical processes and the measurement or calculation of
critical physical parameters that allows us to address fundamental
questions in astronomy and astrophysics.  

The field of laboratory astrophysics comprises both theoretical and
experimental studies of the underlying physics that produce the
observed astrophysical processes.  We have identified six areas of
physics as relevant to astronomy and astrophysics.\footnote{The authors
  comprise past and current members of the American Astronomical
  Society Working Group on Laboratory Astrophysics.}
Astronomy is an observational
science focused primarily on detecting photons generated by atomic,
molecular and condensed matter physics.  Chemistry is implicitly
included here as part of molecular and condensed matter physics.  Our
understanding of the universe also relies on knowledge of the
evolution of matter (nuclear and particle physics) and of the
dynamical processes shaping it (plasma physics).  Planetary science,
involving {\it in-situ} measurements of solar system bodies, requires
knowledge from atomic, molecular, condensed matter and plasma physics.
Hence, our quest to understand the cosmos rests firmly on scientific
knowledge in six areas: atomic, molecular, condensed matter, plasma,
nuclear and particle physics.

Here we review recent advances in our astrophysical understanding of
the cosmos arising from work in laboratory astrophysics.  We focus
primarily on the past decade.  Our work complements that of previous
reviews on laboratory astrophysics in atomic physics (Beiersdorfer
2003; Kallman and Palmeri 2007; International Astronomical Union [IAU]
Commission 14), molecular physics (Salama 1999; Tielens 2005; Herbst
and van Dishoeck 2009; IAU Commision 14), condensed matter physics
(Draine 2003; Whittet 2003), plasma physics (Drake 1999; Remington
\etal 2006; Zweibel and Yamada 2009; Yamada \etal 2010), nuclear
physics (K\"appeler \etal 2011; Wiescher \etal 2010; Adelberger \etal
2011) and particle physics (Grupen 2005; Aprile and Profumo 2009).

Because laboratory astrophysics, as implied by its name, is
astrophysically motivated, we have structured our report into five
broad categories which blanket the field of astronomy and
astrophysics.  This helps to bring out the synergy between the
various subareas of laboratory astrophysics.  The specific categories
are: planetary systems and star formation (section~\ref{sec:PSF}),
stars and stellar evolution (section~\ref{sec:SSE}), the galactic
neighborhood (section~\ref{sec:GAN}), galaxies across time
(section~\ref{sec:GAT}) and cosmology and fundamental physics
(section~\ref{sec:CFP}).  This structure parallels the scientific
divisions used by the rececent U.S.\ National Research Council Astro
2010 Survey on Astronomy and Astrophysics (Blandford \etal 2010a).
These five sections are further subdivided into relevant subareas of
laboratory astrophysics.  Space limitations prevent these subsections
from being exhaustive.  Rather they are aimed at giving the reader an
overview of recent successes in the field and appropriate citations to
provide entry into the relevant research.  We conclude with a brief
discussion and outlook for the future in section~\ref{sec:end}.

\section{Planetary systems and star formation}
\label{sec:PSF}

Planetary systems and star formation encompass ``solar system bodies
(other than the Sun) and extrasolar planets, debris disks, exobiology,
the formation of individual stars, protostellar and protoplanetary
disks, molecular clouds and the cold ISM\footnote{A list of acronyms
  used throught the text is given in appendix~A.}  [interstellar
  medium], dust, and astrochemistry'' (Blandford \etal 2010a).

\subsection{Atomic physics}

\subsubsection{Young late-type stars.}

\begin{figure}
\begin{center}
\includegraphics[angle = 0, height=0.35\textheight]{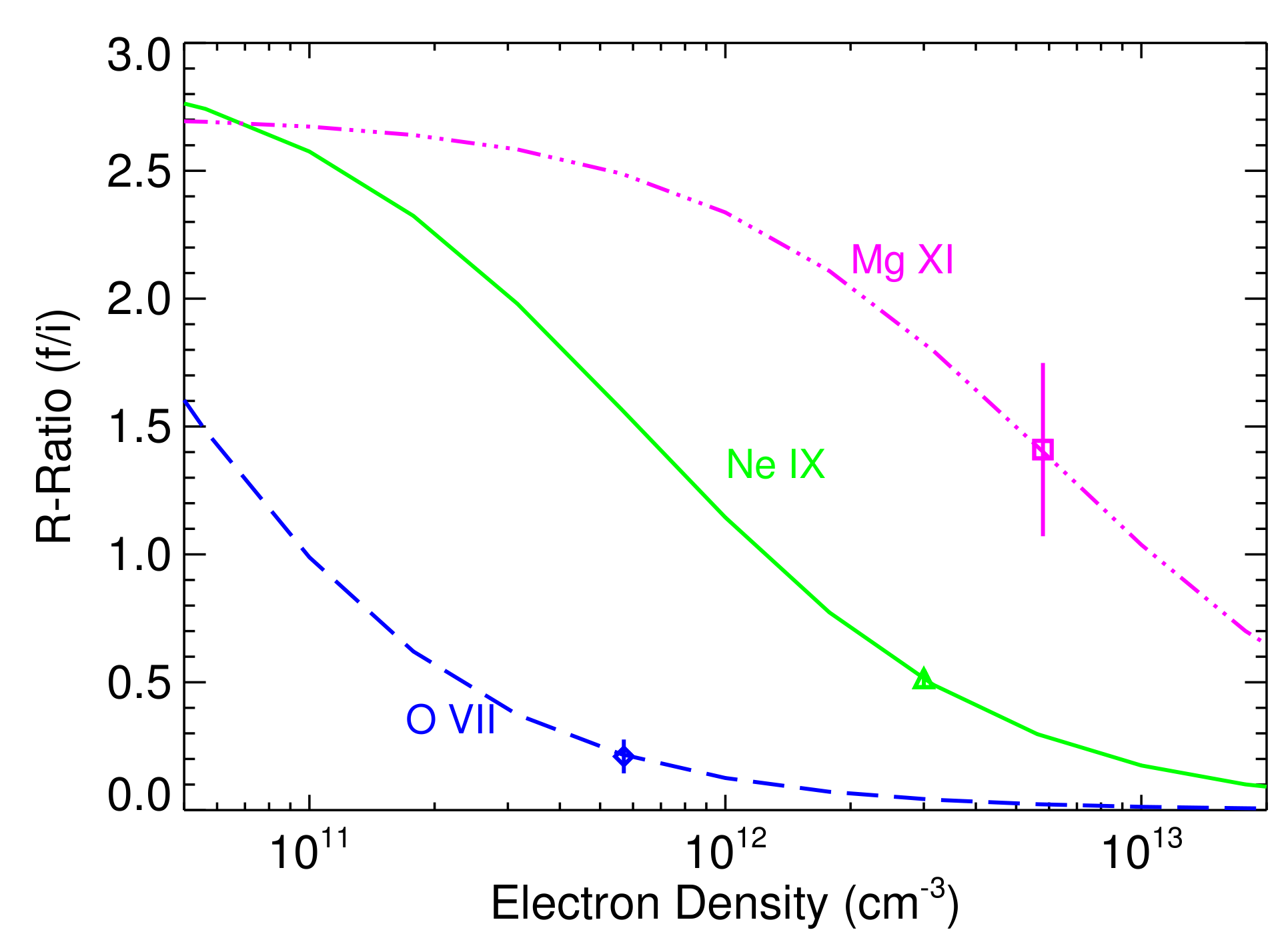}
\caption{\label{fig:PSF:At:You} Theoretical He-like ion forbidden to
  intercombination line R-ratios (f/i) as a function of density
  (curves) overplotted with the observed line ratios from the {\it
    Chandra} spectrum of the young star TW Hya (points with 1$\sigma$
  error bars).  As discussed in the text, the electron temperature and
  density have been determined using accurate atomic data.  Accretion
  shock models are in good agreement with the Ne IX and Mg XI
  densities and temperatures at the shock front.  However, the models
  fail to match the observed O VII density, from Brickhouse \etal
  (2010).}
\end{center}
\end{figure}

In accreting stellar objects with strong magnetic fields (such as
young late-type stars, X-ray binaries with neutron stars and magnetic
cataclysmic variables [CVs]), the stellar magnetic field truncates the
accretion disk and channels the accreting material towards a ``hot
spot'' near the pole of the star (Konigl 1991).  This material
accelerates in the gravitational field of the star, reaching
supersonic velocities and producing a shock which emits in X-rays.
For low mass young stars, the free-fall velocity (the maximum velocity
obtained by material accelerated from infinity) is $\sim
500$~km~s$^{-1}$ and the expected shock temperatures are around a few
MK (Calvet and Gullbring 1998; Kastner \etal 2002).  High electron
densities of $\sim 10^{13}$~cm$^{-3}$ are also expected at the shock,
assuming the ram pressure of the gas balances the stellar atmospheric
pressure.  Electron temperature and density diagnostics are available
using He-like lines observed in X-ray spectra from O~\textsc{vii},
Ne~\textsc{ix} and Mg~\textsc{xi}.  However, atomic theoretical models
of these diagnostic lines have only recently become accurate enough to
test shock models (Chen \etal 2006; Smith \etal 2009).  Applying the
new atomic data to a long observation (500 ks) of TW Hya with the {\it
  Chandra X-ray Observatory} High Energy Transmission Grating,
Brickhouse \etal (2010) showed that the shock models work well at the
shock front.  But, again using accurate diagnostics, the standard
model fails to describe the spectra of the post-shock cooling gas.  In
the standard model, the electron density increases as the shocked gas
cools and recombines, but instead the opposite is observed: the
observed density of the cooler O~\textsc{vii} is lower than that of
the hotter Ne~\textsc{ix} by a factor of 4
(figure~\ref{fig:PSF:At:You}) and lower than the model prediction by a
factor of 7.  In contradiction to the post-shock models of cooling and
``settling'' gas, the shock heats a significant mass of stellar
atmosphere to soft X-ray emitting temperatures.  This discovery has
implications for coronal heating and wind driving in the presence of
accretion.

\subsubsection{Cometary X-ray emission.}

The discovery of X-ray and extreme ultraviolet emission from comet
C/Hyakutake (Lisse \etal 1996) was a great surprise.  The subsequent
identification of the emission mechanism as charge exchange with the
highly charged ions of the solar wind (Cravens 1997; Krasnopolsky
\etal 1997) has led to tremendous progress in understanding the solar
system (see Bhardwaj \etal 2007).  High spectral resolution
observations revealed the classic signature of charge exchange, namely
dominant features from high angular momentum states and thus high
principal quantum levels (Kharchenko and Dalgarno 2000; Krasnopolsky
and Mumma 2001; Lisse \etal 2001).  Calculations and experiments of
charge exchange are now incorporated into X-ray studies of the
interaction between the solar wind and planets, comets and the
heliosphere.  Cravens \etal (2001) predicted that charge exchange of
solar wind ions in the heliosphere and geocorona could produce half
the soft X-ray background.  The long-standing mystery of the soft
X-ray background (and one of the key goals of {\it Chandra}) is now
being solved: perhaps all or most of this background comes from charge
exchange of the solar wind within the heliosphere (Koutroumpa \etal
2006), with important implications for the interstellar environment
surrounding the solar system.  Experimental measurements continue to
be important for quantitative analysis of charge exchange spectra
(e.g., Beiersdorfer \etal 2000; Greenwood \etal 2000; Beiersdorfer
\etal 2003; Otranto and Olson 2011).  Dennerl (2010) provides a
good review of this field.

\subsubsection{Exoplanetary discovery.}

Nearly 500 planets around other stars have been discovered to date
using a variety of techniques, with many more expected from the {\it
  Kepler} Mission (Borucki \etal 2010).  The $\sim 100$ exoplanets
that transit their host stars are scientifically invaluable since both
the mass and radius of the planet can be determined (e.g., Maxted
\etal 2010).  Transit searches involve two main stages: repeated
photometric detection of transits of acceptable depth and duration,
followed by spectroscopic confirmation.  The first exoplanet
discovered by the transit method exploited a detailed stellar
atmosphere model of the star, cross-correlated with the observed
spectra, in order to determine radial velocities (Konacki \etal 2003,
2004; Sasselov 2003).  This approach has now become a standard tool in
the field, with many refinements added (Torres \etal 2011).  These
atmosphere models incorporate an enormous database of atomic and
molecular line transitions (Kurucz and Bell 1995; Castelli \etal
1997).  The precision in radial velocity that can be achieved depends
strongly on the fraction of spectral lines in the model that match the
observation; hence, ongoing efforts to improve the line lists go hand
in hand with continuing discoveries in this field.

\subsection{Molecular physics}

\subsubsection{Molecular clouds: diffuse interstellar bands.}

\begin{figure}
\begin{center}
\includegraphics[angle = 0, height=0.35\textheight]{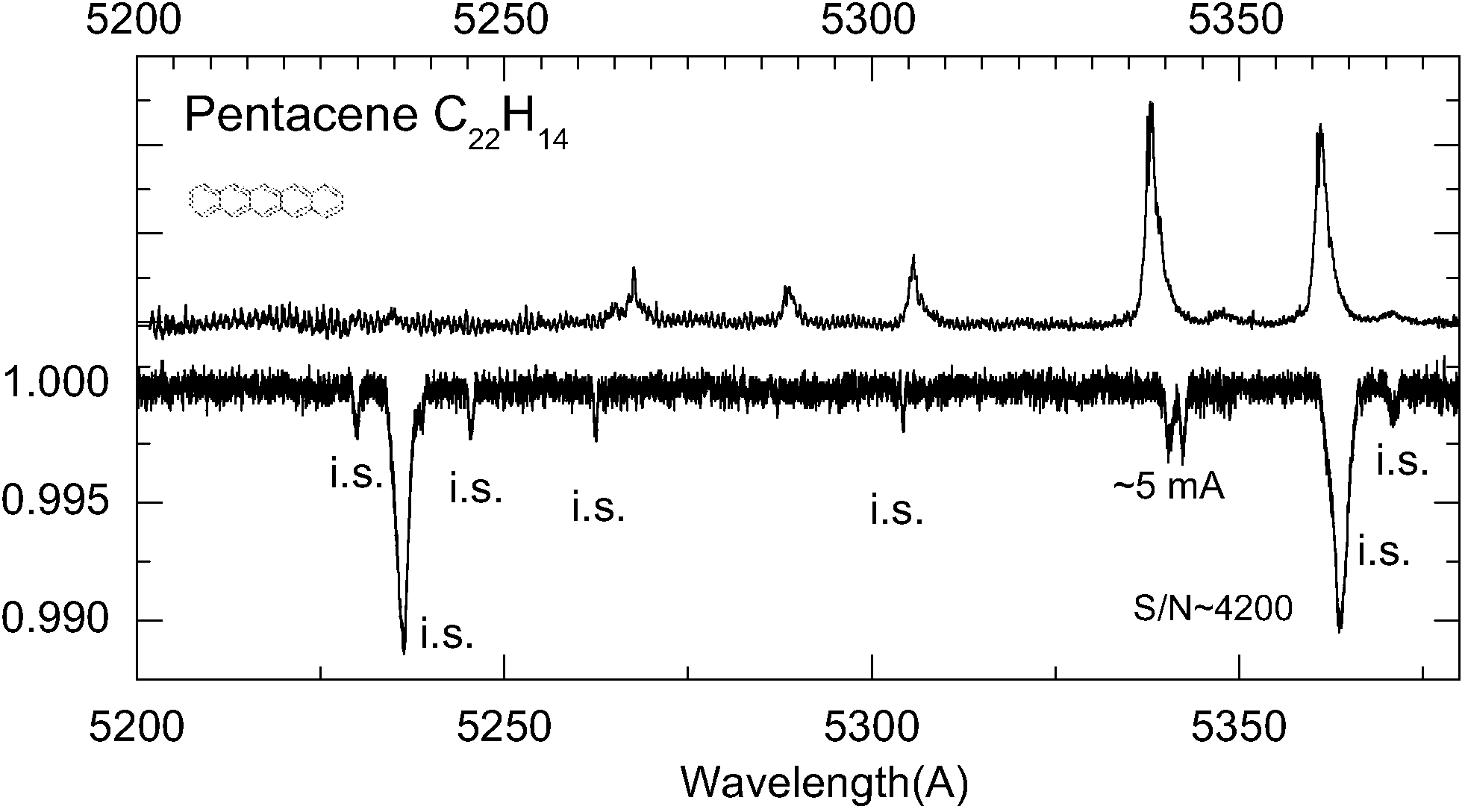}
\caption{\label{fig:Pentacene} The top trace shows the (inverted)
  laboratory absorption spectrum (in arbitrary units) of the neutral
  PAH molecule pentacene (C$_{22}$H$_{14}$) prepared in a cold
  supersonic free jet expansion.  The lower trace shows the average
  absorption spectrum of interstellar translucent clouds (in
  normalized flux units) providing, for the first time, accurate upper
  limits for the abundances of interstellar PAHs in the optical. S/N
  refers to Signal-to-Noise, i.s.\ to interstellar, and 5 m{\AA} is the
  resolution (from Salama \etal 2011).}
\end{center}
\end{figure}

The diffuse interstellar absorption bands (DIBs) are ubiquitous
absorption features observed in the line of sight to stars that are
obscured by diffuse or translucent interstellar clouds.  Close to 500
bands have been reported to date in local and extragalactic
environments spanning from the near ultraviolet (UV) to the near
infrared (IR; Snow and McCall 2006).  Various candidates have been
proposed as carriers for the bands, ranging from impurity-doped dust
grains, to molecules, to atoms.  Today the DIBs are widely thought to
be associated with carbon molecules and ions (polycyclic aromatic
hydrocarbons [PAHs], carbon chains, fullerenes) that are part of an
extended size distribution of interstellar dust (Sarre 2006; Snow and
McCall 2006).  Astronomers are very interested in the molecules that
carry the DIBs, because these molecules may make up the largest cache
of organic material in the universe.  Recent advances in laboratory
techniques have made it possible to measure the spectra of cold
molecules and ions under conditions that are relevant to astrophysics
(Salama 2008).  As a result, accurate upper limits for the abundances
of PAH molecules along the lines of sight of translucent clouds have
been reported for the first time (Salama \etal 2011;
figure~\ref{fig:Pentacene}), while coincidences with naphthalene
(C$_{10}$H$_8^+$) and anthracene (C$_{14}$H$_{10}^+$) cation bands
have been tentatively reported for DIBs in the line-of-sight of Cernis
52 (BD +31 640), an early type reddened star behind the Perseus
supernova remnant that shows anomalous microwave emission
(Iglesias-Groth \etal 2008, 2010).  A near coincidence between a DIB
and a weak absorption feature of the diacetylene cation (C$_4$H$_2^+$)
was also detected in the average spectrum of 11 reddened stars
(Krelowski \etal 2010), while a coincidence was tentatively reported
between a weak DIB observed in the lines of sight of two objects and a
band associated with propadienylidene (H$_2$C$_3$) by Maier \etal
(2011).  All coincidences reported to date are tentative and point to
hydrocarbon molecules.

\subsubsection{Molecular clouds: molecular anions.}

Molecular anions were predicted many years ago to be abundant in the
interstellar medium (Sarre 1980; Herbst 1981).  Subsequent chemical
considerations by Terzieva and Herbst (2000) indicated that efficient
electron attachment occurs once a carbon chain reaches six atoms.
Molecular anions, however, have only recently been detected in space
through a combination of spectroscopic laboratory measurements and
observations of the molecular envelope of the star IRC$+$10216 and of
the dense Taurus molecular cloud TMC-1 (McCarthy \etal 2006).  McCarthy \etal
(2006) showed that the unidentified harmonic sequence found by
Kawaguchi \etal (1995) in IRC$+$10216 was C$_6$H$^-$.  The number of
detected molecular anions has increased dramatically as a result of
laboratory studies since then.  These include C$_4$H$^-$ in
IRC$+$10216 by Cerncharo \etal (2007), C$_8$H$^-$ in TMC-1 by
Br\"{u}nken \etal (2007) and in IRC$+$10216 by Remijan \etal (2007),
CN$^-$ in IRC$+$10216 (Ag\'undez \etal 2010), and C$_3$N$^-$ in the
same object (Thaddeus \etal 2008).  Sakai \etal (2010) detected
C$_4$H$^-$, C$_6$H$^-$ and C$_8$H$^-$ in a starless core of a
molecular cloud (Lupus-1A).  This wealth of observational data has
renewed interest in the effects of molecular anions on interstellar
chemistry (e.g., Walsh \etal 2009).

\subsubsection{Molecular clouds: polyaromatic hydrocarbons.}

These emission features, known as the unidentified infrared (UIR)
bands, were first discovered by Gillet \etal (1973) and attributed to
$\sim 10$~\AA\ size grains by Sellgren (1984).  These UIR bands are
now generally attributed to PAHs (Salama 2008 and references therein).
The features of this universal spectrum provide information on the
physical conditions in the emitting regions and the nature of the
molecular carriers.  Puget and Leger (1989) and Allamandola \etal
(1989) have proposed a model dealing with the UIR interstellar
emission features where PAHs are present as a mixture of radicals,
ions and neutral species.  The ionization states reflect the
ionization balance of the medium while the size, composition and
structure reflect the energetic and chemical history of the medium.
The proposed excitation mechanism of the IR bands is a one-photon
mechanism that leads to the transient heating of the PAHs by stellar
photons.  The IR emission bands are associated with the molecular
vibrations of molecular PAH species (discrete bands) and larger
carbonaceous grains (continuum-like structures).  In this model, PAHs
constitute the building blocks of interstellar carbonaceous dust
grains and play an important role in mediating energetic and chemical
processes in the interstellar medium. However, exploitation of these
features as astrophysical probes has been slow in developing because
the IR properties of PAHs under interstellar conditions were largely
unknown for at least twenty years after the bands were discovered.
During the past two decades, advanced experimental and computational
laboratory astrophysics programs have been developed to collect data
to test and refine the PAH hypothesis.  The information for hundreds
of PAH molecular species is now compiled in databases that allow
astronomers to quantitatively interpret their observations for a
variety of environments in our local galaxy and in extragalactic
environments (Malloci \etal 2007; Bauschlicher \etal 2010).

\subsubsection{Dark interstellar clouds.}

The chemistry that occurs in dark interstellar clouds, and especially
in the denser regions of such clouds known as cold cores, is an
unusual one.  Although organic molecules appear to grow in these
regions, they are very unsaturated and consist mainly of bare carbon
clusters, radicals of the C$_n$H and species with two hydrogen
atoms, such as c-C$_{3}$H$_{2}$.  This unusual pattern of growth
exists despite the fact that molecular hydrogen is the dominant
species in the gas and might be expected to hydrogenate the molecular
species into more saturated forms.  Although the basic mechanism for
the growth of unsaturated species in these cold regions was worked out
in the early 1970's (Watson 1973; Herbst and Klemperer 1973), the last
ten years have witnessed some very important laboratory work in
rounding out the picture.  Before then, it was thought that all
organic neutrals are produced via syntheses based entirely on
ion-molecule reactions, which synthesize precursor organic ions that
do not react with H$_{2}$, but instead come apart following
dissociative recombination reactions with electrons.  This picture was
incomplete because (a) there was little evidence concerning the actual
products of dissociative recombination and (b) the growth of neutral
species via reactions involving radicals and regular neutral species
was not considered because it was assumed not to occur at low
temperatures.  Thanks to laboratory astrophysics, the picture has
changed.  The products of dissociative recombination have now been
studied in the laboratory mainly by the use of storage rings in
Denmark, Sweden and Germany (Geppert and Larsson 2008, Petrignani \etal 
2009) in which molecular ions can be cooled down before reaction
with electrons.  Rapid radical neutral reactions have been studied
with Laval nozzles to temperatures down to near 10 K in laboratories
in Rennes, France and Birmingham, UK (Chastaing \etal 2001; Sims
2006).  Between these two sets of experiments, our knowledge of the
chemical mechanism of molecular growth in cold clouds has become much
more complete down to near 10~K.

\subsubsection{Pre-stellar cores.}

Pre-stellar cores have begun the evolutionary journey to form low- and
medium-mass stars.  They have temperatures of around 10~K and a gas
density of approximately $10^4$~cm$^{-3}$.  At this stage, the
collapse is isothermal because any heat developed is radiated away by
atoms and molecules.  The gaseous cores are dominated by hydrogen,
helium and deuterium, as many, if not most, of the heavier molecules
are depleted onto dust particles.  For example, the abundance of CO
drops precipitously towards the center of pre-stellar cores (Bacmann
\etal 2002, 2003).  The evidence is not as clear cut for other heavy
species but their low abundance is determined indirectly by detailed
simulations of the deuterium fractionation chemistry, which show a
huge fractionation effect in which deuterated isotopologues (e.g.,
H$_{2}$D$^{+}$) can be very abundant (Roberts \etal 2004).  Such a
large effect can only occur in the near absence of heavy reactive
species (Vastel \etal 2006).  The chemical simulations are based
heavily on experimental measurements of rate coefficients involving
deuterated species, such as those obtained in an ion trap (Schlemmer
\etal 2006).  The extent of depletion of species such as CO is
confirmed by measurements on the rate of desorption of this species
from dust particles, which is not rapid enough to keep a large amount
of material in the gas (\"{O}berg \etal 2009a).

\subsubsection{Hot cores and corinos.}

Hot cores and corinos are warm objects ($100-300$~K) associated with
low-mass protostars or young stellar objects of high mass.  In these
objects, the inventory of gas-phase organic molecules is quite
different from what it is in cold interstellar clouds, where the
molecules are mainly unsaturated (hydrogen-poor).  Instead, in hot
cores and corinos the organic molecules are much more
terrestrial-like and consist of simple alcohols, esters, ethers and
nitriles.  For many years, it was thought that gas-phase reactions
might produce these molecules, but laboratory experiments (Horn \etal 
2004) show that some of the reactions suggested do not occur or
are inefficient.  A new school of thought has arisen that the
molecules can be produced on the surfaces of dust particles and then
desorbed or evaporated into the gas.  Several suggestions were made
including the production of organic molecules on cold grains, mainly
via atomic addition reactions, and the production of these molecules
via radical-radical association reactions during the actual heating up
of a cold cloud into a hot core because of star formation (Herbst and
van Dishoeck 2009).  The production of radicals in this latter view
comes from photon bombardment of simple surface species such as
methanol, produced during the cold era (Garrod and Herbst 2006).
Although laboratory experiments have not completely ruled out the
idea that more complex species can be produced on cold surfaces, new
experiments seem to confirm the radical-radical hypothesis
(\"{O}berg \etal 2009b).

\subsubsection{Protoplanetary disks.}

Protoplanetary disks are dense objects of gas and dust that rotate
around newly-formed low-mass stars and may be the precursors of
solar-type systems.  Astronomers have obtained both rotational and
vibrational spectra of molecules in these disks and the molecular
inventory is a strong function of how far the molecules lie from the
central star and how high they lie off the midplane of the disk.  The
chemical models used to simulate the chemistry of these complex
objects owe much to laboratory astrophysics.  One recent success has
been an understanding of how some CO can be in the gaseous form at
temperatures well below its sublimation point despite the high density
of dust particles, which should guarantee that all CO should be in the
form of ice mantles.  Recent experiments on the photodesorption of CO
indicate that the efficiency per photon of photodesorption for UV
radiation is approximately 10$^{-3}$, which under the conditions of
protoplanetary disks can explain why CO can be detected in the gas
phase (\"{O}berg \etal 2009a; Hersant \etal 2009).  The recent
detection of acetylene (C$_{2}$H$_{2}$) and HCN in hotter regions near
the central star can be explained by chemical models that make use of
numerous laboratory studies of reactions at temperatures much higher
than 300 K (Ag\'{u}ndez \etal 2008a; Harada \etal 2010).

\subsubsection{Metal Hydride Spectra of L and T type Stars.}

Refractory hydrides such as FeH, CrH, CaH and MgH have recently been
found to be abundant in the atmospheres of M, S and L sub-dwarf-type
stars (Kirkpatrick 2005), as deduced from optical spectroscopy of
these objects.  In fact, the shift from prominent spectra of metal
oxides to metal hydrides is dramatic in the transition from M type to
L and T type sub-dwarfs (Burrows \etal 2002).  These brown sub-dwarfs,
especially the L types, are extremely important for the understanding
of planet formation, as they trace the intermediate stage between
stars that undergo nucleosynthesis and those that do not, i.e.,
planets.  Hydride spectra such as that of CrH are also excellent
tracers of very cool stellar atmospheres (Burrows \etal 2002) and may
be an important key in identifying planets.  None of this work would
have occurred without laboratory spectroscopic measurements, conducted
across a broad spectral range (e.g., Harrison \etal 2006).  Laboratory
studies of CrH, for example, have been carried out using a variety of
spectral techniques, including laser-induced florescence (Chowdhury
\etal 2005), Fourier transform infrared spectroscopy (Bauschlicher
\etal 2001) and millimeter/sub-mm direct absorption methods (Halfen
and Ziurys 2004).  Such hydrides are not stable under terrestrial
conditions and must be created by unusual synthetic techniques
involving laser ablation, hollow cathode sources and Broida-type
ovens.  Such work has provided not only wavelengths for spectral
identification, but other important physical properties such as
electronic state terms, energy levels and Einstein A coefficients,
which are essential for astrophysical interpretation of
stellar/planetary atmospheres (e.g., Burrows \etal 2005).  Not all
hydrides have been as well-characterized as CrH, however, and much lab
work needs to be done for species such as FeH and TiH.

\subsubsection{Comets}

Comets offer a unique opportunity to study organic astrochemistry,
knowledge of which till recently has largely been obtained from remote
astronomical observations and from laboratory simulations of the
formation and evolution of organic molecules in various
cosmically-relevant environments.  Comets are considered as the most
primitive objects in the solar system.  The composition and the
structure of cometary nuclei contain a record of the primordial solar
nebula at the time of their formation.  Cometary nuclei are made of
refractory solids and frozen volatiles.  The composition of the
volatile component is similar to that observed in dense molecular
clouds reflecting the close relationship between cometary materials
and interstellar icy grain mantles.  Hence, in comets the composition
of the volatile ices is largely dominated by H$_2$O ice (about
70-90\%) while other major components include CO, CH$_3$OH, CO$_2$
and H$_2$CO (Salama 1998; Bockel\'ee-Morvan \etal 2004; Fink 2009).

Comets are also thought to have been a major source for the volatile
ices on planetary bodies.  Thus, cometary ices constitute a link
between interstellar and solar system materials.  The captured
materials from sample return missions provide new insight into the
formation of our solar system.  The {\it Stardust} mission flew
through the near-nucleus coma of comet 81P/Wild 2 on 2 January 2004,
swept up material using aerogel collectors and returned these samples
to Earth on 15 January 2006.  {\it Stardust} is the first space
mission to bring back solid material from a known body other than the
Moon.  One of the key questions that the {\it Stardust} samples
addressed is the origin of primitive organic matter in the solar
system.  After the recovery of the Sample Return Capsule, the returned
material from {\it Stardust} was examined in the laboratory with the
goal to determine the nature and amount of the returned samples
(Brownlee \etal 2006; H\"orz \etal 2006; Sandford \etal 2006; McKeegan
\etal 2006; Keller \etal 2006; Flynn \etal 2006; Zolensky \etal 2006).

Laboratory astrophysics played a crucial role in the optimization of
the knowledge gained from the return of these extraterrestrial
samples.  An impressive battery of advanced laboratory astrophysics
techniques was called upon to help decipher the information contained
in the returned samples.  The techniques involved transmission
electron microscopy (TEM), Raman and Fourier transform infrared (FTIR)
spectroscopy, time of flight secondary ion mass spectrometry
(TOF-SIMS) and scanning electron microscopy using energy-dispersive
X-ray (SEM-EDX) analyses, among others.  These laboratory studies show
the highly heterogeneous nature of the collected cometary grains and
reveal an interesting distribution of organic material, including the
detection of amide, carboxy, and alcohol/ethers groups (e.g., Cody
\etal 2008; Clemett \etal 2010) and the amino acid glycine (Elsila
\etal 2009).  While concerns remain as to the organic purity of the
aerogel collection medium and the thermal effects associated with
hypervelocity capture, the majority of the observed organic species
appear indigenous to the impacting particles and are hence of cometary
origin.  Additionally, though the aromatic fraction of the total
organic matter present is believed to be small, it is notable in that
it appears to be N rich.  Spectral analysis in combination with
instrumental detection sensitivities suggest that N is incorporated
predominantly in the form of aromatic nitriles (R-CN; Clemett \etal
2010).

\subsubsection{Exoplanetary atmospheres.}

In addition to mass and radius, other properties of an exoplanet
(e.g., temperature and composition) can be determined using spectral
changes during eclipses.  Since the first thermal emission from an
exoplanet was discovered (Charbonneau \etal 2005; Deming \etal 2005),
a number of other firsts have been reported.  One was the discovery of
strong evidence for water vapor in the atmosphere of an exoplanet
(Tinetti \etal 2007).  Signatures of water and carbon dioxide are now
observed both in absorption and emission in a number of exoplanet
atmospheres (Charbonneau \etal 2008; Knutson \etal 2008; Grillmair
\etal 2008).  The measurement of temperature differences between the
night-side and day-side of a tidally-locked close-in hot Jupiter has
emphasized the role of stellar radiation on the planetary atmosphere
(Knutson \etal 2007).  While the composition of exoplanets at or above
Jupiter in mass is not in doubt (they must be gas giants composed of
hydrogen and helium), spectroscopy is needed to determine the
composition of the smallest planets discovered to date (the so-called
"Super-Earths"), since they may also be rocky (like the Earth) or icy.
Near infrared spectroscopy of one such Super-Earth has ruled out
hydrogen gas, unless there are thick clouds (Bean \etal 2010).  These
observations are also consistent with the presence of hot water vapor
(steam), in which case the planet might have an icy rather than rocky
core.  All these discoveries rely heavily on spectroscopic modeling of
the stellar atmosphere (Hauschildt \etal 2009), as well as the
exoplanet atmosphere (Seager \etal 2005; Miller-Ricci \etal 2009;
Kaltenegger and Sasselov 2010) and thus on the supporting laboratory
astrophysics and atomic and molecular line data (Castelli and Kurucz
2004; Rothman \etal 2005).



\subsection{Condensed matter physics}





\subsubsection{Outer solar system ice.}

The connection between the ISM and solar systems profoundly influences
our understanding of the birth and death cycles of stars in our
Galaxy.  Present models of star formation suppose that interstellar
amorphous ice grains accreted to form the outer rim of the solar
system from Oort Cloud to Kuiper Belt Objects (KBOs; Jewitt 1999).
Based on these models, outer solar system icy bodies with surface
temperatures $<100$~K form amorphous ices.  At these temperatures
amorphous ices remain stable over the lifetime of our star ($4.5
\times 10^9$~yr).  Galilean icy satellites like Europa at 5~AU with
surface temperatures $\sim 120$~K are crystalline.  Beyond Jupiter,
the rest of the outer solar system icy bodies have equilibrium surface
temperatures $<100$~K and hence are expected to contain amorphous
ices.  These are: Saturnian icy moons and rings at 10~AU ($\sim
100$~K), Uranian satellites around 20~AU, trans Uranian objects and
KBOs ($\sim 50$~K) at 40~AU from the Sun and the Oort Cloud ($\sim
30$~K) spanning up to several thousands of AU towards the local ISM.

\begin{figure}
\begin{center}
\includegraphics[angle = 0, height=0.35\textheight]{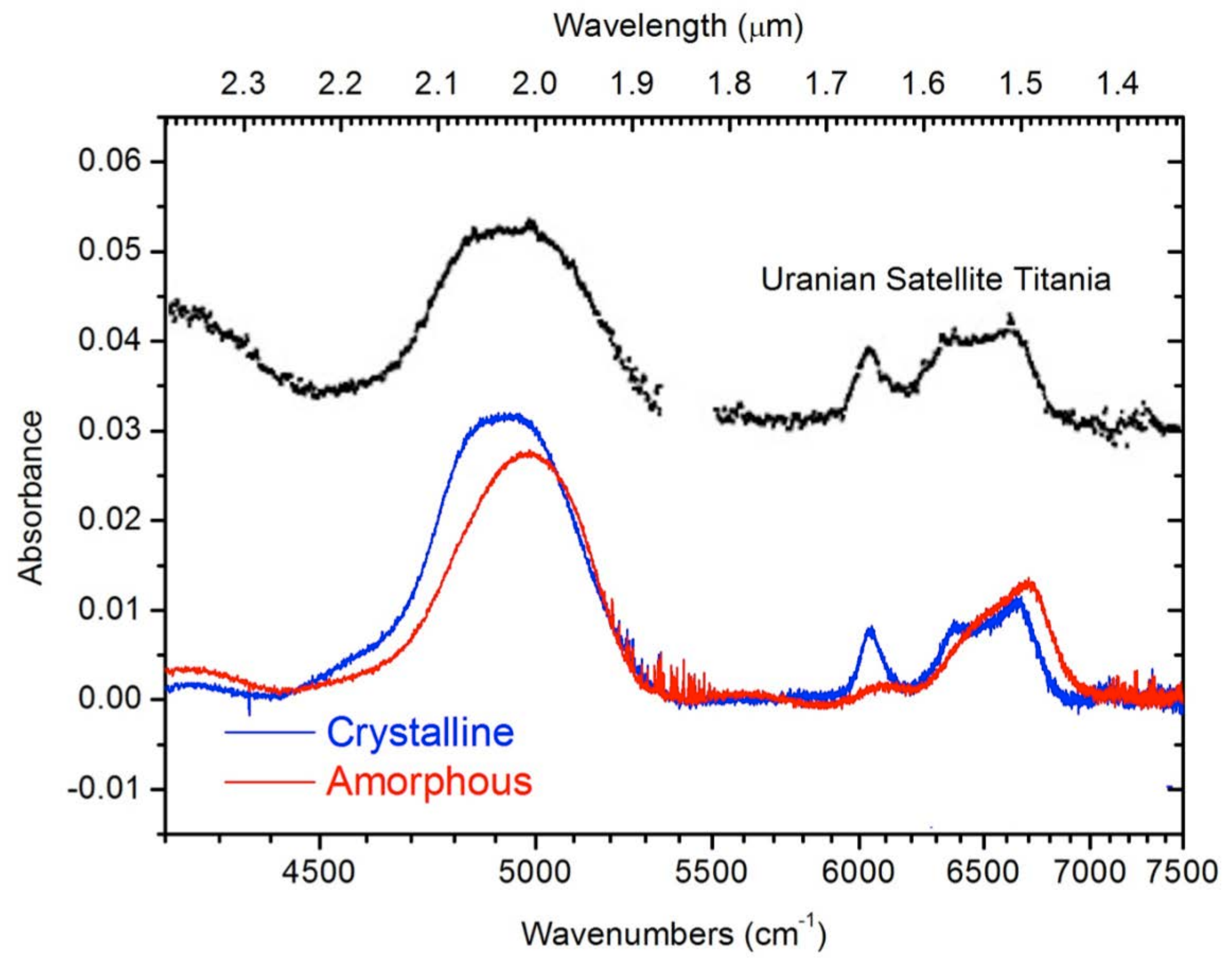}
\caption{\label{fig:Uranian} Observed near infrared spectrum (black)
  of the Uranian satellite Titania (Grundy \etal 2006), compared with
  the laboratory spectra of amorphous (red) and crystalline (blue)
  water ice at 10~K (from M. Gudipati, private communication, and in
  excellent agreement with the published results of Grundy and Schmitt
  1998 and Mastrapa and Brown 2006).  The absorption around 6000
  cm$^{-1}$ (1.65~$\mu$m) is the most prominent feature of the
  crystalline ice that is not present in amorphous ices.  Other subtle
  spectral differences can also be seen in the laboratory spectra.}
\end{center}
\end{figure}

Near infrared spectroscopic studies carried out in the laboratory
(Grundy and Schmitt 1998) revealed that amorphous ices show
significantly different absorption features in this region compared to
the crystalline ices, as shown in the lower part of the
figure~\ref{fig:Uranian}.  Recent spectrally resolved observations
showed that the surface ices of trans Uranian icy bodies (Grundy \etal
2006), trans Neptunian objects (TNOs; Trujillo \etal 2007; Demeo \etal
2010) and KBOs (Jewitt and Luu 2004) are significantly crystalline,
based on comparison of these spectra with the laboratory data.  Some
recent models attribute the surface crystallinity to micrometeorite
impacts (Porter \etal 2010).  This counter-intuitive observation,
supported by laboratory data, has opened up a new chapter in our
understanding of the evolution of icy bodies in the solar system and
in the ISM.

Recently it has also been shown in the laboratory (Zheng \etal 2009)
that the crystallinity of ice at $>50$~K is not destroyed or altered
to amorphous-like form by electron irradiation under conditions
similar to those that exist on KBOs and comets originating from them.
However, it is still unclear how the amorphous ice grains in the
interstellar medium are converted into the crystalline surface layer
of KBOs and whether the subsurface of KBOs is amorphous or crystalline
and hence the comets originating from them.  More laboratory studies
are needed in order to resolve this amorphous-crystalline puzzle that
connects the ISM with the Outer Solar System.

\subsubsection{Cometary ice, chemistry and the origins of life?}

One of the working postulates of the origins of life is that cometary
impacts brought organic chemicals and water to Earth (Whittet 1997;
McClendon 1999; Matthews and Minard 2006).  Comets are expected to
retain the interstellar amorphous ice structure.  Organic rich comets
have been found to be highly porous (Richardson \etal 2007; A'Hearn
2008).  One of the outstanding questions is whether the delicate
building blocks of life survived the comet impacts on Earth.  With
very similar ice grain composition between comets (Crovisier \etal
2004) and interstellar ice grains (Gibb \etal 2004), these ices are
dominated by H$_2$O, followed by CO$_2$, CO, methanol (CH$_3$OH),
hydrocarbons, nitrogen containing molecules (NH$_3$ and derivatives)
and sulfur containing molecules such as OCS, as well as minerals such
as silicates.  All these ingredients (H, C, N, O, S and minerals
containing these elements) are essential for all forms of life on
Earth as is also phosphorous (P) which is yet to be positively
detected in comets.  Laboratory studies using the primitive molecules
mentioned above and simulating the composition of comet and
interstellar ice grains have shown that radiation processing of these
ices indeed produced building blocks of life upon subsequent heating
to evaporate ice (Dworkin \etal 2001; Bernstein \etal 2002; Deamer
\etal 2002; Mu\~{n}oz Caro \etal 2002; Elsila \etal 2007; Nuevo \etal
2009).  These laboratory studies are critical, corroborating one of
the possible origins of life on Earth.

Recent laboratory studies have also enhanced our understanding of the
mechanisms involved in the radiation processing of organic molecules
in ices that result in the formation of complex building blocks of
life (Gudipati 2004; Gudipati and Allamandola 2004, 2006; Bouwman
\etal 2010).  Using PAHs as probes embedded in ices, these laboratory
studies have shown that radiation induced ionization of PAHs is an
important first step, forming electron and PAH radical cation pairs in
ice, which subsequently lead to the formation of oxidized PAHs.  These
laboratory studies have opened up a new understanding of chemistry in
ices, involving charged species, bringing us one step closer to
understanding how ices evolve under irradiation.  Charged ice grains
behave differently compared to their neutral counterparts due to
strong long-range Coulomb forces.  The implications of these studies
to astrophysics and planetary sciences are slowly unfolding (Kalvans
and Shmeld 2010).

\subsection{Plasma physics}

\subsubsection{Accretion disks and magnetorotational instability.}

Accretion disks form in various astrophysical systems including young
stars, protostars and some CVs.  The accretion disk forms because the
accreting matter brings substantial angular momentum, which must be
transported away in order for the matter to move inward.  Physical
viscosity is far too small and it is generally believed that
magnetohydrodynamic (MHD) turbulence is responsible for the angular
momentum transport.  At present, the leading candidate to drive such
turbulence is magnetorotational instability (MRI; Balbus and Hawley
1991, 1998), with the turbulence itself produced by secondary
instabilities that convert the structures generated by the MRI into
multiscale turbulent fluctuations (Pessah 2010; Pessah and Goodman
2009).  A major challenge in coming to understand the MRI comes from
the limitations of various approaches.  Analytic and semi-analytic
theories have made great progress (Julien and Knobloch 2010) but
always struggle to define turbulent states.  The astrophysical systems
have very large Re and Rm, where Re is the usual viscous Reynolds
number and Rm is the analogous magnetic Reynolds number,
characterizing how slowly the magnetic structures are dissipated by
resistive heating of the plasma.  Numerical simulations cannot reach
the astrophysical regime, being very limited in Re and having values
of Rm that can be larger but remain limited.  The past decade has seen
laboratory experiments that reported observation of the MRI (Stefani
\etal 2006) and a helical variant (Sisan \etal 2004).  These
experiments complement the simulations, having larger values of Re
than the simulations can produce but smaller values of Rm.
Experiments to date have been performed with a liquid metal conducting
fluid, a system well described by MHD theory.  The combination of
experiments, simulations and observations now provides a more complete
set of information for theoretical work that seeks to identify the
important scaling parameters and to provide a unified understanding of
MRI across all regimes.

\subsubsection{Young stellar objects: jet structure.}

Many open questions remain in the study of jets emanating from young
stellar objects (YSOs; Reipurth and Bally 2001).  These
non-relativistic beams of hypersonic plasma are likely magnetized and
are known to cool effectively via radiation losses.  Of particular
interest for astrophysics are issues related to the internal jet
structure.  Are the hypersonic beams of plasma (hyperfast mode in the
case of MHD jets) structurally smooth or inherently inhomogeneous?

Depending on the stability conditions of the jets this question speaks
directly to the launch mechanisms of the jets as structurally smooth
jets, implying time independent conditions at the central engine
launching the jet.  Recent observations using {\it Hubble Space
  Telescope} and other high-resolution platforms indicate that jets
may contain significant sub-radial structure ($\delta x < r_{\rm
  jet}$), which implies that jets may be inherently heterogeneous or
``clumpy'' phenomena (Hartigan and Morse 2007; Hartigan \etal 2011).

Recent experimental studies have attempted to explore this issue by
developing platforms that can create steady jet beams as a starting
point for further work.  Of particular note have been the pulsed power
studies of Lebedev \etal (2002) who were able to develop stable
hypersonic radiative jets. These jets have high Mach numbers ($M \sim
20$) and have been shown to propagate without disruption over long
distances, achieving aspect ratios of 10 or more.  Shorter duration
jets have also been produced in a number of studies (Foster \etal
2005).  In some cases these experimental platforms have allowed
researchers to explore the interaction of jets with large-scale
obstacles (Hartigan \etal 2009).  This is an astrophysically relevant
issue as jets from young stars are observed, in some cases, to be
deflected by clumps or clouds in their path.  Deflection of jets by
winds induced by the motion of the jet source through a background has
also been observed and this process has been studied in the laboratory
as well (Lebedev \etal 2004).

Thus experimental studies to date have shown that stable hypersonic
jets can propagate over long distances and that even when interacting
with side winds the jets are not fully disrupted.  Future studies
should focus on the generation and propagation of ``clumps'' within the
beams.

\subsubsection{Young stellar objects: magnetized jets.}

Astrophysical jets are believed to form via a combination of
accretion, rotation and magnetic fields (Pudritz \etal 2007).  The
central engine may be a star, a compact object like a black hole or
surrounding accretion disk.  YSO jets are also believed to form via
magnetized accretion disks and many open issues remain concerning both
the magneto-centrifugal launch processes and the propagation of the
magnetized jet at large distances from the central engine.  In general
theorists expect the fields to be strong to moderate as characterized
by the plasma beta which is the ratio of gas (g) to magnetic (B) 
pressures $\beta = P_{\rm g}/P_{\rm B} \leq 1$.

Using a planar magnetized coaxial plasma gun, Bellan (2005) and Bellan
\etal (2005) have developed a platform to study MHD jet launching.
The premise behind the experiments is that the basic magnetic dynamics
near a star-disc system, namely the winding up of poloidal field lines
generated by the central disc$+$star rotation, can be simulated in the
laboratory by applying a voltage across coaxial electrodes in the
presence of a background colloidal field.  The magnetic helicity
injection with these boundary conditions leads naturally to collimated
unstable plasmas whose dynamics may be indicative of disc driven jets
and plasmoids.

A second approach to the study of magnetized YSO jets comes from
experiments using radial plasma sources, which consist of a pair of
concentric electrodes connected radially by thin metallic wires or a
thin foil (Lebedev \etal 2005; Ciardi \etal 2009).  Resistive
heating of the wires or foil produces a plasma. If wires are used,
when they break, toroidal flux from below drives a magnetic bubble
($\beta < 1$) and a collimated jet forms on axis.  The jet goes
unstable due to kink modes and evolves into a series of hypersonic
clumps. When a foil disk is used, the process becomes episodic with a
series of magnetic bubbles and jets forming one after the other.

Laboratory studies of magnetized jets relevant to YSOs have offered a
new window into the three dimensional (3-D) dynamics of magnetized
plasma systems.  Helicity injection and kink mode instabilities have
been followed in ways that already demonstrate new pathways of jet
evolution not previously considered in analytic or computational
studies.

\subsubsection{Young stellar objects: radiative jets.}

Along with magnetic fields, radiative cooling is another important
process occurring in YSO jets.  In this context radiative cooling
means that optically thin emission from shock excited atoms and ions
will carry away a significant amount of energy from the system.
Systems are radiatively cooling when the timescale for energy loss
($t_{\rm cool} = e/\dot{e}$) is less than the characteristic
hydrodynamic timescale ($t_{\rm h} =L/c$) where $e$, $\dot{e}$, $L$
and $c$ are thermal energy, thermal energy loss rate, system scale and
speed of sound, respectively.  As has been shown in numerous studies,
radiative cooling will produce dramatic differences in the evolution
of jet systems compared with adiabatic flows (Blondin \etal 1990).  In
particular, the collapse of bow shocks onto the jet body will occur
when thermal pressure generated at the shock is removed via the
radiative cooling.  Resolution issues hamper numerical simulations of
jet dynamics with radiative cooling.  A detailed understanding of
instabilities at cooling bow shocks, for example, has not yet been
achieved.

Experiments have produced radiative jets by creating radially
imploding plasmas having an axial velocity component.  In early work
lasers were used to irradiate conically shaped targets (Farley \etal
1999). Work using wire arrays (Lebedev \etal 2002) created stagnation
of plasma flow on the axis of symmetry, forming a standing conical
shock effectively collimating the flow in the axial direction. This
scenario is essentially similar to that discussed by Canto \etal
(1988) as a purely hydrodynamic mechanism for jet formation in
astrophysical systems.  In both types of experiments, the diameter of
the jet decreased with increasing atomic number, providing direct
evidence of radiative cooling.  In a more recent experiment, a
ring-shaped laser spot was employed to produce an imploding Cu plasma,
generating a jet that penetrated into adjacent gas (Tikhonchuk \etal
2008).  Analysis showed the experimental parameters to be rigorously
well scaled to astrophysical cases. Structure was seen in the shocked
ambient medium, providing evidence relevant to the instabilities at
cooling bow shocks.

Thus experiments have produced radiative hypersonic jets in laboratory
settings, allowing existing theories of jet dynamics to be explored
and opening up new domains of investigation beyond the reach of
existing analytic methods and simulations.

\subsubsection{Hydrodynamic stability of protoplanetary accretion disks.}

It is widely accepted that MRI plays an important role in generating
turbulence that transports angular momentum outward in accretion disks
(Balbus and Hawley 1998).  The electrical conductivity of portions of
protoplanetary disks is thought to be so low, however, that the
magnetic field is not well coupled and that MRI cannot operate.  It
was proposed that hydrodynamic Keplerian flow can be unstable to
finite amplitude perturbations and that this could lead to angular
momentum transport.  Recent laboratory experiments of hydrodynamic
Keplerian flow between two cylinders have found no evidence of such
instability, up to Reynolds numbers of $2 \times 10^6$ (Ji \etal
2006; Schartman \etal 2011).  This negative result weighs against
instability of Keplerian flow as an angular momentum transport
mechanism in accretion disks and encourages us to look for other
mechanisms.

\subsubsection{Equation of state for planetary interiors.}

Present-day observations of planets can determine only their mass,
size and perhaps surface composition.  One wants to know much more
such as the structure of the planet, the properties of the interior
matter and whether gas giant planets required an ice and rock core to
nucleate their formation.  One seeks a self-consistent model in which
the local density at some radius is determined by the materials
present and the local pressure, while the integrated density profile
within the observed planetary radius corresponds to the mass of the
planet (see the recent review by Fortney and Nettelmann 2010).  

The relations between density, pressure and other thermodynamic
quantities are the equations of state (EOS).  For the specific case of
Jupiter, Saumon and Guillot (2004) have shown that the uncertainties
in the EOS are the dominant limitation to understanding the structure
of the planet.  Laboratory measurements are essential to advance this
field; the relevant EOS theory is difficult, both intrinsically and
with regard to choosing appropriate assumptions.  The first two
first-principles models, using very similar methods, implied different
amounts and distributions of heavy elements in Jupiter (Fortney \etal
2009).  Laboratory data have been used to adjust other EOS models
(Fortney and Nettelmann 2010) and researchers are actively acquiring
more data (Eggert \etal 2008; Hicks \etal 2009).

\begin{figure}
\begin{center}
\includegraphics[angle = 0, height=0.35\textheight]{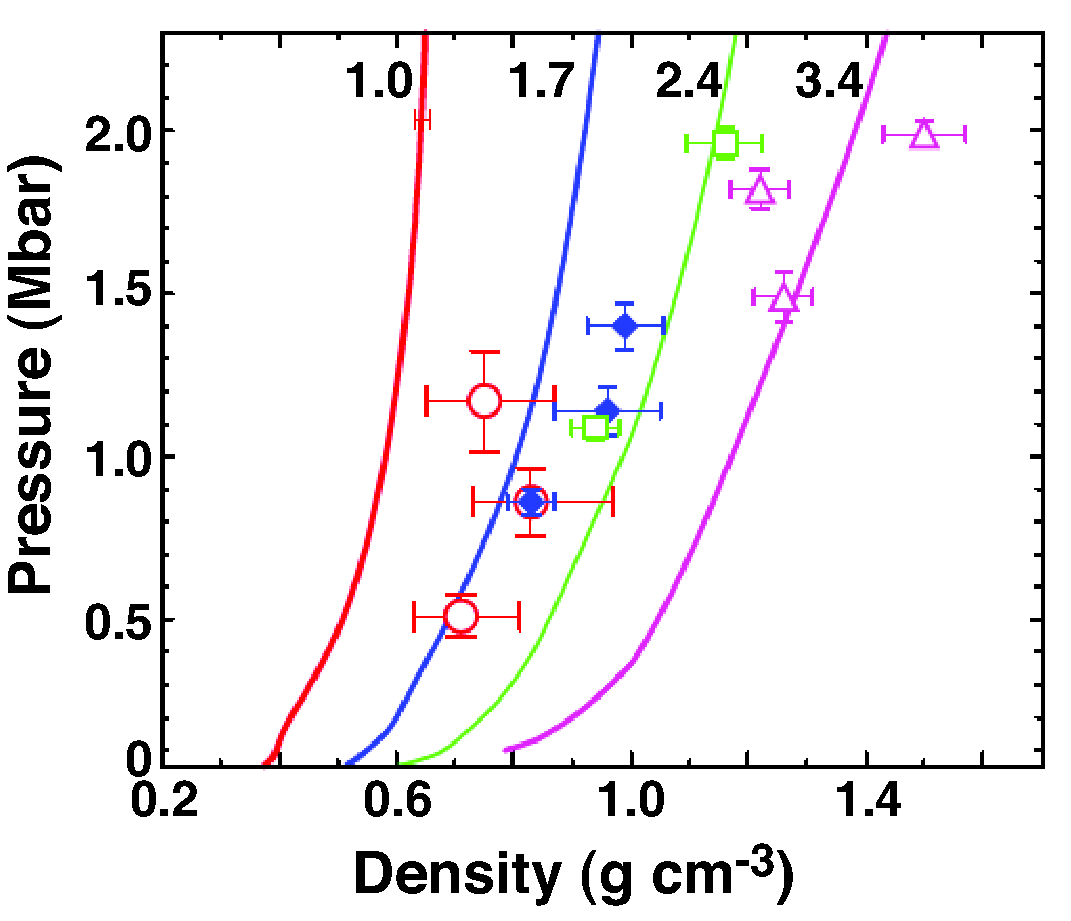}
\caption{\label{fig:PSF:Pla:EOS} Equation of state data and theory for
  He.  The He is first precompressed, in a diamond-anvil cell, by the
  factor that labels each curve and then is shocked to high pressure,
  which allows one to access densities and pressures relevant to
  gas-giant planets. The density of the He before precompression is
  the zero-pressure density of cryogenic He (0.123~g~cm$^{-3}$).  This
  figure, adapted from Fortney \etal (2009), shows experimental data
  from Eggert \etal (2008) and first-principles theory from Militzer
  (2009).}
\end{center}
\end{figure}

Figure~\ref{fig:PSF:Pla:EOS} shows a comparison of state-of-the-art
data and theory for the pressure and density produced by a strong
shock wave in cryogenic He after compression to some initial density
(adapted from Fortney \etal 2009).  The theory curves were produced by
first-principles calculations using a combination of
path-integral-Monte-Carlo and density-functional-theory,
molecular-dynamics calculations (Militzer 2009).  The data points were
inferred from direct measurements of shock velocity in He and in
Quartz, using standard techniques (Eggert \etal 2008).  There is
reasonable agreement between data and theory for high
pre-compressions, but not for low pre-compressions.  This indicates
that more work is needed to fully understand the compression of He.

Further advances are needed in order to obtain a
fully validated account of the properties of He at relevant densities
and pressures.  Progress in these areas will complement improved
measurements of the abundance of oxygen and of the detailed
gravitational field structure by the {\it Juno} orbiter (Bolton 2006).
Other space missions will identify hundreds of additional Neptune-like
to Jupiter-like planets.  EOS research during the last decade has
provided data that constrain planetary models and demonstrated
methods that will produce further data going forward.

\section{Stars and stellar evolution}
\label{sec:SSE}

Stars and stellar evolution covers ``the Sun as a star, stellar
astrophysics, the structure and evolution of single and multiple
stars, compact objects, supernovae, gamma-ray bursts, solar neutrinos,
and extreme physics on stellar scales'' (Blandford \etal 2010a).

\subsection{Atomic physics}

\subsubsection{Solar and stellar abundances of rare earth elements.}
\label{subsubsec:Solar}

Accurate heavy-element abundances have recently been determined
for the rare earth elements in the Sun and in old, metal-poor Galactic
halo stars.  These abundances provide insight into the nature of the
earliest stellar generations and element formation in the Galaxy.  The
updated values are the result of extensive new laboratory data for
atomic transition probabilities.  Data have been published for
numerous spectra including:
La~\textsc{ii} (Lawler, Bonvallet and Sneden 2001a), 
Ce~\textsc{ii} (Palmeri \etal 2000; Lawler \etal 2009), 
Pr~\textsc{ii} (Ivarsson, Litzen and Wahlgren 2001), 
Nd~\textsc{ii} (den Hartog \etal 2003), 
Sm~\textsc{ii} (Xu \etal 2003; Lawler \etal 2006), 
Eu~\textsc{i, ii, }and\textsc{ iii} (den Hartog \etal 2002; 
Lawler \etal 2001c), 
Gd~\textsc{ii} (den Hartog \etal 2006),
Tb~\textsc{ii} (den Hartog \etal 2001; Lawler \etal 2001b), 
Dy~\textsc{i} and \textsc{ii} (Curry \etal 1997; Wickliffe \etal 2000), 
Ho~\textsc{i} and \textsc{ii} (den Hartog \etal 1999; Lawler \etal 2004), 
Er~\textsc{ii} (Lawler \etal 2008),  
Tm~\textsc{i} and \textsc{ii} (Anderson \etal 1996; 
Wickliffe and Lawler 1997), 
Lu~\textsc{i, ii, }and \textsc{iii} (den Hartog \etal 1998; 
Quinet \etal 1999; Fedchak \etal 2000), 
Hf~\textsc{ii} (Lawler \etal 2007), 
Os~\textsc{i} and Ir~\textsc{i} (Ivarsson \etal 2003), 
Pt~\textsc{i} (den Hartog \etal 2005), 
Th~\textsc{ii }and\textsc{ iii} (Bi\'{e}mont \etal 2002; 
Nilsson \etal 2002a) and
U~\textsc{ii} (Lundberg \etal 2001; Nilsson \etal 2002b).

\begin{figure}
\begin{center}
\includegraphics[angle = 0, height=0.35\textheight]{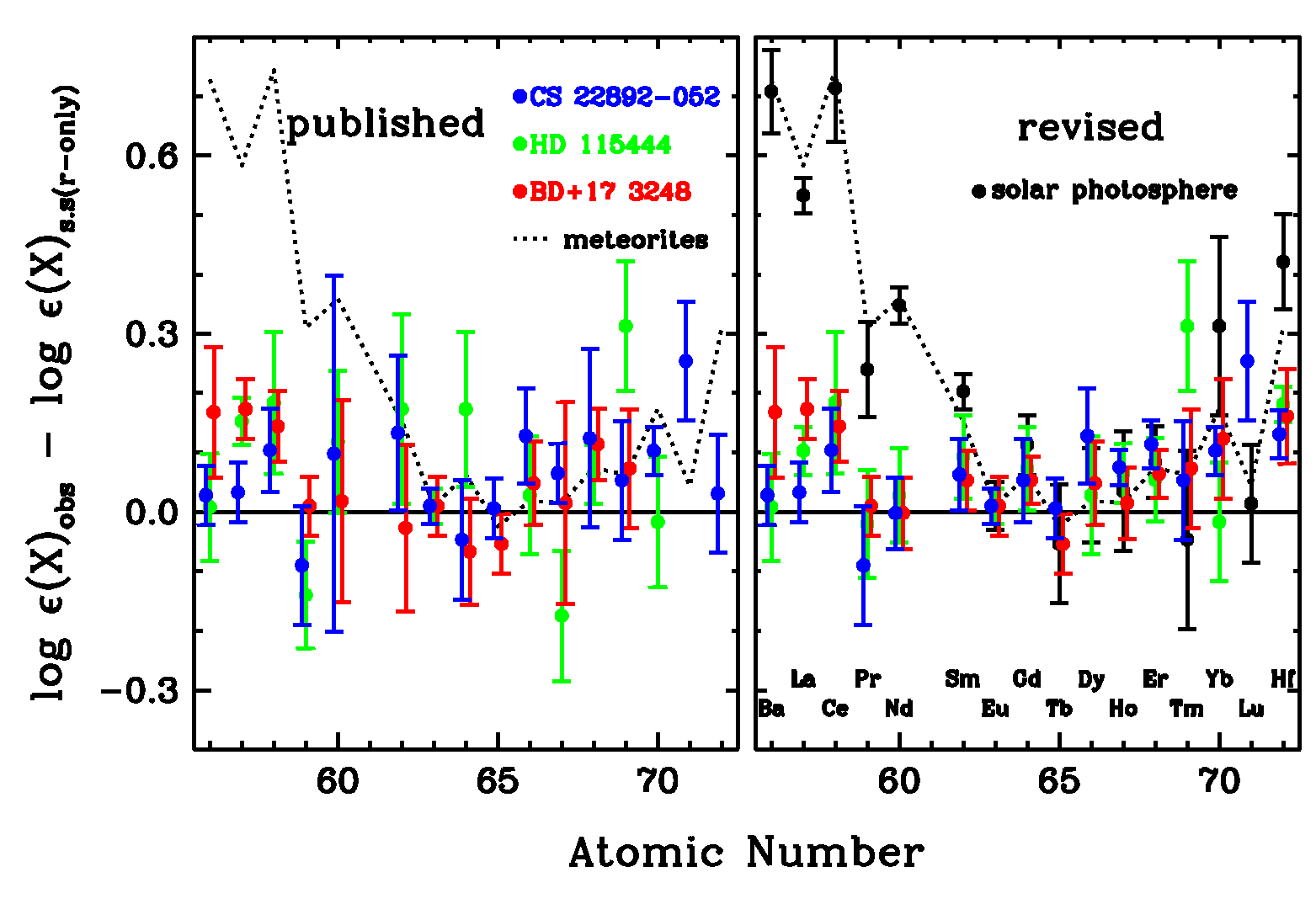}
\caption{\label{fig:SSE3-1Fig1} Shown are the observed abundances
  $\epsilon({\rm X})_{\rm obs}$ relative to the solar system r-process
  only abundances $\epsilon({\rm X})_{\rm s.s.(r-only)}$ as a function
  of atomic number for element X.  The left panel results are based
  upon older published atomic data.  A large scatter is readily
  apparent from star-to-star and with respect to the dashed horizontal
  line, representing the solar system r-process only line (normalized
  to the element Eu).  The right panel shows the relative abundances
  with the newly determined laboratory data (labeled ``revised'').
  For those elements discussed in section~\ref{subsubsec:Solar} with
  new atomic data, the star-to-star scatter has largely disappeared
  and the new abundances are also consistent with the solar r-process
  only abundances.  Adapted from den Hartog \etal (2006) and Sneden
  \etal (2008).}
\end{center}
\end{figure}

These new transition probabilities have culminated in more precise
solar and stellar abundances of Pr, Dy, Tm, Yb and Lu (Sneden \etal
2009).  As a result, it is now conclusively demonstrated that the
abundance pattern for the heaviest elements in the oldest metal-poor
halo stars is consistent with the relative solar system abundances for
rapid neutron capture (r-process) only elements.  This indicates that
the r-process that operated in the early Galaxy, soon after the first
stars formed, must share some common features with -- and perhaps is
identical to -- the r-process that operates now.  Thus, the
star-to-star relative abundances of these elements should be the same
and also consistent with the solar system values.  This can be seen in
figure~\ref{fig:SSE3-1Fig1} where the abundances of three metal-poor
halo stars (CS 22892-052, BD +17 3248 and HD 115444) are compared with
meteoritic and solar system r-process abundances (den Hartog \etal
2006, Sneden \etal 2008).  Additional elements have been measured
since the publication of that figure and the abundance analyses have
now been extended to more stars (see e.g., Sneden \etal 2009).

\subsubsection{The solar abundance problem.}
\label{SSE:At:SolAbu}

Three-dimensional, time-dependent, hydrodynamical solar atmosphere
models are a remarkable computational achievement of the past decade.
These models require significantly lower abundances of C, N, O and Ne
to match photospheric spectra (Asplund \etal 2004), compared with
previous results based on one-dimensional, static non-
local-thermodynamic-equilibrium (non-LTE) models (e.g., Vernazza \etal
1976).  However, the new abundances do not agree with helioseismology
observations (Bahcall \etal 2005a).  Christensen-Dalsgaard \etal
(2009) and others have suggested that increased opacity could bring
the helioseismology models back into agreement with observations, but
that would require about a 30\% increase in atomic abundances at the
base of the convection zone and a few percent in the solar core.  The
convergence of various opacity calculations over the past decade
(including large contributions from atomic theory and experiment) is a
considerable success and thus an opacity increase as large as 30\% may
not be reasonable (Christensen-Dalsgaard \etal 2009).  Furthermore,
new Z-pinch experimental tests of the iron opacity, at conditions
approximating the base of the solar convection zone, show good
agreement with the most recent and advanced opacity models (see
section~\ref{SSE:Pla:Opa}; Bailey \etal 2007; Mancini \etal 2009).
Additional experiments are needed to test the opacity models under
other relevant physical conditions (see Bailey \etal 2009).  For now,
opacities alone do not appear to resolve all the problems fitting
helioseismology data and the solution may well lie elsewhere, such as
in the EOS (Lin \etal 2007; Basu 2010).

\subsubsection{The solar corona.}

Despite decades of research, we still do not understand how the
temperature of the solar atmosphere rises from $\sim 6000$~K at the
photosphere to more than $10^6$~K in the corona.  Fe~\textsc{xvii} is
an important system for studying the corona, producing some of the
strongest lines seen.  It is formed near the peak temperatures of
active regions and emits a number of useful diagnostic line ratios for
temperature, density and opacity.  Resonant line scattering in the
strongest solar coronal X-ray line (Fe~\textsc{xvii} $3d~^1P_1$ to
ground $^1S_0$, known as 3C) has long been thought to contribute to
its observed weakness relative to the nearby Fe~\textsc{xvii}
$3d~^3D_1$ to $^1S_0$ line (known as 3D).  Even at relatively low
optical depth, resonant line scattering could in principle also
account for morphological effects in images of loop structures (Wood
and Raymond 2000).  If this were the case, efforts to increase spatial
resolution of solar coronal imaging instruments to $\sim 0.1$~arcsec
might not be worthwhile.  Theoretical calculations of the 3C/3D line
ratio have until recently been significantly larger than any of the
solar observations.  Over the past decade a number of experimental
measurements (Brown \etal 1998; Laming \etal 2000; Brown \etal 2001;
Gillaspy \etal 2011) and ongoing theoretical work (e.g., Doron and
Behar 2002; Loch \etal 2006; Chen 2008) have produced convergence on
the appropriate line ratio for comparison with observations.  With the
Fe~\textsc{xvii} 3C/3D line ratios on solid ground, Brickhouse and
Schmelz (2006) showed that the solar X-ray corona is optically thin in
Fe~\textsc{xvii} 3C and, by extension, in all the coronal lines.  The
blurring seen in some images (e.g., Fe~\textsc{xv} from the {\it
  Transition Region and Coronal Explorer} [{\it TRACE}] satellite) is
thus the result of unresolved spatial structure near the peak
temperature.  Efforts to observe the solar corona at still higher
spatial resolution are thus warrented.

\subsubsection{O star winds.}

Advances in our understanding of the elemental evolution of the cosmos
has come about from spectroscopic observations of O stars carried out
using {\it Chandra} and the {\it X-ray Multi-Mirror Mission - Newton}
({\it XMM-Newton}) coupled with new laboratory astrophysics data.  The
powerful radiatively driven winds of O stars are important sources of
chemical enrichment in the universe.  Recent analyses of UV P Cygni
profiles and X-ray emission line profiles have been used to determine
mass loss rates (Fullerton \etal 2006; Cohen \etal 2010).  These
studies used the best available wavelengths (accurate to a few m\AA)
and a relatively complete database of important X-ray emission lines
coupled with data on relative line strengths in coronal plasmas, in
order to accurately account for blended complexes of Doppler broadened
emission lines.  The mass-loss rate from O stars was found to be a
factor of a 3 to 6 less than previously thought (Cohen \etal 2010), a
result deriving from recent improvements in atomic data from
laboratory and theoretical calculations.  This changes our
understanding of chemical enrichment of galaxies, especially during
their early starburst phase.

\subsubsection{Type Ia supernovae.}

Type Ia supernovae (SNe) are used as standard candles to study dark
energy and the expansion of the universe.  {\it Chandra} and {\it
  XMM-Newton} X-ray observations of young supernova remnants (SNRs)
have deepened our understanding of these standard candles.  X-ray
observations of young SNRs in the Milky Way and the Magellanic Clouds
offer a detailed view of Type Ia supernova (SN) ejecta and provide
invaluable constraints on the physics of these explosions and the
identity of their progenitor systems.  Utilizing public domain atomic
data, it is now possible to model this X-ray emission and distinguish
SNRs resulting from bright and dim Type Ia SNe.  This technique has
been validated by the detection and spectroscopy of SN light echoes
for the Tycho SNR (Badenes \etal 2006; Krause \etal 2008) and SNR
0509-67.5 in the Large Magellanic Clouc (LMC; Badenes \etal 2008a;
Rest \etal 2008).  A key advantage of these X-ray studies of nearby
SNRs over optical studies of extragalactic SNe is that the SNRs are
close enough to examine the circumstellar medium sculpted by the
progenitor systems (e.g., the Kepler SNR, Reynolds \etal 2007) and
also to study the resolved stellar populations associated with them
(Badenes \etal 2009).  Recent X-ray spectroscopic observations have
also discovered emission from Mn and Cr in young Type Ia SNRs which
can be used to measure the metallicity of the progenitor system
(Badenes \etal 2008b), one of the key variables that might affect the
cosmological use of Type Ia SNe and which cannot be determined for
extragalactic SNe.

\subsection{Molecular physics}

\subsubsection{Evolved star envelopes: characterizing gas and dust 
chemistry.}

\begin{figure}
\begin{center}
\includegraphics[angle = 0, height=0.8\textheight]{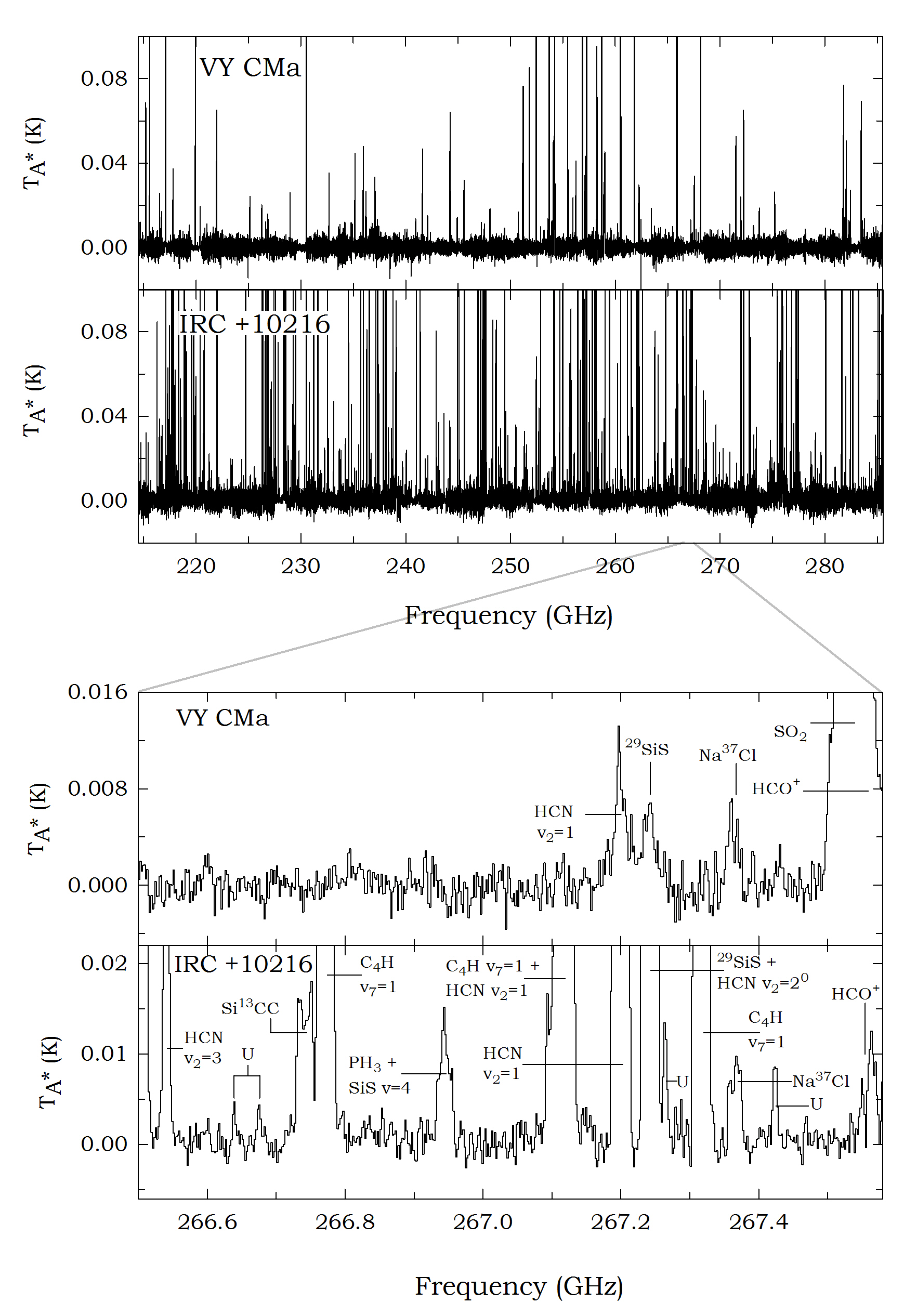}
\caption{\label{fig:VYCMa} The composite spectra of VY CMa and
  IRC$+$10216 across the entire 214.5-285.5 GHz frequency region,
  showing the rich chemical complexity of both sources (Tenenbaum
  \etal 2010a).  The intensity scale is the same for both sources and
  has been truncated to show the weaker lines.  The inset panel
  displays a select 1 GHz section, centered at 267 GHz.  The inset
  panel highlights some of the molecular identifications such as HCN
  $v_2 = 1^{1d} (J = 3 \to 2)$, $^{29}$SiS $(J = 15 \to 14)$,
  Na$^{37}$Cl $(J = 21 \to 20)$ and HCO$^+$ $(J = 3 \to 2)$.  The
  IRC$+$10216 spectrum at 267 GHz shows a tentative line of PH$_3$ and
  various other vibrationally excited HCN features.  Without
  laboratory spectroscopic studies, not a single line in these spectra
  could be securely identified.}
\end{center}
\end{figure}

Mass loss from evolved stars (asymptotic giant branch [AGB], red
giants and supergiants) contributes about 85\% of the material to the
ISM (Dorschner and Henning 1995).  Such mass loss creates large
envelopes of dust and gas surrounding the central star, extending to
$\sim 1000$ stellar radii.  Establishing the chemical content of
stellar envelopes is important in evaluating the overall composition
of the ISM.  These envelopes can either be oxygen-rich (O$>$C) or
carbon-rich (C$<$O).  Such shells also have large temperature and
density gradients (e.g., Ziurys 2006; Kim \etal 2010; Maercker \etal
2008; Patel \etal 2011; Tenenbaum \etal 2010a; Polehampton \etal 2010;
Schoier \etal 2011).  Close to the stellar photosphere, chemical
species, as well as dust condensates, form under thermodynamic
equilibrium.  As the material flows from the photosphere, abundances
become ``frozen-out'', but then are altered by photochemistry at the
shell edge (e.g., Cordiner and Millar 2009).  Circumstellar envelopes
are consequently unusual chemical factories.  The C-rich shell of the
AGB star IRC$+$10216, for example, has been found to contain over 70
different chemical compounds (Ziurys 2006; Ag\'undez \etal 2008b;
Tenenbaum \etal 2010a,b).  Oxygen-rich stars also have complex
chemistries, as observations of the envelope of VY Canis Majoris have
demonstrated (Tenenbaum \etal 2010a,b).  The chemical richness of
circumstellar envelopes is illustrated in figure~\ref{fig:VYCMa}.
Here composite spectra of the envelopes of IRC$+$01216 and VY Canis
Majoris are shown at 1 mm in wavelength.

Most circumstellar species have been detected on the basis of their
pure rotational spectra, which occur at millimeter and sub-mm
wavelengths.  Infrared spectra have also been important for species
with no dipole moments, such as HCCH.  The input of laboratory
spectroscopy, which supplies the critical ``rest frequencies'' for
line identification, has been crucial in this regard.  Recent examples
include the identification of negative ions, such as C$_4$H$^-$ and
C$_3$N$^-$ (Thaddeus \etal 2008) and KCN (Pulliam \etal 2010).

\subsubsection{Evolved star envelopes: refractory-element-bearing 
species.}

Condensation models predict that dust in circumstellar shells take on
a variety of forms, depending on whether the environment is oxygen or
carbon-rich (Lodders and Fegley 1999).  Almost all the refractory
elements (Si, P, and metals) are predicted to be in some sort of
mineral grain.  In C-rich shells, silicon is contained primarily in
SiC, but in O-rich objects in oxide condensates.  Phosphorus is
thought to be in the form of schreibersite, (Fe,Ni)$_3$P.  Magnesium
is contained in silicon and aluminum oxides or MgS.

Refractory elements in circumstellar environments are not all
contained in dust grains, however, as millimeter observations have
clearly shown.  Nine molecules have been found in circumstellar shells
that contain silicon and eleven that contain metals, in the chemist's
sense.  In C-rich envelopes, the metals are either found in halides
(NaCl, KCl, AlF and AlCl) or metal cyanides (MgCN, AlNC, MgNC, KCN and
NaCN; Pulliam \etal 2010); in oxygen-rich shells, oxides and
hydroxides such as AlO and AlOH dominate (Tenenbaum and Ziurys 2010).
Aluminum is thought to be condensed into Al$_2$O$_3$ in O-rich stars
(Lodders 2003).  The presence of AlO and AlOH indicates that
photospheric shocks are likely disrupting grains.

In addition, phosphorus-containing molecules appear to be prevalent in
circumstellar shells, as evidenced by the recent discoveries of CCP,
PN, HCP, PO and, tentatively, PH$_3$ (e.g., Ag\'undez \etal 2007;
Tenenbaum \etal 2007; Milam \etal 2008; Tenenbaum and Ziurys 2008).
Phosphorus is an important biogenic element and has consequences for
the origin of life.  Until very recently, few molecules containing
this relatively rare element had been observed in the ISM.

Gas-phase, high resolution, laboratory spectroscopy has been
absolutely crucial in establishing the presence of metal and
phosphorus-bearing species in circumstellar gas.  Recent discoveries,
such as CCP or AlOH, have relied on such work, in particular those
employing millimeter direct absorption and Fourier transform microwave
methods (e.g., Apponi \etal 1993; Halfen \etal 2009).  Many potential
species of this type are highly reactive, and require unusual,
non-equilibrium synthesis methods.

\subsubsection{Evolved star envelopes: contributions to the interstellar
medium.}

The matter lost from evolved stars becomes part of the ISM via
planetary nebulae.  It has usually been assumed that the molecular
content of circumstellar shells is returned to the atomic state as the
central star becomes a strong UV emitter, defining the planetary
nebula stage.  Yet, observations towards the Helix Nebula, which is
10,000 years old, have demonstrated that CO is present in a large
clumpy shell surrounding the central star and that HCO$^+$, HCN, HNC
and CN exist as well (Bachiller \etal 1997; Young \etal 1999).
Studies conducted recently by Tenenbaum \etal (2010) have resulted in
the detection of CCH, C$_3$H$_2$ and H$_2$CO in the Helix.
Furthermore, mapping of HCO$^+$ and H$_2$CO in the Helix suggests that
these species are as widespread as CO (Zack \etal 2011).  All of these
molecules have been studied via their mm rotational spectra.  Higher
energy transitions are likely to be found in these regions; recent
laboratory spectroscopy (e.g., Lattanzi \etal 2007) will make such
studies possible.

The discovery of complex molecules in old planetary nebulae is
surprising; such species have been subjected to intense UV radiation
for thousands of years.  Theoretical calculations have shown that
instabilities in the stellar wind can create finger-like clumps with
densities as high as 10$^5$ cm$^{-3}$ (Redman \etal 2003).  {\it
  Spitzer Space Telescope} IR images show the presence of finger-like
dust structures in the Helix (Hora \etal 2006).  Such clumps, composed
of gas-phase molecules mixed with dust, become self-shielding.  It
could be that these clumps survive on sufficient timescales to bring
molecular material to the diffuse ISM.  Recent observations by Liszt
and collaborators (e.g., Liszt \etal 2006) have demonstrated that
polyatomic molecules such as H$_2$CO, HCO$^+$ and C$_3$H$_2$ are
abundant in diffuse clouds.

The cycling of molecular material in the ISM has yet
to be fully evaluated.  Without the knowledge of the gas-phase
rotational spectra, our understanding of the molecular ISM
would be negligible (Ziurys 2006).

\subsubsection{Evolved stellar envelopes: fullerenes.}

Fullerene molecules such as C$_{60}$ and C$_{70}$ have been prime
observational targets ever since their discovery in laboratory
experiments designed to simulate the chemistry of carbon star outflows
(Kroto \etal 1985).  However it is only recently that observations
with {\it Spitzer} have revealed for the first time the spectroscopic
signatures of C$_{60}$ and C$_{70}$ in a variety of astronomical
environments.  The detections would not have been possible without
spectroscopic data from laboratory measurements (Kr\"{a}tschmer \etal
1990; Frum \etal 1991; Martin \etal 1993; Nemes \etal
1994; Fabian 1996; Sogoshi \etal 2000).  The laboratory data provided
the wavelengths and line strengths to confirm the astronomical
detections.

C$_{60}$ and C$_{70}$ were first detected in a hydrogen-poor planetary
nebula (Cami \etal 2010).  The hydrogen-poor conditions were thought
to be necessary in light of laboratory measurements on fullerene
production (De Vries \etal 1993; Wang \etal 1995).  Other {\it
  Spitzer} observations reveal the presence of C$_{60}$ in the
reflection nebulae NGC~7023 and NGC~2023 (Sellgren \etal 2010) and in
planetary nebulae in our Galaxy and the Small Magellanic Cloud (SMC;
Garc\'{i}a-Hern\'{a}ndez \etal 2010).  The latter work shows that the
fullerenes are present in a variety of environments, including
hydrogen-rich ones.  Garc\'{i}a-Hern\'{a}ndez \etal (2010) suggest
that the photochemistry of hydrogenated amorphous carbon plays a key
role.

\subsection{Condensed matter physics}

\subsubsection{Carbonaceous dust in outflows of late type stars.}

Cosmic dust particles span a continuous size distribution from large
molecules to $\mu$m-sized particles and play an essential role in the
evolution of the ISM (Tielens 2005).  Carbonaceous dust particles are
primarily formed in the outflow of carbon stars, through a
combustion-like process where small carbon chains form PAHs that
nucleate into larger-size PAHs and, ultimately, into nanoparticles
(Henning and Salama 1998).  According to this model, nucleation occurs
for temperatures above 2,000~K, followed by the growth of amorphous
carbon on the condensation nuclei in the 1,500~K temperature range.
As the temperature falls to around 1,100~K, aromatic molecules begin
to form in the gas phase and condense onto the growing particles
forming graphitic microstructures that will ultimately aggregate into
larger structures such as seen in soot formation (Pascoli and Polleux
2000).  Very little was known until recently about the formation of
cosmic dust due to the difficulty of forming and isolating these large
species and in tracking their evolutionary path under realistic
astrophysical laboratory conditions.  Efforts have been attempted in
this direction leading to new laboratory tools and breakthrough
results (Jager \etal 2007; Ricketts \etal 2011).  Carbon pyrolysis and
plasma-induced combustion experiments on mixtures of small
hydrocarbons indicate that the product distribution is dominated by
PAHs and partially hydrogenated PAHs.  The condensates produced in the
experiments consist of soot particles with graphene layers and PAHs.
The formation process starts with small molecules recombining to form
aromatic benzene rings, followed by the growth of larger PAHs through
subsequent C$_2$ addition to the aromatic rings and the final growth
of grains by the condensation of large PAHs on the surfaces of the
nuclei.  These results demonstrate that low-temperature condensation
is a very likely formation process of soot and PAHs in AGB stars,
confirming the model predictions.

\subsubsection{Silicates in envelopes of late type stars.}

Silicates are an important component of cosmic matter.  Silicates form
in the winds of AGB stars and are processed in the diffuse ISM.  They
are also an important component of dust in protoplanetary and debris
disks where they help regulate thermal exchanges (Henning 2010).  The
detection at IR and millimeter wavelengths of silicate dust grains
containing O, Si, Fe and Mg, as well as some Ca and Al, provides an
important constraint on dust chemical composition and on grain size
(Bouwman \etal 2001; van Boekel \etal 2005; Chiar and Tielens 2006;
Sargent \etal 2009; Juhasz \etal 2009).  Cosmic silicates are mostly
found in the amorphous state, characterized by broad and structureless
IR bands at 10 and 18 $\mu$m that can be attributed to Si-O
stretching and O-Si-O bending modes, respectively (Draine 2003).
In circumstellar environments, however, evidence for crystalline
silicates is found both around (post-)AGB stars and in disks around
Herbig Ae/Be stars, T Tauri stars and brown dwarfs (Henning 2010;
Molster and Waters 2003).  Silicates are also found in cometary
environments (Crovisier \etal 1997; Kelley and Wooden 2009; Hanner
and Zolensky 2010), in spectra from asteroids (Emery \etal 2006) and
in interplanetary dust particles (Bradley 2010).  These findings have
only been possible thanks to vigorous laboratory programs that have
helped characterize the basic properties of silicates that are needed
to detect their signature in astronomical spectra.  A vast amount of
data resulting from laboratory studies dealing with both amorphous and
crystalline silicates is now available in the literature, making it
possible to derive information on topics as diverse as the evolution
of cosmic dust, transport in protoplanetary and debris disks and
redshifts in high-$z$ objects (for recent reviews see Henning 2010 and
references therein).

\subsection{Plasma physics}

\subsubsection{Ion heating in the solar corona and solar wind.}

UV spectroscopy of the solar corona has revealed that ion
temperatures vary among species and that ion distribution functions
are non-Maxwellian and anisotropic.  These effects are most pronounced
in certain minor ions and in particular increase with particle mass
(Kohl \etal 2006; Cranmer \etal 2008; Landi and Cranmer 2009).  These
anisotropies may be a signature of heating by high frequency
turbulence, possibly driven by magnetic reconnection.  Similar effects
have been observed in laboratory plasmas.  Brown \etal (2002) reported
an energetic ion population associated with 3-D magnetic reconnection
in the Swarthmore Spheromak Experiment device.  Recently, ion heating
associated with magnetic reconnection events in the Madison Symetric
Torus (MST) has been studied, revealing similar anisotropies and mass
dependences (Tangri \etal 2008; Fiksel \etal 2009).  The physical
mechanism may be related to the reconnection driven turbulent cascade
also recently studied on MST (Ren \etal 2009).  Thus, the experiments
have shown that ions can be heated anisotropically, in a mass
dependent way, by MHD turbulence generated in reconnection events.
This suggests that turbulent heating is responsible for the species
dependent temperature and anisotropy observed in the solar corona and
that the turbulence could be generated by reconnection.

\subsubsection{Reconnection in stars.}

Magnetic reconnection is a key process in stellar astrophysics.  It is
the leading candidate for the energy release mechanism in flares and
may be an important mechanism for coronal heating.  It must also occur
in stellar interiors, as part of the magnetic dynamo.  Laboratory
experiments have made essential contributions to reconnection studies.
Two recent review articles discuss these contributions in depth
(Zweibel and Yamada 2009; Yamada \etal 2010).  Highlights include
laboratory studies of flux rope dynamics, including reconnection in
line tied plasmas and relaxation to a lower energy state (Bergerson
\etal 2006; Cothran \etal 2009; Sun \etal 2010), a criterion for the
onset of fast collisionless reconnection mediated by the Hall effect
(Yamada 2007) and studies of the electron diffusion layer, which
clarify the mechanisms responsible for breaking the fieldlines and the
apportionment of energy in the reconnection region (Ren \etal 2008).
These studies thus suggest a possible mechanism for triggering fast
reconnection in solar flares and provide detailed information on how
energy is apportioned among thermal and nonthermal electron and ion
populations in solar reconnection.

\subsubsection{Stellar dynamos.}

Although magnetic cycles are well established on the Sun and other
stars, a theoretical explanation of stellar dynamos is still lacking
and experimental confirmation is sparse.  For many years, dynamo
theory has been dominated by kinematic studies in which a mean field
is built up from infinitesimal values by small scale turbulence and
large scale shear.  Recently, dynamo action has been reported in a
number of liquid sodium experiments (Gailitus \etal 2000; Monchaux
\etal 2007; Spence \etal 2007).  Liquid sodium, like stellar interior
plasmas, is much more resistive than it is viscous.  These experiments
are being used to understand saturation mechanisms, the surprising
role of turbulence in suppressing the growth of large scale magnetic
fields and the electromotive forces produced by large scale and
small scale turbulent flows.  These experiments are influencing the
development of a new dynamo paradigm, in which dynamos are essentially
nonlinear and maintained by large scale flows rather than small scale
turbulence.

\subsubsection{Stellar opacities.} 
\label{SSE:Pla:Opa}

Heating by fusion reactions deep within stellar cores produces thermal
X-ray radiation and the outward transfer of this radiation is an
essential element of stellar dynamics and evolution.  The rate of
attenuation of such radiation is the opacity. One can calculate
opacities from fundamental atomic physics, but in even moderately
complex elements such as iron this involves the interaction of many
millions of transitions.  Because of this complexity, calculations of
opacity are uncertain and experimental measurements are essential to
determine which calculations are correct.  This has led to a quest to
produce in the laboratory conditions present in stellar interiors so
as to measure relevant opacities (Bailey \etal 2009).  During the
1990s, laboratory research and atomic theory resolved the issue of
understanding pulsations in Cepheid variable stars (Rogers and
Iglesias 1994; Springer \etal 1997; see also the review by Remington
\etal 2006).  More recently, research has turned to the challenging
issue of understanding solar structure.  The Sun has an inner,
radiative heat conduction zone that gives way to a convective zone
nearer the surface.  Solar models typically find a location of the
boundary between these zones that differs from the measured one by
more than 13 standard deviations (Basu and Antia 2008).  One possible
cause of this is knowledge of the energy-averaged opacity, which
indeed must be accurate to $\sim 1\%$ in order to fix the boundary to within
the uncertainties of the observation (Bahcall \etal 2004).  By
producing conditions of the stellar interior and measuring the
detailed spectral structure of the opacity, researchers are now able
to address this issue (Bailey \etal 2007; Bailey \etal 2009;
Mancini \etal 2009).  These measurements showed that while very
recent opacity models were nearly accurate enough under the conditions
studied, previous opacity models were much less accurate.  Challenges
going forward are to produce accurate measurements in hotter, denser
plasmas, in effect moving deeper into the Sun, while also addressing 
the other uncertainties discussed in section~\ref{SSE:At:SolAbu}.

\subsubsection{Photoionized gas.}

CVs are binary star systems composed of a white dwarf and (most often)
a normal star.  Mass from the normal star falls towards the white
dwarf, producing a wide variety of phenomena.  Recent laboratory work
has focused on shock phenomena (Falize \etal 2009a,b, 2011) and on
photoionization.  CVs emit X-ray radiation from the accreting matter.
Such emission is also important in other accreting systems, for
example, neutron stars, black holes and star forming regions.  The
radiation photoionizes the nearby matter, producing plasma that is
``overionized'' (ionized far beyond the level that would be produced
by collisional ionization at the local electron temperature).  One
needs experiments to assess the accuracy of radiative rates across a
wide range of transitions. An early effort in this direction (Foord
\etal 2004) used the radiation pulse produced by imploding a
cylindrical array of metal wires to vaporize and then photoionize very
thin foils of Fe and NaF.  They later compared the measured charge
state distributions to those calculated by photoionization codes
(Foord \etal 2006), finding broad agreement but some differences.  To
obtain more uniform photoionized plasma, present-day experiments use
radiation from an imploding wire array to create plasma in a gas cell
(Bailey \etal 2001; Mancini \etal 2009).  The radiation from a Z pinch
machine has also been used to photoionize a gas cell (Cohen \etal
2003).  In an alternative approach, a laser source is used to heat a
gold cavity whose emission produces a moderately overionized plasma in
a gas cell (Wang \etal 2008).  More recently, Fujioka \etal (2009)
used a laser-driven implosion to produce a $\sim 5$~MK blackbody
radiation source, which in turn photoionized a laser-ablated, Si
plasma.  The photoionization experiments to date have shown that
detailed comparisons of code results with laboratory data can improve
our understanding of photoionized systems.

\subsubsection{Instabilities in type II supernovae.}
\label{SSE:Pla:Ins}

Core-collapse SNe (ccSNe) involve much uncertain physics.
Their complete physics and full range of dynamics are far beyond what
can now be simulated.  As a result, theories or simulations of these
events must employ reduced physics, creating a need to test those
simplified models.  The potential for discovery is high, as
unanticipated interactions of the physical processes may arise.
Laboratory work relevant to ccSNe is currently limited to the ``late''
phase of explosion, after the initial core collapse and after the
shock wave forms that blows apart most of the star.  It is now widely
accepted that unstable mixing of stellar materials occurring during
that phase is essential to explain observations of supernova SN 1987A
(Arnett \etal 1989; Chevalier 1992), but early simulations including
these effects failed to do so (Muller \etal 1991).  This, combined
with the observed asymmetry of SN ejecta, led to the hypothesis that
such explosions were jet-driven (Wang \etal 2001), although the
mechanism that would cause this remains unidentified.  Meanwhile, and
quite unexpectedly, simulations employing improved traditional
explosion models produced relevant levels of asymmetry (Kifonidis
\etal 2000, 2003, 2006; Guzman and Plewa 2009).  This seems to be a
nice story with an endpoint, yet all its elements remain uncertain
without experimental evidence that other unanticipated coupling does
not exist.  Simulations cannot for example test the hypothesis that
small-scale dynamics may feed back on the large-scale hydrodynamic
evolution (Leith 1990).

Experiments have been developed to examine unstable hydrodynamics in a
regime relevant to late-phase ccSNe dynamics.  Work through 2005 is
reviewed by (Remington \etal 2006).  Such experiments can be well
scaled in detail (Ryutov \etal 1999) to local conditions in a ccSNe.
To date the large-scale behavior they have seen has been consistent
with a variety of simulations (Kuranz \etal 2009), showing that on a
large scale our understanding of instabilities in the late phase of
ccSNe is correct.  However, to explain the observations requires only
that $\sim 1$\% of the inner material in the star finds a way to reach
its outer layers with high velocity, and a number of small details
have not been consistent between simulations and experiments (Calder
\etal 2002; Miles \etal 2004; Kuranz \etal 2010).  Further work is
seeking to understand the origin of the differences between
observations and simulations, and to develop experiment designs
relevant to the global dynamics of the explosion (Grosskopf \etal
2009).

\subsubsection{Radiative shocks in type II supernovae.}
\label{SSE:Pla:Rad}

During the explosion, the radiation pressure in the shocked matter
produced by a type II supernova exceeds the material pressure, but
because the mean free path for thermal radiation is small compared to
other scale lengths in the system, the shock wave behaves as a
hydrodynamic shock with a polytropic index $\gamma = 4/3$ (Ryutov
\etal 1999).  This changes as the shock wave breaks out of the star
and radiation can escape ahead of the shock.  The shock enters a
regime in which the thermal energy produced by the shock is mostly
radiated away even though the layer behind the shock is many
mean-free-paths thick. A dense shell forms, which may be unstable (see
sections~\ref{GAN:Pla:SNR:rad} and \ref{GAN:Pla:SNR:Vis}).  Current
astrophysical instruments are beginning to observe such shock-breakout
events (Calzavara and Matzner 2004; Chevalier and Fransson 2008;
Soderberg \etal 2008). Experiments have begun to produce and study
shock waves in the same radiation-hydrodynamic regime as the
shock-breakout events, with strong radiation emission, escape of the
radiation ahead of the shock, and trapping of the radiation behind the
shock (Bouquet \etal 2004; Reighard \etal 2006; Doss \etal 2009,
2010).  Such experiments are a subset of radiative-shock experiments
more broadly (Bozier \etal 1986; Grun \etal 1991; Edwards \etal 2001;
Edens \etal 2005; Koenig \etal 2006; Hansen \etal 2006; Busquet \etal
2007). They typically involve producing a low atomic number $Z$ plasma
``piston'' moving at $\gax 100$~km~s$^{-1}$ and using it to drive a
shock wave in Xe or some other high-$Z$ gas.  In the experiments the
radiation transport is dominated by broadband thermal emission and
absorption, while that in the star is more complex.  Even so, the
experiments are a vehicle to better understand the
radiation-hydrodynamic behavior of this type of system and they have
the potential to discover unanticipated behavior.  To date, the
experiments have shown that the Vishniac-like instability to which
such dense shells are subject is not so virulent as to greatly distort
the shock.  Ongoing work is developing scaling connections to SNe and
SNRs (Doss 2011) and simulating the observed behavior (van der Holst
\etal 2011).  These experiments constrain astrophysical simulation
models, which cannot be expected to correctly model the SN if they
cannot model these data.





\subsubsection{Compact objects and gamma ray bursts: relativistic 
collisionless shocks.}

Most astrophysical shocks are collisionless shocks, in which
electromagnetic turbulence randomizes the motion of the incoming
particles, replacing the role of collisions in ordinary
shocks. Astrophysical observations often imply that relativistic
collisionless shocks must be present, as for example in gamma ray
bursts (GRBs; Piran 1999; Waxman 2006).  Yet the observed emission
from GRBs, attributed to synchrotron emission by electrons, required
magnetic fields orders of magnitude larger than could be produced by
mechanisms known to be present in the 1990s.  The shocks involved are
too complex to be fully described by a first-principles analytic or
semi-analytic theory.  The past decade has seen an explosion of work on
such shocks (only some of which can be cited here), made possible
primarily by the application of Particle-In-Cell (PIC) simulation
methods on ever-larger computers.  The first 3-D PIC simulation of
colliding electron-positron plasmas (Silva \etal 2003), which are
thought to occur in GRBs and elsewhere, found that the 3-D Weibel
instability produces both long-lived magnetic fields whose energy
density is near that of the ions and nonthermal particle acceleration.
This supported the theory (Medvedev and Loeb 1999) that the Weibel
instability was the key process, which previously was only one of many
theories.  Further simulations studied initially unmagnetized
(Spitkovsky 2008) and initially magnetized (Murphy \etal 2010)
electron-ion shocks, both also considered important in GRBs.  In both
cases one also sees the Weibel instability and the generation of
strong magnetic fields, in addition to significant electron heating.
The results of Murphy \etal (2010) in combination with observations of
polarization in emission from GRBs provide evidence for a significant
primordial magnetic field in such events.  Applying a similar model to
relativistic electron-positron jets, Nishikawa \etal (2009) found that
the gamma-ray emission should come primarily from the shocked jet
material rather than from the shocked ambient medium, confirming this
interpretation of the observations of those objects.  In this way
large PIC simulations have become an important tool to advance
understanding of relativistic astrophysical systems.

\subsection{Nuclear physics}

\subsubsection{Nuclear synthesis via neutron capture.}

Elements beyond the iron peak are produced primarily by neutron (n)
capture in the s- (slow) and r- (rapid) processes.  The main s-process
occurs in low mass AGB stars while the weak s-process takes place in
the He- and C-burning shells of massive stars.  There is uncertainty
about site or sites of the r-process, with Type II SNe (and
their associated neutrino-driven winds) and neutron star mergers being
leading candidates (Qian and Wasserburg 2007).

The fractional contributions of the weak and main s-processes have
been deduced from studies of solar system (including meteorite)
abundances.  The r-process must account for ``shielded'' or other
n-capture isotopes off the s-process path, and for other differences
between observed abundances and those attributable to the s-process.
Discussions can be found in Raiteri \etal (1993), Arlandini \etal
(1999), The \etal (2007), and Heil \etal (2008).  These studies
attribute the light n-capture elements (e.g., Sr and Zr) with
high-mass stars and heavier s-process elements, such as Ba, with
low-mass stars (Busso \etal 1999).  Recent laboratory data for
s-process cross sections are summarized by K\"appeler \etal (2011),
updating K\"appeler \etal (1989).

Although the sites of the r-process are uncertain, data from
metal-poor stars show that an r-process operated in the early galaxy
at a frequency typical of ccSNe (Cowan and Thielemann 2004; Sneden et
al. 2008; see also section~\ref{subsubsec:Solar}).  While properties
of lighter r-process nuclei have been determined in the laboratory
(Kratz \etal 2000; Pfeiffer \etal 2001; Moller \etal 2003), much of
the r-process path is through short-lived, very neutron-rich nuclei
that are difficult to produce.  Future facilities (e.g., the Facility
for Rare Isotope Beams [FRIB]) will allow more extensive measurements
of relevant masses and $\beta$-decay rates.

Interstellar abundances, however, do not appear to match solar system
values.  The abundances of Ga and Ge are 25\% of the meteoritic value
for low density, warm gas, where depletion onto interstellar grains is
expected to be minimal (Cartledge \etal 2006; Ritchey \etal 2011).
The inferred Rb abundance is about 35\% of the meteoritic value
(Federman \etal 2004; Walker \etal 2009).  The noble gas Kr, which
does not deplete onto grains, has an average abundance of 50\% of the
solar system value (Lodders 2003; Cartledge \etal 2003).  Ga, Ge, Kr
and Rb are predicted to form primarily in high-mass stars.  In
contrast, Cd and Sn, which are mainly synthesized in low-mass stars,
are not depleted for low density lines of sight (Sofia \etal 1999),
despite similarities between Ga, Ge, Rb, Cd and Sn condensation
temperatures (Lodders 2003).  The observed depletion patterns cannot
be attributed to imprecise oscillator strengths, which are well known
from laboratory and theoretical work (Morton 2000, 2003; Schectman
\etal 2000; Alonso-Medina \etal 2005; Oliver and Hibbert 2010).
Additional interstellar studies of other n-capture elements are
needed.

\subsubsection{Stellar nuclear fusion: pp chain.}

The proton-proton or pp chain is the principal mechanism by which
low-mass hydrogen-burning stars like the Sun produce energy through
$4{\rm p} \rightarrow {}^4\mathrm{He} + 2e^+ + 2 \nu_e$ where $e^+$
represents a positron and $\nu_{\rm e}$ an electron neutrino.  The
competition between the three cycles of the pp chain (ppI, ppII and
ppIII) depends sensitively on the stellar core temperature, as the
reactions require Coulomb barrier penetration, and on the specific
rates of the reactions, which are conventionally given in terms of the
astrophysical $S$-factor, from which the highly energy-dependent
$S$-wave Coulomb behavior of the cross section has been removed
(Adelberger \etal 2011).  Laboratory measurements of $S$-factors are
important to both solar neutrino physics and helioseismology.  The
uncertainties in laboratory $S$-factor measurements limit the
precision of standard solar model (SSM) neutrino flux and sound speed
predictions (Bahcall \etal 2005b).  Associated astrophysics challenges
include demonstrating through neutrino spectrum distortions that
matter effects influence neutrino oscillations, detecting day-night
effects and resolving discrepancies discussed in
section~\ref{SSE:At:SolAbu} between the SSM and helioseismology
measurements related to solar metallicity (Haxton and Serenelli 2008;
Aharmin \etal 2010; Abe \etal 2011).

Recent key advances in laboratory astrophysics include a series of
precise measurements of the reactions $^3$He($^3$He,2p)$^4$He (Bonetti
\etal 1999) and $^3$He($^4$He,$\gamma$)$^7$Be (Singh \etal 2004;
Bemmerer \etal 2006a; Brown \etal 2007; Confortola \etal 2007; Gyurky
\etal 2007; di Leva \etal 2009) which control the ratio of ppI solar
neutrino flux to that of the ppII and ppIII.  There have also been
several new and precise measurements of $^7$Be(p,$\gamma$)$^8$B
(Hammache \etal 1998, 2001; Strieder \etal 2001; Junghans \etal 2002,
2003, 2010; Baby \etal 2003a,b), until recently the limiting nuclear
physics uncertainty in predicting the flux of ppIII solar neutrinos.
These measurements will have an important impact on the analysis of
the currently running Borexino experiment (Arpsella \etal 2008) which,
in conjunction with the Sudbury Neutrino Observatory (SNO; Aharmin
\etal 2010) and Super-Kamiokande (Abe \etal 2011), will provide a
direct test of matter effects on neutrino oscillations.  They also
impact the comparison between the total SSM $^8$B flux and that
measured in SNO, which is sensitive to SSM parameters such as core
metallicity.  Recent progress in $S$-factor determinations came from
technological advances like the Laboratory for Underground
Nuclear Astrophysics (LUNA; Costantini \etal 2009; Broggini
\etal 2010), a specialized low-energy accelerator operating at great
depth, allowing nearly background-free measurements of important cross
sections.

\subsubsection{Stellar nuclear fusion: CNO cycle.}

Heavier hydrogen-burning stars produce their energy primarily through
the carbon-nitrogen-oxygen (CNO) cycles, where nuclear reactions are
characterized by higher Coulomb barriers.  Hence energy production
rises steeply with temperature, $\epsilon_\mathrm{CNO} \sim T^{18}$,
compared to $\epsilon_\mathrm{pp} \sim T^4$ at solar temperatures.
Unlike the pp chain, the CNO cycle requires pre-existing metals (in
the astronomer's sense meaning all elements heavier than He).  These
serve as catalysts for hydrogen burning, with the energy production at
fixed temperature proportional to metallicity.  The CNO cycle is
responsible for about 1\% of solar energy generation, but dominates
hydrogen burning in stars with central temperatures $\gax 2 \times
10^7$~K.

\begin{figure}
\begin{center}
\includegraphics[angle = 0, height=0.35\textheight]{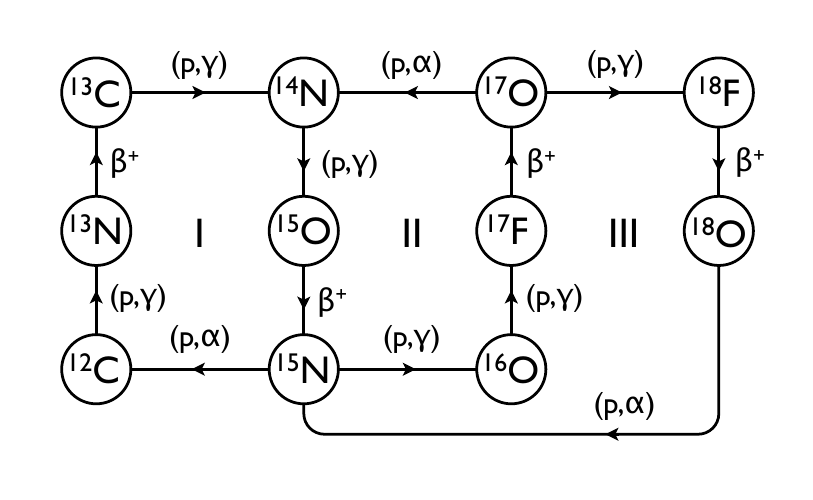}
\caption{\label{fig:CNOnew} The three pathways that dominate the CNO
  cycles at lower temperatures.  Recent experiments on the
  branch-point reactions involving $^{15}$N and $^{17}$O are discussed
  in the text.  Adapted from Wiescher \etal (2010).}
\end{center}
\end{figure}

The rate-controlling step in the carbon-nitrogen (CN) cycle, denoted
by I in figure~\ref{fig:CNOnew}, is ${}^{14}$N(p,$\gamma$)${}^{15}$O.
The nuclear physics of this reaction is complex, with contributions
from several $^{15}$O resonances both above and below threshold.  Work
on this reaction over the past decade has been intense.  New
measurements have been made with both direct methods (Formicola \etal
2004; Imbriani \etal 2005; Runkle \etal 2005; Bemmerer \etal 2006b;
Lemut \etal 2006; Marta \etal 2008) and indirect methods, covering the
energy range between 70 and 480 keV (see Wiescher \etal 2010 and
Adelberger \etal 2011 for summaries).  After summing all transitions
$S_{1~14}^\mathrm{tot}(0)=1.66 \pm 0.12$~keV~barn was obtained
(Adelberger \etal 2011), a value nearly a factor of two below
previously recommended values.  This has had significant consequences
for astrophysics, such as increasing the age estimate for the oldest
globular cluster stars by nearly a billion years (Runkle \etal 2005).
The increased precision of the $S$-factor will be critical to the
analyses of data from the neutrino detector SNO+ ({\tt
  http://snoplus.phy.queensu.ca/Home.html}) now under construction.
By measuring the CNO neutrino flux, SNO+ may directly determine the
carbon and nitrogen content of the solar core.

While $^{14}$N(p,$\gamma$) controls the cycling rate of the CN cycle,
other reactions determine the flow rate out of this cycle towards
oxygen and heavier metals.  The turn-on of these branches influences
the opacity evolution and temperature profile of hydrogen-burning
stars.  Competition between $^{15}$N(p,$\alpha$)$^{12}$C and
$^{15}$N(p,$\gamma$)$^{16}$O governs the division of the flow between
the left two cycles illustrated in figure~\ref{fig:CNOnew}.  Here
$\alpha$ represents an $^4$He nucleus.  Recent work on the second
reaction (Bemmerer \etal 2009) has led to corrections in earlier
results (Adelberger \etal 2011).  The new measurements were done at
novae energies and reduce the final nucleosynthesis yield of $^{16}$O
by up to 22\%, depending on the nova temperature (Bemmerer \etal
2009).  There is a similar competition between
${}^{17}$O(p,$\alpha$)$^{14}$N, which closes CNO cycle II of
figure~\ref{fig:CNOnew}, and ${}^{17}$O(p,$\gamma$)$^{18}$F, which
leads either to the more complicated reaction network of the hot CNO
cycles or to CNO~I and II via $^{18}$F$(\beta^+
\nu){}^{18}$O(p,$\alpha$)$^{15}$N.  Recent work has led to significant
cross section changes affecting the flow towards the hot CNO cycle in
novae (Fox \etal 2005; Chafa \etal 2007).

\subsubsection{Stellar nuclear fusion: hot CNO burning.}

At temperatures greater than approximately a few times $10^8$~K a more
complicated set of reactions allows mass flow to heavier nuclei
(Wiescher \etal 2010).  In addition, the equilibrium abundances
characterizing previously described cycles change: the rates of key
radiative capture reactions increase to the point that they match or
exceed those of the $\beta$ decays of $^{13}$N and $^{15}$O, for
example, so that weak rates now govern the cycling time and rate of
energy production, while rapid (p,$\gamma$) reactions competing with
$\beta$ decay open up new pathways.  The hot CNO network involves
reactions on unstable nuclei that require laboratory tools that only
recently have become available, with the development of radioactive
ion beam facilities.  The resulting advances include:
\begin{itemize}
\item The cycle
  $^{12}$C(p,$\gamma$)$^{13}$N(p,$\gamma$)$^{14}$O($\beta^+
  \nu$)$^{14}$N(p,$\gamma$)$^{15}$O($\beta^+\nu$)$^{15}$N(p,$\alpha$)$^{12}$C
  opens up when the radiative capture rate on $^{13}$N exceeds the
  $\beta$ decay rate.  The key resonance governing the capture was
  measured in inverse kinematics, using an intense $^{13}$N
  radioactive beam of $3 \times 10^8$~s$^{-1}$ (Decrock \etal 1991;
  Delbar \etal 1993).  The direct capture contribution was recently
  determined from the asymptotic attentuation coefficient (Li \etal
  2006; Guo and Li 2007).  These measurements together have led to a
  substantial increase in the recommended low-energy cross section
  (Wiescher \etal 2010), lowering the ignition temperature for the hot
  CNO cycle.  This cross section impacts models of novae, including
  the $^{13}$C/$^{12}$C ratio in nova ejecta, as well as the predicted
  abundance of $^{13}$C, an important s-process neutron source
  (Arnould \etal 1992).
\item A critical branching in the hot CNO cycle depends on the
  competition between $^{18}$F(p,$\alpha$)$^{15}$O and
  $^{18}$F(p,$\gamma$)$^{19}$Ne.  These destruction channels for
  radioactive $^{18}$F are also the largest nuclear physics
  uncertainty affecting $\gamma$-ray emission from novae (Hernanz
  \etal 1999).  The development of high intensity $^{18}$F beams of
  $\sim 10^5$~particles~s$^{-1}$ have allowed experimenters to
  determine the dominant $^{18}$F reaction rates for temperatures
  characterizing ONeMg novae (Bardayan \etal 2000, 2002; Chipps \etal
  2009; Murphy \etal 2009).
\end{itemize}

\subsubsection{Core-collapse supernovae.}

Nuclear physics governs three important aspects of ccSNe, the core
bounce (and ultimately the structure of the neutron star), energy
transport and nucleosynthesis.

The core bounce depends on the nuclear EOS at densities that could
range up to six times that of ordinary nuclear matter, at temperatures
of tens of MeV and at extremes of isospin.  The conditions at maximum
compression are beyond the direct reach of experiment, but are
constrained by astrophysical observations, including the stability of
the 1.396 ms pulsar Terzan 5 (Hessels \etal 2006) and the recent
determination of a two-solar-mass neutron star measured by Shapiro
delay (Demorest \etal 2010) as well as by laboratory measurements of
nuclear compressibilities.  Laboratory measurements of giant monopole
resonance energies in nuclei with and without neutron excesses
constrain the compressibility for isospin symmetric matter and the
symmetry energy $K_\tau$ critical to neutron-dominated matter
(Piekarewicz 2010).  New measurements, carried out in Sn isotopes, has
led to $K_\tau = -395 \pm 40$~MeV (Garg \etal 2007), increasing the
error bar on compressibility estimates (Piekarewicz 2010).

In a core-collapse supernova explosion the energy released through
gravitational collapse must be preferentially transferred to the
mantle of the star, to enable ejection.  This is thought to be
accomplished through the combined effects of the shock wave and
neutrino heating.  The neutrino heating and associated physics --
neutrino opacity, neutrino cooling, $\beta$ decay rates important to
lepton number emission and nucleosynthesis -- are governed in part by
nuclear Gamow-Teller and first-forbidden responses (Langanke and
Martinez-Pinedo 2003).  The Gamow-Teller responses have been mapped in
the laboratory using forward-angle (p,n) and (n,p) scattering
(Rapaport and Sugarbaker 1994) and then incorporated into nuclear
models used in supernova simulations.  The resulting modern electron
capture and $\beta$ decay rates have been found to increase the
electron mass fraction $Y_{\rm e}$ throughout the iron core.  As the
size of the homologous core and thus the shock radius is proportional
to $Y_{\rm e}^2$, this has significantly increased calculated shock
wave strengths (Heeger \etal 2001; Bronson-Messer 2003).  These
improvements have also led to changes in neutrino ($\nu$) process
nucleosynthesis yields for key isotopes such as $^{11}$B and $^{19}$F
(Heger \etal 2005).

Recent studies of metal-poor halo stars (Cowan and Sneden 2006) have
associated early Galaxy r-process events with ccSNe, which provide in
their $\nu$-driven winds and mantles conditions under which an
r-process might occur.  The rate of nucleosynthesis is controlled by
weak interactions, as new neutrons can be captured only after neutron
holes are opened by $\beta$ decay.  Thus the rate of $\beta$ decay is
critical to determining which supernova zones might be able to sustain
the necessary nucleosynthetic conditions for the requisite time.
Recent laboratory $\beta$ decay measurements for very-neutron-rich
isotopes near mass number $A=100$ have demonstrated that half-lives
are a factor of two or more shorter than previously believed, which
significantly relaxes constraints on the r-process time scale
(Nishimura \etal 2011).

\section{The galactic neighborhood}
\label{sec:GAN}

The galactic neighborhood includes ``the structure and properties of
the Milky Way and nearby galaxies and their stellar populations and
evolution, as well as interstellar media and star clusters''
(Blandford \etal 2010a).

\subsection{Atomic physics}

\subsubsection{Galactic chemical evolution.}

The early chemical evolution of the Galaxy can be studied from
abundances of the iron-peak elements.  These elements are synthesized
in supernova explosions and the stellar abundance trends with
metallicities (i.e., [Fe/H]) provide important constraints on the
explosion mechanisms of type~II and Ia events.  Early work by
McWilliam (1997) demonstrated that as [Fe/H] decreased below -2.4,
Cr/Fe decreased while Co/Fe increased, leading to a rising trend of
Cr/Co with decreasing Fe/H.  This behavior provides clues to synthesis
from SNe in the Galaxy as a function of metallicity. For
example, models with $\alpha$-rich conditions tend to produce more
elements heavier than Fe, such as Co, in contrast to lighter elements
such as Cr. It is also possible to reproduce these abundance trends by
varying such effects as the explosion energies, neutron excess, mass
cut position and progenitor masses in explosive supernova
nucleosynthesis.  Additional studies have recently been completed,
focusing on the iron peak elements Ti, V, Cr, Mn, Fe, Co and Ni as a
function of [Fe/H] (Henry \etal 2010).  The derived abundance trends
have been based upon utilizing neutral (and less abundant) species for
the Fe-peak element species and assuming LTE conditions.

Recent laboratory determinations of atomic data (e.g., oscillator 
strengths) have been obtained for 
Cr~\textsc{i} (Sobeck \etal 2007),
Cr~\textsc{ii} (Nilsson \etal 2006),
Mn~\textsc{i} and \textsc{ii} (den Hartog \etal 2011),
Co~\textsc{i} (Nitz \etal 1999) and 
Co~\textsc{ii} (Crespo Lopez-Urrutia \etal 1994).  
These new experimental data have led to increasingly more accurate
abundance values for the iron-peak elements in old halo stars. As a
result of these new precise values, we are getting a clearer picture
of the nature, and sources, of the earliest element formation in the
Galaxy.  In addition these new abundance values are providing
increasingly stringent constraints on models (e.g., mass cut,
energies, progenitor masses, elemental content of the ejecta, etc.)
of supernova explosions and nucleosynthesis.  Finally, an examination
of the abundance trends of the iron-peak elements over different
stellar metallicities is providing direct insight into the chemical
evolution of the Galaxy.

\subsection{Molecular physics}

\subsubsection{Interstellar medium chemical complexity.}

Recent developments in detector technology for ground-based
measurements and the launch of the {\it Herschel Space Observatory}
provide new opportunities to improve our understanding of interstellar
chemistry.  This has been particularly true for molecular ions and
radicals which are important intermediate species in chemical networks
describing the molecular evolution of interstellar clouds.
Intermediates which have been detected spectroscopically include
SH$^+$ (see section~\ref{GCT:AP:AGNs:Chem}), H$_2$Cl$^+$, OH$^+$,
H$_2$O$^+$ and CH$^+$.  Accurate transition frequencies are required
for observational searches of these species, many of which have
transitions at sub-millimeter and far-IR wavelengths.  For SH$^+$ a
combination of laser (Hovde and Saykally 1987), microwave (Savage
\etal 2004) and infrared (Brown and M\"{u}ller 2009) measurements
provided the needed accuracy.  The measurements in the THz range for
H$_2$Cl$^+$ by Araki \etal (2001) yielded the required transition
frequencies.  The frequencies for OH$^+$ come from the study of Bekooy
\etal (1985).  The H$_2$O$^+$ frequencies are given by M\"{u}rtz \etal
(1998).  CH$^+$ data come from the spectroscopic work of Amano (2010).
Lastly, there are numerous spectroscopic studies of NH and NH$_2$,
which have been compiled into the Cologne Database for Molecular
Spectroscopy (M\"{u}ller \etal 2005).

{\it Herschel} has detected many of the above species.  Lis \etal
(2010) discovered H$_2$Cl$^+$ in absorption towards the star-forming
region NGC~6334I in both $^{37}$Cl and $^{35}$Cl isotopologues.  They
found that the HCl/H$_2$Cl$^+$ ratios are consistent with chemical
models, but the H$_2$Cl$^+$ column densities greatly exceeded model
predictions.  The OH$^+$ and H$_2$O$^+$ ions, which lead to H$_2$O in
ion-molecule chemical schemes, were seen in several star-forming
clouds and the intervening diffuse clouds (e.g., Gupta \etal 2010;
Neufeld \etal 2010; Schilke \etal 2010).  For example, Neufeld {\it et
  al} (2010) detected these ions in absorption towards the cloud W49N.
The OH$^+$/H$_2$O$^+$ abundance ratio indicated that the ions formed
in clouds with small fractions of H$_2$.  Since these ions are
produced by cosmic-ray ionization of atomic and molecular hydrogen, an
ionization rate could be inferred.  The values are consistent with
other recent determinations.  Falgarone \etal (2010) observed
absorption from $^{12}$CH$^+$ and $^{13}$CH$^+$.  As the absorption
from $^{12}$CH$^+$ is optically thick, they were only able to set a
lower limit of 35 on the isotope ratio.  This value is consistent with
other determinations of the $^{12}$C/$^{13}$C ratio in ambient gas.
Lastly, we note that Persson \etal (2010) detected NH and NH$_2$ in
absorption in diffuse gas.  Neither gas-phase nor grain-surface
chemical models adequately explain the data; clearly further
investigations into nitrogen chemistry are required.

\subsubsection{Cosmic ray measurements.}

Energy input from Galactic cosmic rays, mainly relativistic protons
and helium ions, drives important processes in the ISM.  Ionization of
H and H$_2$ heats the gas and initiates chemical reactions.  Cosmic
rays interacting with the gas break apart ambient C, N and O nuclei in
a process called spallation, producing significant quantities of
stable Li, Be and B isotopes.  The interactions with H and H$_2$ also
lead to $\gamma$-ray production through the decay of neutral pions.
Many of these processes are dominated by low energy cosmic rays (tens
of MeV), which are shielded from the Earth by the magnetic field of
the Sun.

One way to obtain the cosmic ray ionization rate involves
measurements of H$_3^+$ in diffuse molecular clouds (Snow and McCall
2006).  The analysis is dependent on an accurate determination of the
dissociative recombination rate coefficient, which until recently was
poorly known.  Measurements using storage rings (McCall \etal 2003;
Kreckel \etal 2005; Tom \etal 2009; Kreckel \etal 2010) and afterglows
(Glosik \etal 2008, 2009; Kotr\'{\i}k \etal 2010), as well as
theoretical calculations (Dos Santos \etal 2007), are now converging
on the most appropriate value for the rate coefficient.  The cosmic
ray ionization rate in diffuse molecular clouds inferred from H$_3^+$
observations is now more secure (e.g., Indriolo \etal 2007).  One
implication of this work is that the shape of the cosmic ray spectrum
may differ from what has commonly been assumed (Indriolo \etal 2009).
	


\subsection{Plasma physics}

\subsubsection{Supernova remnants: radiative shock thermal instabilities.}
\label{GAN:Pla:SNR:rad}

During the supernova phase, a contact surface forms at the change in
density gradient where the stellar envelope gives way to the stellar
wind, between the driven forward shock and an eventual reverse shock.
This contact surface is unstable and is subject to instabilities such
as those discussed in sections~\ref{SSE:Pla:Ins} and
\ref{SSE:Pla:Rad}.  Other issues that arise in SNRs involve the
predicted role of radiation.  Here and in
section~\ref{GAN:Pla:SNR:Vis} we discuss two of these issues.

As the shocks produced by SNe or other circumstances propagate across
the ISM, the newly shocked material cools by the emission of
radiation.  The rate of cooling varies with temperature and there are
regimes in which linear theory and simulations find that this produces
an instability, causing oscillations in the shock velocity (Chevalier
and Imamura 1982; Innes \etal 1987; Kimoto and Chernoff 1997).
Observational evidence of cooling that might be part of such an
instability has been reported (Raymond \etal 1991).  The instability
also would be expected to occur in accreting systems such as TW Hydrae
and other T Tauri stars (Koldoba \etal 2008), but recent observations
find no evidence of it (Drake \etal 2009; Gunther \etal 2010).  This
creates a focused need for the observation of such instabilities in a
laboratory environment, to show if they can in fact exist.  This was
accomplished (Hohenberger \etal 2010) by the production of cylindrical
shock waves by focusing a 1.4~ps laser pulse into a medium composed of
Xe gas clusters (Moore \etal 2008; Symes \etal 2010).  Measurements of
the shock trajectory clearly showed velocity oscillations attributed
to this instability. Future experiments can proceed towards systems
that are more closely scaled to specific astrophysical cases.

\subsubsection{Supernova remnants: Vishniac instabilities.}
\label{GAN:Pla:SNR:Vis}

SNRs at times produce very thin dense shells of material by radiative
cooling, driven outward by the pressure within the SNR and
decelerating as they accumulate more mass.  Vishniac (1983) showed
such shells to be unstable.  Ryu and Vishniac (1991) showed that
blast waves producing a density increase above about 10 to 1 are
likewise unstable.  This instability also may operate in other
contexts where one finds a thin, dense shell, such as shocks emerging
from SNe (see section~\ref{SSE:Pla:Ins} and \ref{SSE:Pla:Rad}).
Clumping in simulations of SNRs is often attributed to this process
(van Veelen \etal 2009).  In observations, it is most often difficult
to tell whether observed clumping is due to this instability as
opposed to inhomogeneity in the medium being shocked (Grosdidier \etal
1998) or to other instabilities such as Rayleigh-Taylor.  However, the
underlying theory is highly simplified, involving several assumptions
including that the shell is infinitesimally thin and an unusual
definition of the sound speed in the shell.  This created the need for
experimental tests.  Experiments have produced the instability by
driving a blast wave through Xe gas, generating the required large
density increase by radiative cooling.  Grun \etal (1991) reported the
first observation attributed to this process, but it was only recently
that Edens \etal (2005) reported a test of the predicted growth rate.
Laming (2004) has discussed the common physics underlying these
instabilities in astrophysical and laboratory systems and the
connection of the Vishniac process with the thermal instability
discussed above.

\subsubsection{Shock-clump interaction.}

High resolution images of astrophysical environments reveal that, in
general, circumstellar and interstellar plasma distributions are
essentially heterogeneous.  Strong density perturbations over the
ambient density, $\delta\rho/\rho_{\rm amb} \ge 1$, exist on a range
of scales.  The origin of such heterogeneity may lie in turbulent
motions which exist in many astrophysical environments or through
thermal or dynamical instabilities.  Any supersonic flows through
these environments will necessarily involve so called {\it shock-clump
  interactions}.  The importance of such clumpy flows cannot be
understated as critical issues such as mixing, transport and global
evolution will all differ in clumpy as opposed to smooth flows.  The
observational literature shows many clump studies addressing these
issues in environments ranging from supernova to active galactic
nuclei (AGNs; Smith and Morse 2004; Chugai and Chevalier 2006; Byun
\etal 2006; Westmoquette \etal 2007; Fesen \etal 2011).

Theoretical studies of shock-clump interactions have relied heavily on
numerical simulations as the problem is essentially multi-dimensional
and nonlinear interactions dominate.  Many studies of adiabatic shocks
interacting with a single clump have been performed (e.g., Klein \etal
1994).  Studies of magnetized and radiatively cooled single shocked
clumps also exist but are fewer in number.  Only a handful of multiple
clump studies have been published (Fragile \etal 2004; Yirak \etal
2008).  Because 3-D simulation studies are often resolution limited
(Yirak \etal 2011) laboratory studies can offer relatively clean
platforms for deeper exploration of shock-clump interactions.  A
robust literature reporting a host of shock-clump high energy density
laboratory astrophysics (HEDLA) studies has emerged over the last
decade.

\begin{figure}
\begin{center}
\includegraphics[angle = 0, width=1.0\textwidth]{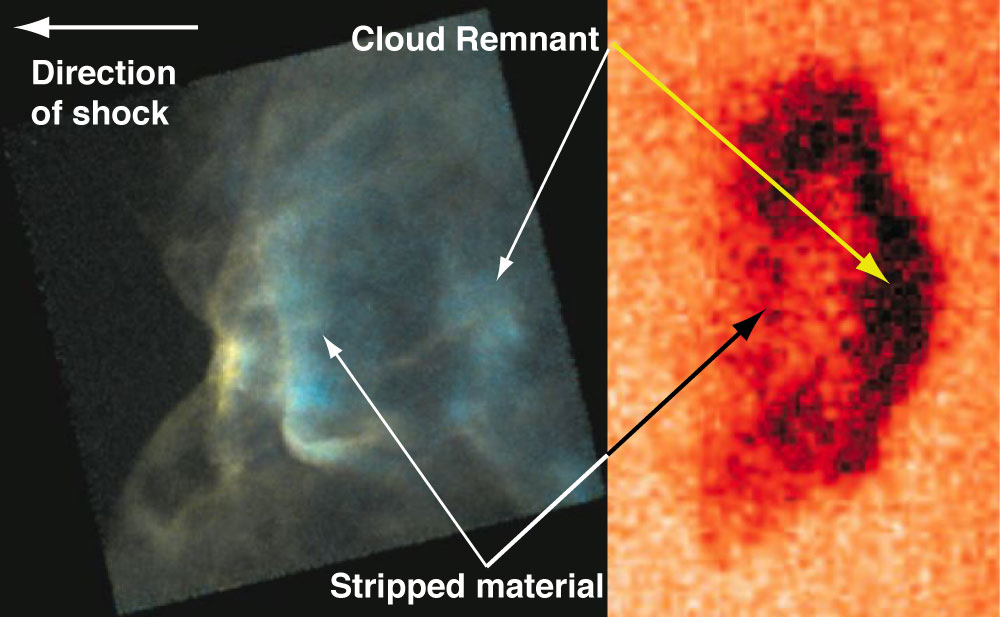}
\caption{Astrophysical data from {\it Chandra} (left) and from the
 laboratory Nova laser (right) showing one phase of shock-clump
 interaction.  Hwang \etal (2005) used the laboratory data of Klein
 \etal (2003) to interpret the astrophysical image.  Adapted from
 Rosner and Hammer (2010).}
\label{fig:puppis}
\end{center}
\end{figure}

The first HEDLA studies of shock-clump interactions focused on single
clumps interacting with a passing shock (Kang \etal 2000; Robey \etal
2002; Klein \etal 2003).  These works, along with simulations and
analytical work, were able to explore key features of shocked clump
evolution including the breakup of downstream vortices by the Widnall
instabilty.  Charateristic density distributions of the clump as it is
flattened by the passage of the shock along with break up of the
vortex ring were well characterzed in both experiments and
simulations.  The data shown in Klein \etal (2003) were used to
interpret the evolutionary stage of an observed structure in Puppis~A
by direct comparison with experimental data (Hwang \etal 2005; see
also figure~\ref{fig:puppis}).  Recent studies have begun focusing on
shock interactions with multiple clumps (Rosen \etal 2009).  Issues
such as the interaction of bow shocks from nearby clumps as well as
the effect of upstream clumps enhancing the breakup of downstream
clumps in their dynamic shadow are currently being explored
(Poludnenko \etal 2004).

\section{Galaxies across time}
\label{sec:GAT}

Galaxies across cosmic time covers ``the formation, evolution, and
global properties of galaxies and galaxy clusters, as well as active
galactic nuclei and [quasi-stellar objects (QSOs)], mergers, star
formation rate, gas accretion, and supermassive black holes''
(Blandford \etal 2010a).

\subsection{Atomic physics}

\subsubsection{Active galactic nuclei warm absorbers.}

Early {\it Chandra} and {\it XMM-Newton} observations of the AGN IRAS
13349+2438 detected a new absorption feature in the $15-17$~\AA\ range
(Sako \etal 2001).  This spectral feature is believed to originate in
the warm absorber material surrounding the central supermassive black
hole in AGNs and has since been observed in a number of other AGNs
(e.g., Pound \etal 2001; Blustein \etal 2002; Kaspi \etal 2002; Behar
\etal 2003; Sako \etal 2003; Steenbruge \etal 2003; Gallo \etal 2004;
Kaspi \etal 2004; Matsumoto \etal 2004; Pounds \etal 2004; Krongold
\etal 2005; Steenbrugge \etal 2005; McKernan \etal 2007).  These
unresolved transition arrays (UTAs) were quickly identified as $2p-3d$
innershell photoabsorption in iron M-shell ions (Sako \etal 2001).
New atomic calculations were soon carried out which demonstrated that
the shapes, central wavelengths and equivalent widths of these
features can be used to diagnose the properties of AGN warm absorbers
such as wind and outflow velocities, ionization and elemental
structure and mass loss rates and relative abundances (Behar \etal
2001; Gu \etal 2006).  However, the ability to diagnose these
properties was initially hindered by a lack of reliable ionization
balance calculations, proper line identification and wavelengths and
accurate absorption strengths.

Initial AGN models which matched absorption features from second- and
third-row elements failed to correctly reproduce the observed
absorption from the fourth-row element iron (e.g., Netzer \etal 2003).
The models predicted too high an iron ionization level.  This was
attributed to an underestimate in the models of the low temperature
dielectronic recombination rate coefficients for the Fe M-shell ions
(Netzer 2004; Kraemer \etal 2004).  This motivated a series of
theoretical calculations (Gu 2004; Badnell 2006a,b; Altun \etal 2006,
2007) and experimental studies (Schmidt \etal 2006, 2008; Luki\'c
\etal 2007; Lestinsky \etal 2009) which found dielectronic
recombination rate coefficients up to orders of magnitude larger than
the data previously available.  These data improved agreement of the
models with observations, though a number of issues still remain
(Kallman 2010).

Comprehensive spectral models of the deep {\it Chandra} observation of
the warm absorber in NGC~3783 suggested two ionization components in
pressure equilibrium (Krongold \etal 2003), with similar kinematic
velocities.  Netzer \etal (2003) found three ionization components
each with two sets of velocities and all three in pressure
equilibrium.  Subsequent theoretical calculations by Gu \etal (2006)
indicated only a single component in the wind, supporting the idea of
pressure equilibrium (see section~\ref{GAT:At:The}).  Until recently,
benchmark measurements capable of testing such bound-bound
photoabsorption calculations did not exist.  This has now become
possible with the use of a portable electron beam ion trap which can
be coupled to third or fourth generation light sources (Epp \etal
2007; Simon \etal 2010).  The results of Simon \etal (2010) largely
verified the calculation of Gu \etal (2006) for Fe~\textsc{xv}.  As a
result of the photoabsorption work described here and the dielectronic
recombination work mentioned above, more reliable models of AGN warm
absorbers are now being developed.  An example of this is discussed in
section~\ref{GAT:At:The}.

\subsubsection{Thermal stability of active galactic nuclei emission
line regions.}
\label{GAT:At:The}

Many models of the origin of the emission lines of AGNs have been
proposed (see chapter~14 of Osterbrock and Ferland 2006, hereafter
AGN3).  Possibilities include winds from stars or the accretion disk,
an ionized layer above the surface of the disk, or distinct clouds
confined by a surrounding hot medium.  If the latter is the case, then
the gas phases where clouds can exist are determined by the thermal
cooling curve.  This is the relationship between the gas temperature
and the cooling rate (AGN3, chapter 3).  If gas pressure equilibrium
applies, then regions with very different kinetic temperatures and
hydrogen densities can exist at the same gas pressure.  This scenario
dates back to early work done on the ISM (Field \etal 1969) and was
revived by Krolik \etal (1981) for AGNs.

The form of the cooling curve results from massive amounts of atomic
data.  Collisional excitation and radiative decay rates are needed for
thousands of lines while collisional and photoionization rates,
together with radiative, dielectronic and charge transfer
recombination rate coefficients, are needed for hundreds of ions.
Dielectronic recombination is the most uncertain of these rates.
Improvements in the dielectronic recombination data, mainly from
storage ring measurements and expanded theory, have greatly affected
our understanding of the stable phases (see the reviews of Schippers
2009 and Schippers \etal 2010).  Atomic theory and experiment are now
in far better agreement for the dielectronic recombination data with
significantly larger low temperature rate coefficients than those of
the previous generation.

Chakravorty \etal (2008, 2009) revisited the thermal stability of AGNs
using an updated version of the spectral simulation code Cloudy
(Ferland \etal 1998).  This code uses, among many data sources, the
compilation of recombination rates from Badnell \etal (2003) and
Badnell (2006c).  Chakravorty \etal (2008, 2009) found that the
updated dielectronic recombination rates produced significant changes
in the predicted distribution of ions.  The shape of the stability
curve changed significantly as a result.  These changes were large
enough that the existence of certain gas phases were affected with
implications for the final spectrum.  However, the modern dielectronic
recombination data do not extend to the low-charge, multi-electron
systems that are needed to fully understand AGN clouds.  This remains
an outstanding need.

\subsection{Molecular physics}

\subsubsection{Chemistry surrounding active galactic nuclei.}
\label{GCT:AP:AGNs:Chem}

Chemical models of X-ray dominated regions (XDRs) surrounding AGNs and
YSOs (Maloney \etal 1996) reveal significant abundances of doubly
charged ions to be cospatial with H$_2$.  The role of doubly-charged
ions as a diagnostic has been actively pursued since then.  Dalgarno
(1976) pointed out the potential importance of reactions involving
these ions.  Recently, Abel \etal (2008) considered the effects on
AGNs.  Laboratory studies show that some ${\rm X}^{2+} + {\rm H}_2$
reactions occur rapidly at elevated temperatures.  Chen \etal (2003)
measured a total rate coefficient for the reaction ${\rm S}^{2+} +
{\rm H}_2$, while Gao and Kwong (2003) studied the reaction ${\rm
  C}^{2+} + {\rm H}_2$.  Neither study, however, determined branching
fractions among the various final chemical channels.  Abel \etal
(2008) estimated what branching fractions would yield an observable
effect on the SH$^+$ chemistry.  They found that as long as the branch
to ${\rm SH}^+ + {\rm H}$ was a few percent, doubly-ionized chemistry
would be the dominant pathway for SH$^+$ production.  They also showed
that S$^{2+}$ was effectively destroyed once H$_2$ forms and that the
S$^{2+}$ abundance remains high in gas dominated by atomic hydrogen
and not only in ionized gas as was previously thought.  A key
consequence of their calculations is that much of the mid-infrared
emission from [S~\textsc{iii}] at 18.7 and 33.5~$\mu$m may come from
the XDR and not the ionized gas associated with an AGN.  Recent
detections of SH$^+$ in our Galaxy (Menten \etal 2011) suggest the
possibility for observing this molecular ion elsewhere and using the
proposed diagnostics of Abel \etal (2008).

\section{Cosmology and fundamental physics}
\label{sec:CFP}

Cosmology and fundamental physics includes ``the early universe, the
microwave background, the reionization and galaxy formation up to
virialization of protogalaxies, large scale structure, the
intergalactic medium, the determination of cosmological parameters,
dark matter, dark energy, tests of gravity, astronomically determined
physical constants, and high energy physics using astronomical
messengers'' (Blandford \etal 2010a).

\subsection{Atomic physics}

\subsubsection{Primordial abundances.}
\label{CFP:At:BBN}

The abundances of the primordial elements H, D, $^3$He, $^4$He, and
$^7$Li provide a key test of Big Bang cosmology.  The data are taken
from neutral gas in the Lyman-alpha forest for D, H~\textsc{ii}
regions both within the Galaxy ($^3$He) and outside ($^4$He), and
observations of metal-poor stars for $^7$Li (Steigman 2011).
Corrections are made for the effects of post Big Bang nucleosynthesis
(BBN) processing.  For example, D and $^7$Li are burned in stellar
environments, and $^7$Li is synthesized in cosmic ray interactions
with nuclei in the ISM.  For a recent discussion, see Charbonnel \etal
(2010).

The primordial $^4$He abundance is usually measured in giant
H~\textsc{ii} regions or dwarf irregular galaxies.  In these
extragalactic emission nebulae, H and He are photoionized.
Corrections for stellar production of $^4$He are determined from
correlations with metallicity.  A recent examination of 93 spectra for
86 low-metallicity extragalactic H~\textsc{ii} regions showed a linear
dependence of $^4$He on O/H, and yielded an extrapolated
zero-metallicity $^4$He mass fraction of $0.2565\pm0.0010({\rm
  stat})\pm0.0050({\rm syst})$ (Izotov and Thuan 2010).  Others have
advocated more conservative errors (Aver \etal 2010).

Accurate $^4$He/H determinations from ratios of optical recombination
lines require precise photo-production rates for electron
recombination with H$^+$ (Storey and Hummer 1995) and $^4$He$^+$.
Recent atomic calculations for the two-electron system
$^4$He~\textsc{i} (Benjamin \etal 1999; Bauman \etal 2005; Porter
\etal 2007) are in good agreement.  Remaining issues include
collisional processes involving the ground or metastable levels,
photoionization cross sections for non-hydrogenic moderate-$n$,
small-$l$ levels and transition probabilities for these levels (Porter
\etal 2009).

The abundance of D is important because of its sensitivity to the
baryon-to-photon ratio $\eta_{\rm B}$, varying as $\eta_{\rm
  B}^{-1.6}$.  From a limited set of high red-shift, low-metallicity
QSO absorption line systems, $\log({\rm D/H}) = -4.55\pm0.04$ was
found (Pettini \etal 2008), in good agreement with the {\it Wilkinson
  Microwave Anisotropy Probe (WMAP)} determinations of $\eta_{\rm B}$.

The observations of $^7$Li in the atmospheres of old halo stars is
constant to within measurement errors of 5\% over a variety of masses
and metallicities.  While lithium is fragile in stellar environments,
a well-formed ``plateau'' is found at low metallicity, yielding an
abundance $^7{\rm Li/H} = (1.23^{+0.34}_{-0.16}) \times 10^{-10}$
(Ryan \etal 2000) that is about a factor of four below BBN predictions
based on the {\it WMAP} $\eta_{\rm B}$.  Any astrophysical explanation
of this anomaly would have to account for the stability of the
plateau.

\subsubsection{Protogalaxy and first star formation.}
\label{subsubsec:AtomicProto}

In the early universe during the formation of protogalaxies and the
first stars, commonly called Population III stars, H$^-$ plays an
important role in the formation of H$_2$, as is described in
section~\ref{subsubsec:MolProto}.  H$_2$ is an important coolant leading
to the formation of structure during this epoch and reliable
predictions of the H$^-$ abundance are critical for reliable
cosmological models.  H$^-$ can be destroyed by photodetachment
\begin{equation}
\label{eq:HgammaPD}
{\rm H}^- + \gamma \to {\rm H} + {\rm e}^-
\end{equation}
and by mutual neutralization
\begin{equation}
\label{eq:H+H-MN}
{\rm H}^+ + {\rm H}^- \to {\rm H} + {\rm H}.
\end{equation}
These processes decrease the H$^-$ abundance, thereby limiting the
amount of H$_2$ which forms and correspondingly reducing the cooling
of the primordial gas.

Recent theoretical work has been carried out for each of these
reactions.  Miyake \etal (2010) have calculated new photodetachment
rates taking into account both the H$^-$ resonance states lying near
11 eV and radiation fields characteristic of Population III stars,
blackbody sources, power-law spectra and the hydrogen Lyman modulated
sawtooth spectra of the high-redshift intergalactic medium.  Stenrup
\etal (2009) have recently calculated new mutual neutralization data
valid for temperatures relevant during protogalaxy and first star
formation.  Their results agree with previous theoretical calculations
to within $30-40\%$ (Bates and Lewis 1955; Fussen and Kubach 1986),
but are about a factor of $2-3$ smaller than the experimental results
of Mosely \etal (1970), suggesting the need for further experimental
work.

\subsection{Molecular physics}

\subsubsection{Protogalaxy and first star formation.}
\label{subsubsec:MolProto}

Ro-vibrational collisional excitation of H$_2$ followed by radiative
relaxation is an important cooling mechanism in the early universe.
H$_2$ is formed during this epoch by the associative detachment
reaction
\begin{equation}
\label{eq:HH-AD}
{\rm H} + {\rm H}^- \to {\rm H}_2^- \to {\rm H}_2 + {\rm e}^-.
\end{equation}
H$_2$ formation, in turn, can be limited by
reactions~(\ref{eq:HgammaPD}) and (\ref{eq:H+H-MN}), both of which
compete with reaction~(\ref{eq:HH-AD}) for H$^-$ anions.

\begin{figure}
\begin{center}
\includegraphics[angle = 0, height=0.35\textheight]{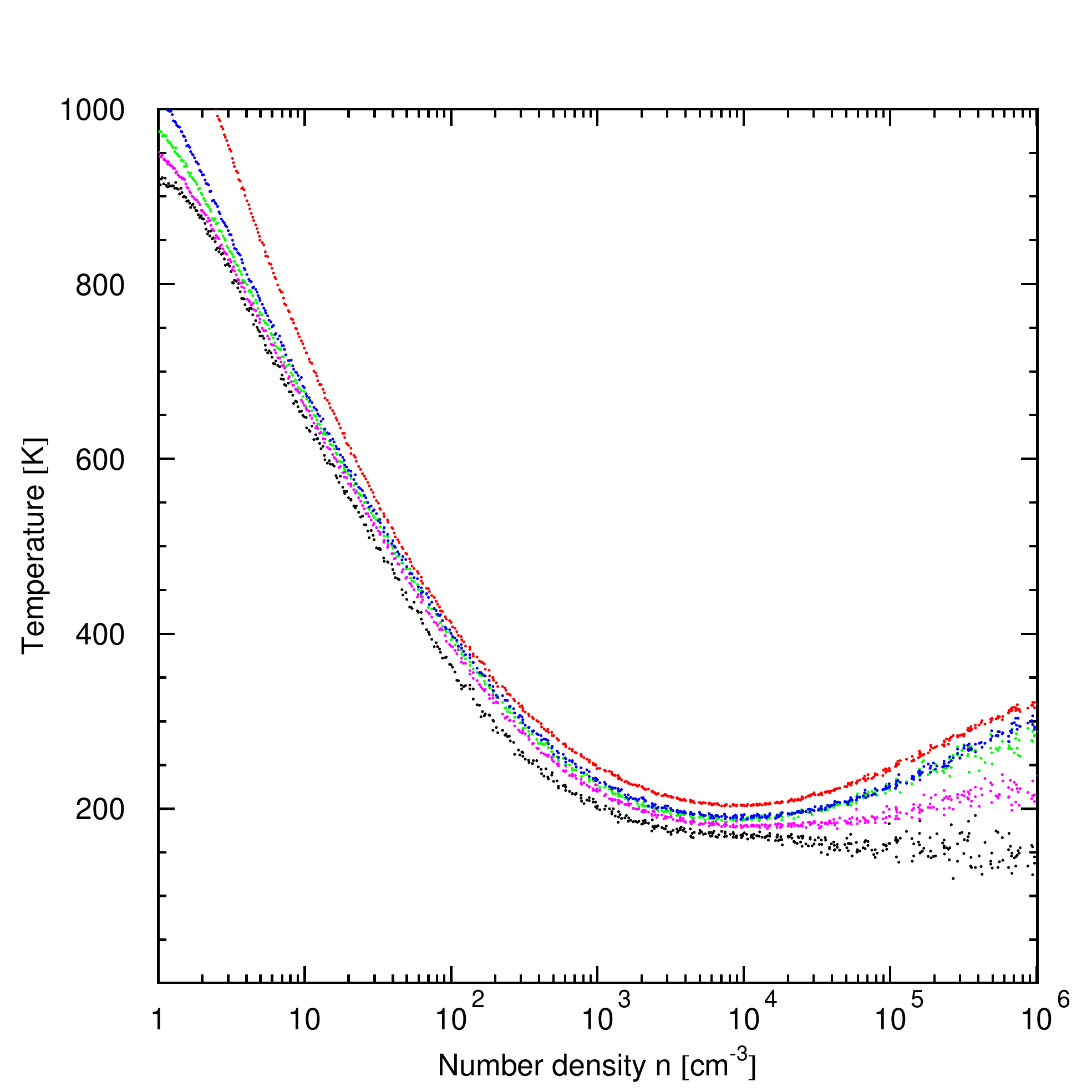}
\caption{\label{fig:PrimoridalCloudTemp} Evolution of a primordial
  cloud evolving in an initially ionized protogalactic halo using
  various associative detachment rate coefficients (Kreckel \etal
  2010).  Each point corresponds to a spherical shell of material
  surrounding the center of the cloud.  The black and red data use the
  previous upper and lower limits for the associative detachment
  reaction as discussed in Glover \etal (2006).  The green data use
  the experimentally benchmarked theoretical results of Kreckel \etal
  (2010), while the magenta and blue data use a rate coefficient,
  respectively, 25\% larger and smaller than this.  During this epoch
  the Jeans mass is set by density at the minimum temperature reached,
  leading to a factor of 20 uncertainty with the old data and a factor
  of only 2 with the new.}
\end{center}
\end{figure}

Until recently, there was nearly an order-of-magnitude uncertainty in
the rate coefficient for reaction~(\ref{eq:HH-AD}) (Glover \etal
2006).  This uncertainty severely limited our ability to model
protogalaxies and metal-free stars forming from initially ionized gas,
such as in ionized regions (i.e., H~\textsc{ii} regions) created by
earlier Population III stars (Glover \etal 2006; Glover and Abel 2008;
Kreckel \etal 2010).  Recently, measurements for this reaction have
been carried out using a merged-beams apparatus leading to an
experimentally-benchmarked theoretical rate coefficient with an
uncertainty of $\pm 24\%$ (Bruhns \etal 2010a; Kreckel \etal 2010;
Bruhns \etal 2010b).  As a result, for example, the uncertainty in the
model-predicted Population III Jeans mass due to errors in the atomic
data has decreased from a factor of 20 to 2 (Kreckel \etal 2010; see also
figure~\ref{fig:PrimoridalCloudTemp}).  As
a result of all the experimental and theoretical work described here
and in section~\ref{subsubsec:AtomicProto}, we are significantly closer
to the point where remaining uncertainties in models for protogalaxy
and first star formation tell us something about cosmology and not
about the underlying chemistry.

\subsection{Nuclear physics}

\subsubsection{Big Bang nucleosynthesis.}
\label{CFP:Nuc:BBN}

The comparison between BBN calculations and primordial abundances is a
cornerstone of modern cosmology, determining $\eta_{\it B}$ (now
confirmed by {\it WMAP}), and limiting the baryonic matter
contribution to the universe to about 4\% of the closure density
(Olive 1999).  Thus most of the dark matter is nonbaryonic.  BBN in
combination with inventories of the matter in stars, inter-cluster
diffuse gas, and the intergalactic medium indicate that a significant
fraction ($\gax 25$\%) of the baryonic matter is nonluminous (Silk
1999; Bregman 2007).

\begin{figure}
\begin{center}
\includegraphics[angle = 0, height=0.35\textheight]{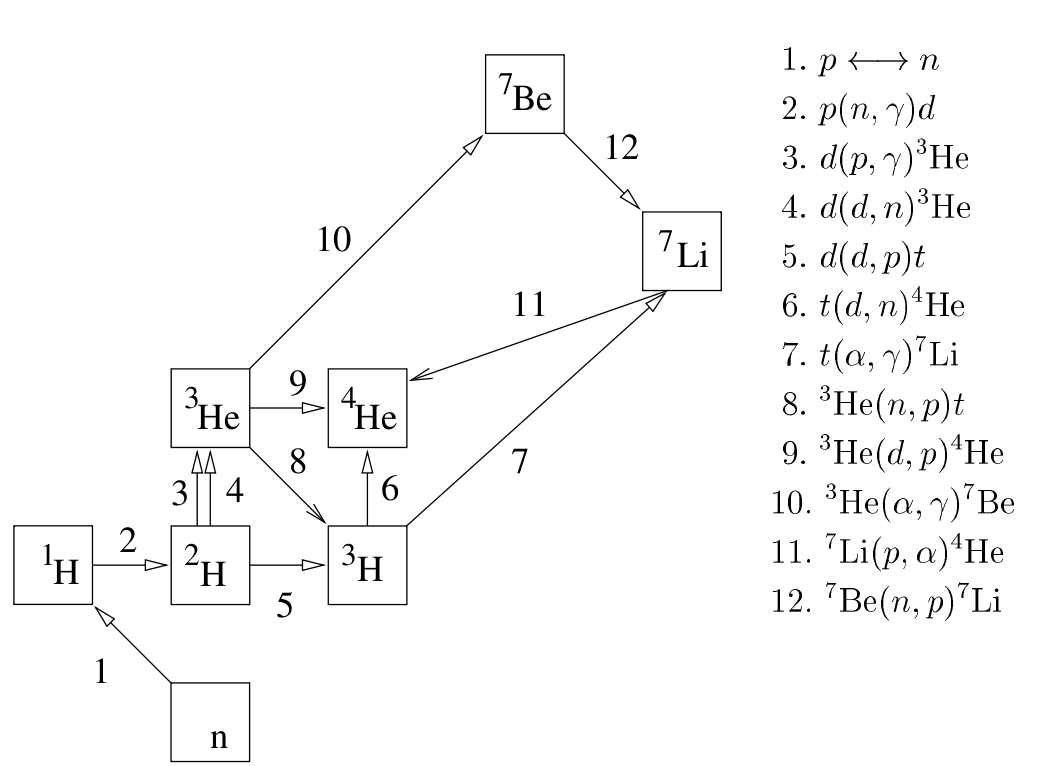}
\caption{The reaction network for Big Bang nucleosynthesis, from 
  Nollett and Burles (2000).} 
\label{fig:BBN}
\end{center}
\end{figure}

BBN calculations depend on the Maxwellian-averaged nuclear cross
sections for the various reactions of Fig.~\ref{fig:BBN}.
Comprehensive efforts have been made to assess the effects of cross
section uncertainties on BBN predictions (see, e.g., Nollett and
Burles 2000 and Coc and Vangioni 2010).  Table~1 of Coc and Vangioni
(2000) describes the impact of key nuclear physics uncertainties on
the abundances of $^4$He, D, $^3$He, and $^7$Li, given the {\it WMAP}
value of $\eta_{\rm B}$.  The $^4$He yield is sensitive to weak rates
now well constrained by neutron $\beta$ decay (Lopez and Turner
1999).  The reaction ${\rm n} + {\rm p} \rightarrow {\rm D} + \gamma$
has a large impact on $^7$Li by competing with
$^7$Be(n,p)$^7$Li(p,$\alpha$)$^4$He for neutrons: $^7$Li is
synthesized as $^7$Be at the {\it WMAP} value for $\eta_{\it B}$.
While there is meager low-energy data on this reaction, calculations
based on effective field theory (EFT) treatments are believed to be
accurate to 1\% (Chen and Savage 1999; Ando \etal 2006).  New
measurements (Tornow \etal 2003) of the inverse reaction, made at
energies of 2.39-4.05 MeV, are in excellent agreement with EFT
predictions.  $^7$Li is also sensitive to the production channel rate
for $^3$He($\alpha,\gamma)^7$Be.  Four new data sets, summarized in
Adelberger \etal (2011), have now determined this cross section to
$\pm 5.2$\%.  Recent measurements (Leonard \etal 2006) of a third
reaction important to $^7$Li, $^2$H(d,p)$^3$H, confirm earlier
parameterizations of this cross section.
 
Recent work has not uncovered a nuclear physics explanation for the
discrepancy between BBN predictions and the $^7$Li abundance
determined from metal-poor stars.  For a discussion, see Chakraborty
\etal (2010) and references therein.

\subsubsection{Ultra-high-energy cosmic rays and neutrinos.}

Recent instrumentation advances in high-energy astrophysics include
the Pierre Auger Observatory (Pierre Auger Collaboration 2010), for
the study of ultra-high-energy (UHE) cosmic rays, the IceCube
Observatory, a South Pole high-energy neutrino detector scheduled for
completion in 2011 (Abbasi \etal 2010), and prototype UHE neutrino
detectors, such as the Antarctic Impulsive Transient Antenna (ANITA)
experiment (Barwick \etal 2006) and the Radio Ice Cherenkov Experiment
(RICE; Kravchencko \etal 2006).
 
The Pierre Auger program includes measurements of the spectrum,
anistropies, and composition of UHE cosmic rays, including at the GZK
cutoff (Greisen 1966; Zatsepin and Kuz'min 1966) of $\sim 5 \times
10^{19}$ eV.  Interactions with the cosmic microwave background (CMB)
limit the distances UHE protons/nuclei can travel.  Interactions with
the CMB and with infrared, optical, and UV background photons are well
constrained by a large database of laboratory nuclear physics.  The
energy-loss mechanisms include single and multiple pion production off
the proton, nuclear reactions such as photodisintegration, photo-pair
processes, and photoabsorption followed by re-emission.  References to
propagation models based on this input physics can be found in Kotera
and Olinto (2011).
 
A key objective of the Pierre Auger science program, determining the
primary energy and mass of UHE cosmic rays, requires a detailed model
of the interactions of cosmic ray protons and nuclei with nuclei in
the upper atmosphere.  Cosmic rays above 10$^{14}$ eV are measured
indirectly, through the cascades of secondary particles that result
from their atmospheric collisions.  The energy and composition of the
incident UHE cosmic ray are determined by comparing the observed
extensive air showers with those predicted by models.  Center-of-mass
energies near the GZK cutoff are two orders of magnitude beyond the
limits of our highest energy machines, the Large Hadron Collidor (LHC)
and the Relativistic Heavy Ion Collidor (RHIC).  Thus significant
extrapolations are required.  For a discussion of the uncertainties,
see Alessandro \etal (2011).  Recent tests of existing codes against
first LHC data are described in d'Enterria \etal (2011).
 
Cosmic ray neutrinos are a tool for probing the universe at asymptotic
energies and distances and for identifying point sources, as neutrinos
are not deflected by magnetic fields.  IceCube was designed to detect
neutrinos with energies between 10$^{10}$ and 10$^{17}$ eV, through
the Cerenkov light emitted by charged particles they produce.  The
extension to higher energies, required to detect the neutrinos from
the nuclear reactions responsible for the GZK cutoff, requires ice
volumes a factor $\sim$ 100 beyond IceCube's km$^3$, as well as new
detection techniques.  Methods under development are based on coherent
radio emission, the Askaryan effect (Askaryan 1962; Askaryan \etal
1979).  Recent laboratory tests of the Askaryan effect using targets
of silica and rock salt confirmed that radio emission provides a means
of detecting UHE neutrinos (Saltzberg \etal 2001; Gorham \etal 2005).

\subsection{Particle physics}

\subsubsection{Baryon number asymmetry: experiment.}

The explanation for the excess of baryons over antibaryons in the
early universe, and thus a nonzero $\eta_{\rm B}$, is a key puzzle in
cosmology.  Baryogenesis requires charge-parity (CP) violation and
baryon number violation.  CP violation arises in the standard model
through the Cabibbo-Kobayashi-Maskawa (CKM) phase and through the
quantum chromodynamics (QCD) $\bar{\theta}$ parameter, and has been
observed in the laboratory in kaon decays and at the B factories.
However, the known CP violation is not sufficient to account for the
baryon number asymmetry.  Baryon number violation has not been seen in
the laboratory, despite intense effort.

Static electric dipole moments (EDMs) require CP-violation.  As there
is a significant gap between current experimental bounds on EDMs and
standard-model predictions based on the CKM phase, detection of an EDM
might indicate a new source of CP violation relevant to baryogenesis.
Current limits come from atomic beam experiments on the electron EDM,
$|d_{\rm e}|<1.6\times 10^{-27}$ e cm (Commins \etal 1994), and from
trap experiments with ultracold neutrons, $|d_{\rm n}|<2.9\times
10^{-26}$ e cm (Baker \etal 2006).  Alternatively, neutron and proton
EDMs as well as CP-violating nucleon-nucleon (NN) interactions can be
probed in neutral atoms.  The $^{199}$Hg vapor-cell experiment,
$|d(^{199}\mathrm{Hg})| < 3.1 \times 10^{-29}$ e cm, provides the most
stringent limits on the proton and quark chromo EDMs, and on the
strength of scalar, pseudoscalar, and tensor CP-violating semileptonic
interactions (Griffith \etal 2009).

Baryogenesis could have arisen from the decays of heavy right-handed
neutrinos, with the symmetry violation communicated to the baryons
through mechanisms within the standard model (so-called
``sphalerons''; Fukugita and Yanagida 1986).  Recent laboratory
discoveries -- nonzero neutrino masses and two large mixing angles --
have made this scenario quite plausible.  The CP-violating observable
is proportional to a product that involves the three mixing angles and
the Dirac CP phase.  A great deal of laboratory effort is now focused
on both short- and long-baseline neutrino oscillation experiments to
measure the third mixing angle and to detect leptonic CP violation at
low energies by comparing neutrino oscillation channels, e.g.,
$\nu_\mu \rightarrow \nu_{\rm e}$ and $\bar{\nu}_\mu \rightarrow
\bar{\nu}_{\rm e}$.  Experiments in progress include the Daya Bay (Lin
2011) and Double Chooz (Palomaries 2009) reactor experiments, and the
Tokai-to-Kamioka (T2K) long-baseline neutrino oscillation experiment
(Rubbia 2011).  FermiLab ``intensity frontier'' plans include a search
for neutrino CP violation (see {\tt
  http://www.fnal.gov/pub/science/experiments/intensity/}).

Laboratory limits on baryon number violation come from proton decay
searches.  The Super-Kamiokande Collaboration (Nishino \etal 2009) has
placed limits on the partial lifetimes for modes favored by minimal
SU(5) Grand Unified Theories (GUTs), ${\rm p} \rightarrow e^+ \pi^0$
and ${\rm p} \rightarrow \mu^+ \pi^0$, of $8.2 \times 10^{33}$ yr and
$6.6 \times 10^{33}$ yr, respectively, at a 90\% confidence level.  The
collaboration has also established (Kobayashi \etal 2005) stringent
limits on modes favored by super-symmetric GUTs, ${\rm p} \rightarrow
\mu^+K^0$, ${\rm n} \rightarrow \bar{\nu} K^0$, ${\rm p} \rightarrow
\mu^+K^0$, and ${\rm p} \rightarrow e^+ K^0$ of $2.23 \times 10^{34}$,
$1.3 \times 10^{32}$, $1.3 \times 10^{33}$, and $1.0 \times 10^{33}$
yr, respectively.

\subsubsection{Baryon number asymmetry: theory.}

In theory, no major paradigm shift has occurred in the last ten years.
(For a review of baryogenesis models see Dine and Kusenko 2003.)
However, considerable progress has been made in refining the
predictions of various scenarios and new possibilities have been
proposed.  In one class of models, the baryon asymmetry is produced at
the electro-weak phase transition, as a result of new physics at the
electro-weak scale, such as supersymmetry.  While the basic scenario
for electro-weak baryogenesis (EWB) was described long ago (Kuzmin
\etal 1985), recent developments include a re-evaluation (Lee \etal
2005) of the relevant source terms which bias the production of a net
baryon number via sphaleron transitions (Huet and Nelson 1996) and of
the associated resonant relaxation effects (Lee \etal 2005).  Also, it
was realized that the supersymmetric parameter which is space
compatible with the production of enough baryon asymmetry possesses a
two-resonances structure (Cirigliano \etal 2006, 2010).  One of the
two resonances corresponds to the scenario of ``bino-driven'' EWB,
where the origin of dark matter is deeply connected with that of the
baryon asymmetry (Li \etal 2009).

As possible experimental EWB tests, it was recently pointed out
that the EDM size for the electron and for the neutron is bounded
from below in EWB (Li \etal 2010), as a result of unavoidable
electro-weak two-loop contributions (Li \etal 2008).  The issue of
producing a strong enough first order phase transition in
supersymmetry (Carena \etal 2009) has also been investigated, together
with the possibility of enhancing the first order character altering
the Higgs sector, for instance adding a singlet scalar field (Pietroni
1993; Apreda \etal 2002; Profumo \etal 2007).  Questions related to
the gauge-dependence of criteria identifying strong enough first
order EW phase transitions have also been recently studied (Patel and
Ramsey-Musolf 2011).

Numerous recent efforts targeted the ``coincidence problem'' given by
the ratio of the baryonic density $\Omega_{\rm b}$ to non-baryonic
dark matter density $\Omega_{\rm DM}$ being of order unity
($\Omega_{\rm DM}/\Omega_{\rm b}\sim5$).  A variety of proposals have
been recently put forward, including darkogenesis (Shelton and Zurek
2010), xogenesis (Buckley and Randall 2010) and hylogenesis (Davoudias
\etal 2010) that for reasons of space we cannot review here.

Remarkable progress have also been made on the front of leptogenesis
models (for a comprehensive review see Giudice \etal 2004).  Recent
developments include the flavordynamics of leptogenesis (Pilaftsis
2005) and resonant leptogenesis near the electroweak phase transition
(Pilaftsis and Underwood 2005).  Some of these models might be
testable with the LHC and with experiments sensitive to lepton-number
and/or lepton-flavor violation (Pilaftsis 2009).

\subsubsection{Direct dark matter detection.}

A wide-spread experimental campaign is afoot to search for signatures
of Galactic dark matter scattering off ordinary matter nucleons.
These efforts are theoretically motivated by various compelling
considerations (Goodman and Witten 1985) and typically target weakly
interacting massive particles (WIMPs), although axion searches have
also been very active in the last decade (Duffy and van Bibber 2009).

WIMPs can undergo elastic or inelastic scattering processes with
nucleons (in the latter case exciting or ionizing the target atom, or
producing the nuclear emission of a photon).  The possibility of WIMPs
transitioning themselves to an excited state has also been envisioned
(Tucker-Smith and Weiner 2001).  We briefly review here elastic dark
matter scattering only, a process that can occur via spin-dependent or
spin-independent interactions.  The nuclear recoil induced by WIMP
scattering can produce light (scintillation), charge (ionization)
and/or phonon (heat) signals.  In practice, current generation direct
detection experiments are typically sensitive to two or more of these
signals, with the aim of achieving the best possible background
rejection.  Experiments that make use of scintillation and ionization
include for instance 
XENON (Aprile \etal 2010) and 
ZEPLIN (ZonEd Proportional scintillation in LIquid Noble gases; 
Akimov \etal 2010); 
among those that use scintillation and heat is
CRESST (Cryogenic Rare Event Search with Superconducting Thermometers;
Angloher \etal 2008), while 
CDMS (Cryogenic Dark Matter Search; Akerib \etal 2006) and
EDELWEISS (Exp\'erience pour DEtecter Les Wimps En Site Souterrain;
Gerbier 2010) make use of both ionization and heat.  Other
experimental setups that make use of one channel only include the
scintillation experiment
DAMA/LIBRA (DArk MAtter/Large sodium Iodide Bulk for RAre processes;
Bernabei \etal 2010) or the ionization experiment
CoGeNT (Contact Germanium Neutrino Telescope; Aalseth \etal 2008).  
Interestingly, the latter two experiments
recently reported controversial signals that have been attributed to
Galactic dark matter (Fitzpatrick \etal 2010).

\begin{figure}
\begin{center}
\includegraphics[angle = 0, height=0.35\textheight]{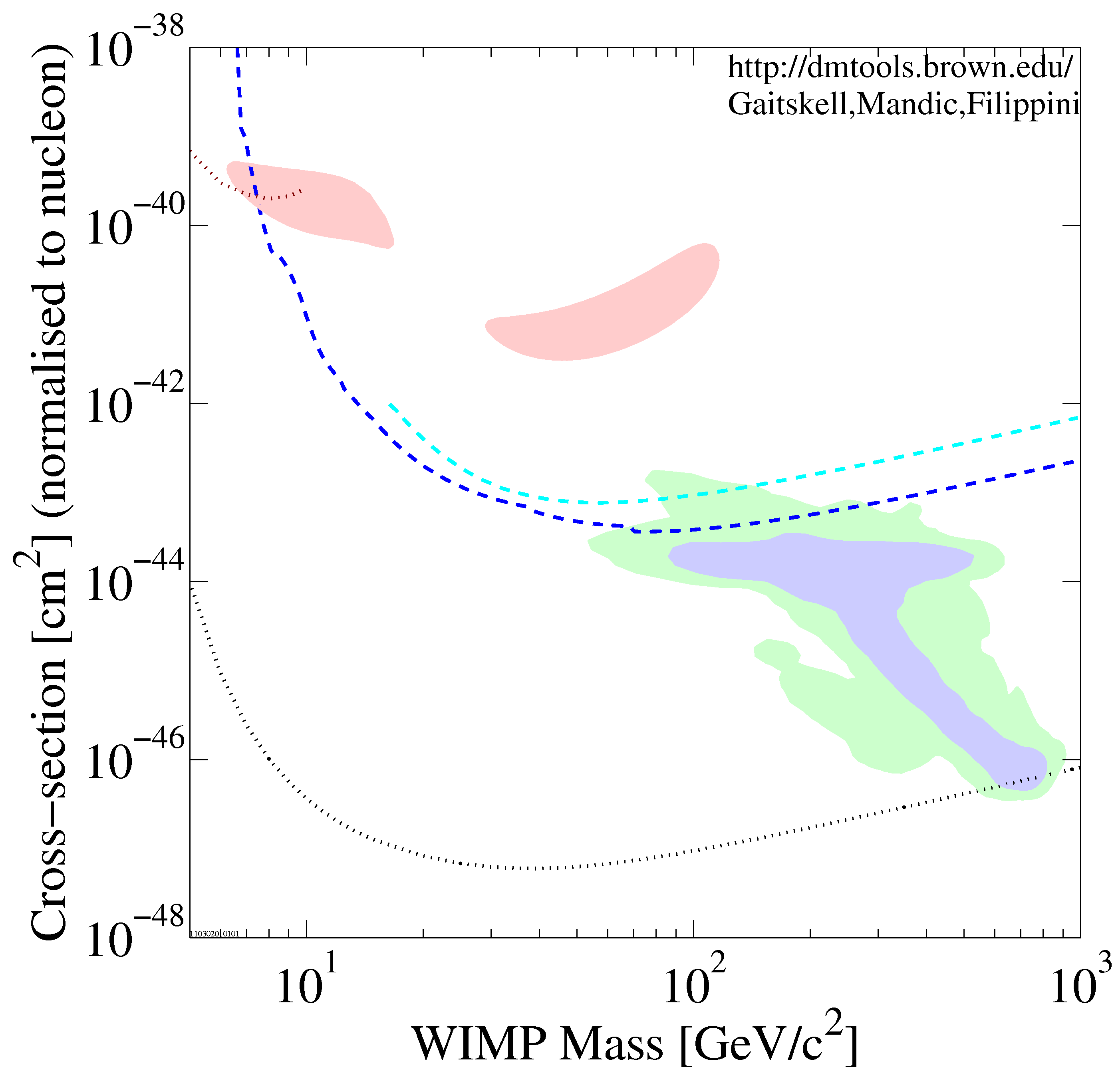}
\caption{\label{fig:DarkMatter} The plane of dark matter-proton
  spin-independent cross section versus mass. The dark and light blue
  lines indicate constraints from CDMS and ZEPLIN, respectively, while
  the red dotted line is from CoGeNT: parameter space points above the
  lines are experimentally excluded. The light red areas represent
  regions compatible with the positive annual modulation signal from
  DAMA/LIBRA (Bernabei \etal 2010). The light green and blue regions
  indicate theoretical predictions for the 95\% ad 68\% confidence
  level regions of CMSSM parameter space as determined in Trotta \etal
  (2008).  Plot obtained through {\tt http://dmtools.brown.edu}}
\end{center}
\end{figure}

The first positive direct detection signal has been reported by the
DAMA collaboration, with a rather impressive total exposure of 1.17
ton-yr (combining DAMA/NaI [DArk MAtter/Sodium-Iodine Target] and
DAMA/LIBRA), which quotes an annual modulation in the recoil energy
range of 2 to 6 keV electron equivalent at the 8.9$\sigma$ confidence
level (Bernabei \etal 2010).  The WIMP elastic-scattering
interpretation of this signal is largely inconsistent with limits
reported by XENON (Angle \etal 2008) and CDMS (CDMS II Collaboration
2010).  The CoGeNT experiment reported an exponential-like excess of
events in the few keV energy range, compatible with a light-mass WIMP
(Aalseth \etal 2011).  Anomalous events have also been reported by
CRESST and CDMS, although with relatively low statistical
significance.  Figure~\ref{fig:DarkMatter} presents a sample of recent
experimental and theoretical results on direct dark matter detection
on the plane defined by the WIMP mass and the spin-independent
WIMP-proton scattering cross section. The regions shaded in light red
are compatible with the DAMA/LIBRA modulation signal (Bernabei \etal
2010), while the theoretical expectation for the scattering cross
section in the Constrained Minimal Supersymmetric Standard Model
(CMSSM) is shown with the blue and green shaded areas (see Trotta
\etal 2008 for more details).  The figure also shows selected
experimental limits, that rule out the corresponding upper-right
corners of the plot.  The limits shown are from the CDMS (dark blue),
ZEPLIN (light blue) and CoGeNT (red dotted) experiments.  We also
indicate the projected reach of ton-size class experiments with a
black dotted line.  In summary, in the last ten years the field of
direct dark matter searches reached a stage of full maturity.  It is
exploring interesting regions of theoretically favored parameter space
and tantalizing signals are emerging from more than one experiment.

\subsubsection{Indirect dark matter detection.}

Many theoretically motivated models for dark matter, including weakly
interacting massive particles WIMPs, predict that dark matter
pair-annihilates into ordinary Standard Model particles. Searches for
the annihilation debris of dark matter are generically dubbed
``indirect dark matter detection''. In the last decade, indirect
detection has been one of the primary science goals of several new
experiments and telescopes looking for high-energy gamma rays,
neutrinos and antimatter.

Most notably, the {\it Fermi gamma-ray space telescope} (Atwood \etal
2009) set significant limits on the pair-annihilation rate of dark
matter from the non-observation, in gamma rays, of nearby dwarf
spheroidal galaxies (Abdo \etal 2010a), of clusters of galaxies
(Ackermann \etal 2010a) and of monochromatic gamma-ray lines (Abdo
\etal 2010b). Atmospheric Cherenkov telescopes, such as the Very
Energetic Radiation Imaging Telescope Array System (VERITAS), the High
Energy Stereoscopic System (H.E.S.S.) and the Major Atmospheric
Imaging Cherenkov (MAGIC) telescope, have also produced interesting
limits, for higher mass dark matter candidates (Aharonian \etal
2006). Construction of the IceCube neutrino telescope at the South
Pole was recently completed and the IceCube collaboration has
delivered the first limits on dark matter annihilation in the Galactic
Center (Abbasi \etal 2010, 2011) and from particles captured from the
center of the Sun (Heros 2010).

Anomalies in the high-energy flux of cosmic-ray positrons, including
the rising positron fraction measured by the {\it Payload for
Antimatter Matter Exploration and Light-nuclei Astrophysics} ({\it
PAMELA}) satellite (Adriani \etal 2009) in the 10-100 GeV range and
the hard positron-plus-electron flux reported by {\it Fermi}
(Ackermann \etal 2010b), triggered a great deal of excitement as
possible signatures of dark matter annihilation (Arkani-Hamed \etal
2009) or decay (Arvanitaki \etal 2009). Astrophysical explanations,
including nearby mature pulsars (Profumo 2008) as well as in-situ
secondary particle acceleration (Blasi 2009), have also been
put forward as plausible counterparts to the cosmic-ray
electron-positron anomalies.

Other signatures that have been associated with Galactic dark matter
annihilation include the WMAP haze (Hopper \etal 2007), a diffuse
radio emission that could be related to electrons and positrons
produced by dark matter and possibly a gamma-ray haze (Dobler \etal
2010). The evidence for the latter has been questioned (Linden and
Profumo 2010).  Recent re-analyses point towards two giant gamma-ray
``bubbles'' whose morphology appears incompatible with a dark matter
origin (Su \etal 2010).

\subsubsection{Dark matter theory.}

The last decade has seen giant leaps in theoretical studies concerning
dark matter.  On one hand, simulation of structure formation in
collisionless cold dark matter cosmologies have achieved unprecedented
resolution and level of detail; on the other hand, model building
inspired by possible experimental signals or by pure theoretical
arguments has triggered interesting new particle physics scenarios.

In the field of $N$-body simulations, which only include
gravitationally interacting dark matter, three milestones, among
several other exciting simulations, have been the Millennium (Springel
\etal 2005), Via Lactea (Diemand \etal 2007) and Bolshoi (Klypin \etal
2011) simulations.  While Millennium, in 2005, provided the basis for
hundreds of studies on statistical properties of dark matter halos and
on models for galaxy formation in a cosmological setting (including
mock catalogues and merger trees), Bolshoi (completed in 2010) uses an
updated set of cosmological parameters and will play a similar role in
the immediate future.  The Via Lactea suite of simulations specialized
on Milky-Way-size dark matter structure, with important implications
for indirect (Diemand \etal 2007) and direct (Kuhlen \etal 2010) dark
matter searches.  An important issue that will dominate future studies
of the dark matter distribution is the effect of baryons on the dark
matter density profiles (Duffy \etal 2010).

On the model-building frontier, numerous studies explored in detail
the phenomenology of supersymmetric models in collider, direct and
indirect detection (Baer \etal 2005).  Several groups focused on
statistical analyses of the supersymmetric (SUSY) parameter space,
based upon, e.g., a bayesian approach (Trotta \etal 2008).  Numerous
theoretical model-building efforts concentrated on explaining observed
anomalies in dark matter search experiments.  These include
leptophilic models (Fox and Poppitz 2009), models with a Sommerfeld
enhancement at low dark matter relative velocities (Pospelov \etal
2008), discussed to account for claimed indirect detection signals,
and inelastic (Tucker-Smith and Weiner 2001) dark matter models,
proposed to interpret direct dark matter signals.

\section{Discussion and Outlook for the Future}
\label{sec:end}

Our astrophysical understanding of the cosmos continues to be
propelled forward by advances in laboratory astrophysics.  This review
has touched on many, but far from all, of the achievements of the past
decade.  The coming decade promises to be equally, if not more,
fruitful.  The Astro 2010 Survey Report and Panel Reports (Blandford
\etal 2010a,b) have laid out a series of exciting scientific
objectives, the achievement of which they point out are going to
require numerous advances in laboratory astrophysics.  We direct the
reader to those reports for a detailed discussion.

Additional in depth discussions about the laboratory astrophysics
needs and opportunities for the coming decade can be found in a number
of White Papers written over the past few years.  These include those
submitted by the Working Group on Laboratory Astrophysics (WGLA) 
to the Astro 2010 Survey (Brickhouse \etal
2009a,b,c,d,e) as well as community input to Astro2010 through the
Science White Papers ({\tt
  http://sites.nationalacademies.org/BPA/BPA\_050603}) and the
Laboratory Astrophysics White Papers ({\tt
  http://sites.nationalacademies.org/BPA/BPA\_051118}).  Another White
Paper is that submitted by the WGLA to the U.S. National Research
Council Planetary Science Decadal Survey: 2013-2022 (Gudipati \etal
2009).  In plasma laboratory astrophysics, there have been a couple of
reports recently released by the community (Prager \etal 2010; Rosner
and Hammer 2010).  And most recently there is the White Paper from the
2010 Laboratory Astrophysics Workshop sponsored by the Astrophysics
Division of the Science Mission Directorate which covered atomic,
molecular, condensed matter and plasma laboratory astrophysics (Savin
\etal 2011).  These all point the way to the future and the richness
of discovery which we can only just begin to guess.

\ack

The authors thank their many colleagues including
J.\ E.\ Bailey,
P.\ Beiersdorfer,
G.\ V.\ Brown,
J.\ R.\ Crespo L\'opez-Urrutia
H.\ Ji,
H.\ Kreckel,
J.\ E.\ Lawler,
M.\ Medvedev,
T.\ Plewa,
D.\ Sasselov, 
R.\ K.\ Smith, 
C.\ Sneden,
B.\ J.\ Wargelin and
S.\ Widicus Weaver 
for stimulating conversations.  
NSB was supported in part by NASA contract NAS8-03060 to the
Smithsonian Astrophysical Observatory for the Chandra X-ray Center.
JJC is supported in part by the National Science Foundation through
grant AST 0707447.  
RPD acknowledges support from DOE/NNSA Defense Sciences and Advanced
Scientific Computing, from DOE/Science Office of Fusion Energy
Sciences and from the Defense Threat Reduction Agency.  
GJF acknowledges support by NSF (0908877), NASA (07-ATFP07-0124,
10-ATP10-0053, and 10-ADAP10-0073) and STScI (HST-AR-12125.01 and
HST-GO-12309).
MSG acknowledges funding from NASA Astrobiology Institute ``Icy
Worlds'' and support from the Jet Propulsion Laboratory, California
Institute of Technology, under a contract with the National Aeronautics
and Space Administration.
WCH was supported in part by the US Department of Energy 
under grant DE-SC00046548 to the University of California at Berkeley.
EH acknowledges the support of NASA through its program in laboratory 
astrophysics and through the {\it Herschel} program.  
SP is partly supported by the US Department of Energy with an
Outstanding Junior Investigator Award and by Contract
DE-FG02-04ER41268 and by NSF Grant PHY-0757911.
FS acknowledges the support of the Astrophysics Research and Analysis 
Program of NASA Science Mission Directorate.  
DWS is supported in part by the NASA Astronomy and Physics
Research and Analysis program, the NASA Solar Heliospheric Physics
program, and the NSF Division of Astronomical Sciences Astronomy and
Astrophysics Grants program.
EGZ was supported in part by the NSF grant PHY-0821899 to the Uiversity
of Wisconsin.

\pagebreak

\appendix

\section{Acronyms}
\label{app:acronyms}

\begin{table}
\caption{\label{tab:acronyms}List of acronyms used in the text.}
\begin{indented}
\item[]
\begin{tabular}{@{}ll}
\br
Acronym & Phrase \\
\mr
AGB         & Asymptotic Giant Branch \\
AGN         & Active Galactic Nucleus \\
AGNs        & Active Galactic Nuclei \\
ANITA       & ANtarctic Impulsive Transient Antenna \\
BBN         & Big Bang Nucleosynthesis \\
ccSNe       & core collapse SuperNovae \\
CDMS        & Cryogenic Dark Matter Search \\
CKM         & Cabibbo-Kobayashi-Maskawa \\
CMB         & Cosmic Microwave Background \\
CN          & Carbon-Nitrogen \\
CNO         & Carbon-Nitrogen-Oxygen \\
CoGeNT      & Contact Germanium Neutrino Telescope \\
CP          & Charge-Parity \\
CMSSM       & Constrained Minimal Supersymmetric Standard Model \\
CRESST      & Cryogenic Rare Event Search with Superconducting Thermometers \\
CVs         & Cataclysmic Variables \\
DAMA/LIBRA  & DArk MAtter / Large sodium Iodide Bulk for RAre processes \\
DAMA/NaI    & DArk MAtter / Sodium-Iodine target \\
DIB         & Diffuse Interstallar absorption Band \\
EDELWEISS   & Exp\'erience pour DEtecter Les Wimps En Site Souterrain \\
EDM         & Electric Dipole Moment \\
EOS         & Equations Of State \\
EWB         & Electro-Weak Baryogenesis \\
FRIB        & Facility for Rare Isotope Beams \\
FTIR        & Fourier Transform InfraRed \\
GRB         & Gamma Ray Burst \\
GUT         & Grand Unified Theory \\
GZK         & Greisen-Zatsepin-Kuz'min \\
HEDLA       & High Energy Density Laboratory Astrophysics \\
H.E.S.S.    & High Energy Stereoscopic System \\
IAU         & International Astronomical Union \\
IR          & InfraRed \\
ISM         & Interstellar Medium \\
KBO         & Kuiper Belt Object \\
LHC         & Large Hadron Collidor \\
LMC         & Large Magellanic Cloud \\
LTE         & Local Thermodynamic Equilibrium \\
LUNA        & Laboratory for Underground Nuclear Astrophysics \\
MAGIC       & Major Atmospheric Imaging Cherenkov \\
MHD         & MagnetoHydroDynamic \\
MRI         & MagnetoRotational Instability \\
MST         & Madison Symetric Torus \\
NN          & Nucleon-Nucleon \\
PAH         & Polycyclic Aromatic Hydrocarbon \\
{\it PAMELA}& {\it Payload for Antimatter Matter Exploration and 
              Light-nuclei Astrophysics} \\
PIC         & Particle-In-Cell \\
QCD         & QuantumChromoDynamics \\
QSO         & Quasi-Stellar Object \\
RHIC        & Relativistic Heavy Ion Collidor \\
RICE        & Radio Ice Cherenkov Experiment \\
\br
\end{tabular}
\end{indented}
\end{table}

\setcounter{table}{0}

\begin{table}
\caption{\label{tab:acronyms2}Continued.}
\begin{indented}
\item[]
\begin{tabular}{@{}ll}
\br
Acronym & Phrase \\
\mr
SEM-EDX     & Scanning Electron Microscopy using Energy-Dispersive X-ray \\
SMC         & Small Magellanic Cloud \\
SN          & Supernova \\
SNe         & SuperNovae \\
SNO         & Sudbury Neutrino Observatory \\
SNR         & SuperNova Remnant \\
SSM         & Standard Solar Model \\
SUSY        & SUperSYmettric \\
TEM         & Transmission Electron Microscopy \\
TMC         & Taurus Molecular Cloud \\
TNO         & Trans Neptunian Object \\
TOF-SIMS    & Time Of Flight Secondary Ion Mass Spectrometry \\
{\it TRACE} & {\it Transition Region and Coronal Explorer} \\
T2K         & Tokai-To-Kamioka \\
UHE         & Ultra-High-Energy \\
UIR         & Unidentified InfraRed \\
UTA         & Unresolved Transition Array \\
UV          & UltraViolet \\
VERITAS     & Very Energetic Radiation Imaging Telescope Array System \\
YSO         & Young Stellar Object \\
WGLA        & Working Group on Laboratory Astrophysics \\
WIMP        & Weakly Interacting Massive Particles \\
{\it WMAP}  & {\it Wilkinson Microwave Anisotropy Probe} \\
XDR         & X-ray Dominated Region \\
{\it XMM}   & X-ray Multi-mirror Mission \\
ZEPLIN      & ZonEd Proportional scintillation in LIquid Noble gases \\
3-D         & 3 Dimensional \\
\br
\end{tabular}
\end{indented}
\end{table}

A complete list of the acronmys used throughout the text is given
in table~\ref{tab:acronyms}.

\pagebreak

\section*{References}
\begin{harvard}
\item[] Aalseth C E \etal 2008 \prl {\bf 101} 251301
\item[] Aalseth C E \etal 2011 \prl {\bf 106} 131301
\item[] Abbasi R \etal 2010 \apj {\bf 732} 18
\item[] Abbasi R \etal 2011 \prd {\bf 84} 022004
\item[] Abel N P, Federman S R and Stancil P C 2008 \apj {\bf 675}
  L81--L84
\item[] Abdo A A \etal 2010a \apj {\bf 712} 147--158
\item[] Abdo A A \etal 2010b \prl {\bf 104} 091302
\item[] Abe K \etal 2011 \prd {\bf 83} 052010
\item[] Ackermann M \etal 2010a \jcap {\bf 1005} 025
\item[] Ackermann M \etal 2010b \prd {\bf 82} 092004
\item[] Adelberger E G \etal 2011 \rmp {\bf 83} 195--245
\item[] Adriani O \etal 2009 \nat {\bf 458}, 607--609
\item[] Ag\'undez M, Cernicharo J and Gu\'lin M 2007 \apj {\bf 662}
  L91--L94
\item[] Ag\'{u}ndez M, Cernicharo J and Goicoechea J R 2008a \aanda
  {\bf 483} 831--837
\item[] Ag\'ndez M, Cernicharo J, Pardo J, Fonfr\'{\i}a Exp\'osito J
  P, Gu\'elin M, Tenenbaum E D, Ziurys L M and Apponi A J 2008b \apss
  {\bf 313} 229--233
\item[] Ag\'undez M \etal 2010 \aanda {\bf 517} L2
\item[] Aharmim B 2010 \etal \prc {\bf 81} 055504
\item[] Aharonian F \etal 2006 \prl {\bf 97} 221102
\item[] A'Hearn M F 2008 \ssr {\bf 138} 237--246
\item[] Akimov D Y 2010 \etal \plb {\bf 692} 180--183
\item[] Akerib D S \etal 2006 \prl {\bf 96} 011302
\item[] Alessandro B \etal 2011 arXiv:1101.1852
\item[] Allamandola L J, Tielens A G G M and Barker J R 1989 \apjs 
  {\bf 71} 733--775
\item[] Alonso-Medina A, Col\'{o}n C and Rivero C 2005 \physcr {\bf
  71} 154--158
\item[] Altun Z, Yumak A, Badnell N R, Loch S D and Pindzola M S 2006
  \aanda {\bf 447} 1165--1174
\item[] Altun Z, Yumak A, Yavuz I, Badnell N R, Loch S D and Pindzola
  M S 2007 \aanda {\bf 474} 1051--1059
\item[] Amano T 2010 \apj {\bf 716} L1--L3
\item[] Anderson H M, den Hartog E A and Lawler J E 1996 \josab {\bf
  13} 2382--2391
\item[] Ando S, Cyburt R H, Hong S W and Hyun C H 2006 \prc {\bf 74}
  025809
\item[] Angloher G \etal 2008 \pos {\bf IDM2008} 014
\item[] Angle J \etal 2008 \prl {\bf 100} 021303
\item[] Apponi A J, Barclay W L Jr and Ziurys L M 1993 \apj {\bf 414}
  L129--L132
\item[] Apreda R, Maggiore M, Nicolis A and Riotto A 2002 \npb {\bf
  631} 342--368
\item[] Aprile E and Profumo S 2009 \njp {\bf 11}, 105002
\item[] Aprile E \etal 2010 \pra {\bf 105} 131302
\item[] Arpesella C \etal 2008 \prl {\bf 101} 091302
\item[] Araki M, Furuya T and Saito S 2001 \jmsp {\bf 210} 132--136
\item[] Arkani-Hamed N, Finkbeiner D P, Slatyer T R and Weiner N 2009
  \prd {\bf 79} 015014
\item[] Arvanitaki A, Dimopoulos, S, Dubovsky S, Graham P W, Harnik R
  and Rajendran S 2009 \prd {\bf 80} 055011
\item[] Arlandini C, K\"{a}ppeler F, Wisshak K, Gallino R, Lugaro M,
  Busso M and Straniero O 1999, \apj {\bf 525} 886--900
\item[] Arnett W D, Bahcall, J N, Kirschner R P and Woolsey S E 1989
  \araa {\bf 27} 629--700
\item[] Arnould M, Paulus G and Jorissen A 1992 \aanda {\bf 254} L9--L12
\item[] Askaryan G A 1962 \JETP {\bf 14} 441--443
\item[] Askaryan G A, Dolgoshein B A, Kalinovsky A N and Mokhov N V
  1979 \nim {\bf 164} 267--278
\item[] Asplund M, Grevesse N, Sauval A J, Allende Prieto C and
  Kiselman D 2004 \aanda {\bf 417} 751--768
\item[] Atwood W B \etal 2009 \apj {\bf 697} 1071--1102
\item[] Aver E, Olive K A and Skillman E D 2010 \jcap {\bf 5} 003
\item[] Baby L T \etal 2003 \prl {\bf 90} 022501
\item[] Baby L T \etal 2003 \prc {\bf 67} 065805
\item[] Bachiller R, Forveille T, Huggins P J and Cox P 1997 \aanda
  {\bf 324} 1123--1134
\item[] Bacmann A, Lefloch B, Ceccarelli C, Castets A, Steinarker J
  and Loinard L 2002 \aanda {\bf 389} L6--L10
\item[] Bacmann A, Lefloch B, Ceccarelli C, Steinacker J, Castets A
  and Loinard L 2003 \apj {\bf 585} L55--L58
\item[] Badenes C, Borkowski K, Hughes J P, Hwang U and Bravo E 2006
  \apj {\bf 645}, 1373--1391
\item[] Badenes C, Hughes J P, Cassam-Chena\"{\i} G and Bravo E 2008a
  \apj {\bf 680} 1149--1157
\item[] Badenes C, Bravo E and Hughes J P 2008b \apj {\bf 680}
  L33--L36
\item[] Badenes C, Harris J, Zaritsky D and Prieto J L 2009 \apj {\bf
  700} 727--740
\item[] Badnell N R, O'Mullane M G, Summers H P, Altun Z, Bautista M
  A, Colgan J, Gorczyca T W, Mitnik D M, Pindzola M S and Zatsarinny O
  2003 \aanda {\bf 406} 1151--1165
  {\tt http://amdpp.phys.strath.ac.uk/tamoc/DR/}
\item[] Badnell N R 2006a \apj {\bf 651} L73--L76
\item[] Badnell N R 2006b \jpb {\bf 39} 4825--4852
\item[] Badnell N R 2006c \apjs {\bf 167} 334--342 \hfill \\
  {\tt http://amdpp.phys.strath.ac.uk/tamoc/RR/}
\item[] Baer H, Mustafayev A, Profumo S, Belyaev A and Tata X 2005
  \jhep {\bf 0507} 065
\item[] Bahcall J N, Serenelli A M and Pinsonneault M 2004 \apj 
  {\bf 614} 464--471
\item[] Bahcall J N, Basu S, Pinsonneault M and Serenelli A M 2005a
  \apj {\bf 618} 1049--1056
\item[] Bahcall J N, Serenelli A M and Basu S 2005b \apj {\bf 621} L85--L88
\item[] Bailey J E \etal 2001 \jqsrt {\bf 71} 157--168
\item[] Bailey J E \etal 2007 \prl {\bf 99} 265002
\item[] Bailey J E, Gochau G A, Mancini R C, Igelsias C A, MacFarlane
  J J, Golovkin I E, Blancard C, Cosse Ph and Faussurier G 2009
  \physplas {\bf 16} 058101
\item[] Balbus S A and Hawley J F 1991 \apj {\bf 376} 214--222
\item[] Balbus S A and Hawley J F 1998 \rmp {\bf 70} 1--53
\item[] Baker C A \etal 2006 \prl {\bf 97} 131801
\item[] Bardayan D W \etal 2000 \prc {\bf 62} 055804
\item[] Bardayan D W \etal 2002 \prl {\bf 89} 262501
\item[] Barwick S \etal 2006 \prl {\bf 96} 171101
\item[] Basu S and Antia M H 2008 \pr {\bf 457} 217--283
\item[] Basu S 2010 \apss {\bf 328} 43--50
\item[] Bates D R and Lewis J T 1955 \ppslsa {\bf 68} 173--180
\item[] Bauman R P, Porter R L, Ferland G J and MacAdam K B 2005 \apj
  {\bf 628} 541--544
\item[] Bauschlicher C W, Ram R S, Bernath P F, Parsons C G and
  Galehouse D 2001 \jcp {\bf 115} 1312--1318
\item[] Bauschlicher C W, Boersma C, Ricca A, Mattioda A L, Cami J,
  Peeters E, S\'anchez de Armas F, Puerta Saborido G, Hudgins D M and
  Allamandola L J 2010 \apjs {\bf 189} 341-–35
\item[] Bean J, Kempton E M-R and Homeier D 2010 \nat {\bf 468}
  669--672
\item[] Behar E, Sako M and Kahn S M 2001 \apj {\bf 563} 497--504
\item[] Behar E, Rasmussen A P, Blustin A J, Sako M, Kahn S M, Kaastra
  J S, Branduardi-Raymont G and Steenbrugge K C 2003 \apj {\bf 598}
  232--241
\item[] Beiersdorfer P 2003 \araa {\bf 41} 343--390
\item[] Beiersdorfer P, Olson R E, Brown G V, Harris C L, Neill P A,
  Schweikhard L, Utter S B and Widmann K 2000 \prl {\bf 85} 5090--5093
\item[] Beiersdorfer P \etal 2003 \sci {\bf 300} 1558--1560
\item[] Bekooy J P, Verhoeve P, Meerts W L and Dymanus A 1985 \jcp
  {\bf 82} 3868--3869
\item[] Bellan P M 2005 \physplas {\bf 12} 058301
\item[] Bellan P M, You S and Hsu S C 2005 \apss {\bf 298} 203--209
\item[] Bemmerer D \etal 2006a \prl {\bf 97} 122502
\item[] Bemmerer D \etal 2006b \npa {\bf 779} 297--317
\item[] Bemmerer D \etal 2009 \jpg {\bf 36} 045202
\item[] Benjamin R A, Skillman, E D and Smits D P 1999 \apj {\bf 514}
  307--324
\item[] Bergerson W F, Forest C B, Fiksel G, Hannum D A, Kendrick R, 
  Sarff J S and Stambler S 2006 \prl {\bf 96} 015004
\item[] Bernabei R \etal 2010 \epjc {\bf 67} 39--49
\item[] Bernstein M P, Dworkin J P, Sandford S A, Cooper G W and
  Allamandola L J 2002 \nat {\bf 416} 401--403
\item[] Bhardwaj A \etal 2007 \pss {\bf 55} 1135--1189
\item[] Bi\'{e}mont E, Palmeri P, Quinet P, Zhang Z G and Svanberg S
  2002 \apj {\bf 567} 1267--1283
\item[] Blandford R \etal 2010a {\it New Worlds, New Horizons in
  Astronomy and Astrophysics} (Washington, D. C.: National Academies
  Press)
\item[] Blandford R \etal 2010b {\it Panel Reports - New Worlds, New
  Horizons in Astronomy and Astrophysics} (Washington, D. C.: National
  Academies Press)
\item[] Blasi P 2009 \prl {\bf 103} 051104
\item[] Blondin J M, Fryxell B A and Konigl A 1990 \apj {\bf 360}
  370--386
\item[] Blustein A J, Branduardi-Raymont G, Behar E, Kaastra J S, Kahn
  S M, Page M J, Sako M and Steenbrugge K C 2002 \aanda {\bf 392}
  453--467
\item[] Bockel\'ee-Morvan D, Crovisier J, Mumma M J and Weaver H A
  2004, in {\it COMETS II} eds M Festou, H U Keller and H A Weaver
  (Tuscon: University of Arizona Press) 391--423
\item[] Bolton S 2006 {\it 36th COSPAR Scientific Assembly, COSPAR,
  Plenary Meeting, Beijing, China} {\bf 36} 3775
\item[] Bonetti R \etal 1999 \prl {\bf 82} 5205--5208
\item[] Borucki W J \etal 2010 \sci {\bf 327} 977--980 
\item[] Bouquet S \etal 2004 \prl {\bf 92} 225001
\item[] Bouwman J, Meeus G, de Koter A, Hony S, Dominik C, Waters L B
  F M 2001 \aanda {\bf 375} 950–62
\item[] Bouwman J, Cuppen H M, Bakker A, Allamandola L J and Linnartz
  H 2010 \aanda {\bf 511} A33
\item[] Bozier J C, Thiell G, Le-Breton J P, Azra S, Decroisette M and
  Schirmann D 1986 \prl {\bf 57} 1304--1307
\item[] Bradley J 2010 in {\it Astromineralogy, Lecture Notes in
  Physics Vol. 815} ed T Henning (Berlin: Springer-Verlag) 259--276
\item[] Bregman J N 2007 \araa {\bf 45} 221--259
\item[] Brickhouse N S and Schmelz J T 2006 \apj {\bf 636} L53--L56
\item[] Brickhouse N S \etal 2009a arXiv:0902.4666
\item[] Brickhouse N S \etal 2009b arXiv:0902.4681
\item[] Brickhouse N S \etal 2009b arXiv:0902.4688
\item[] Brickhouse N S \etal 2009d arXiv:0902.4747
\item[] Brickhouse N S \etal 2009e arXiv:0902.4882
\item[] Brickhouse N S, Cranmer S R, Dupree A K, Luna G J M and Wolk S
  2010 \apj {\bf 710} 1835--1847
\item[] Broggini C, Bemmerer D, Guglielmetti A and Menegazzo R 2010
  \arnps {\bf 60} 53
\item[] Bronson-Messer O E, Hix W R, Liebend\"orfer M and Mezzacappa A
  2003 \npa {\bf 718} 449--451
\item[] Brown G V, Beiersdorfer P, Liedahl D A, Widmann K and Kahn S M
  1998 \apj {\bf 502} 1015--1026 (erratum 2000 {\bf 532} 1245)
\item[] Brown G V, Beiersdorfer P, Chen H, Chen M H and Reed K J 2001
  \apj {\bf 557} L75--L78
\item[] Brown J M and M\"{u}ller H S P 2009 \jmsp {\bf 255} 68--71
\item[] Brown M R, Cothran C D, Landreman M, Schlossberg D and
  Matthaeus W 2002 \apj {\bf 577} L63--L66
\item[] Brown T A D, Bordeanu C, Snover K A, Storm D W, Melconian D,
  Sallaska A L, Sjue S K L and Triambak S 2007 \prc {\bf 76} 055801
\item[] Brownlee \etal 2006 \sci {\bf 314} 1711--1716
\item[] Bruhns H \etal 2010a \rsi {\bf 81} 013112
\item[] Bruhns H, Kreckel H, Miller K A, Urbain X and Savin D W 2010b
  \pra {\bf 82} 042708
\item[] Br\"{u}nken S, Gupta H, Gottlieb C A, McCarthy M C and
  Thaddeus P 2007 \apj {\bf 664} L43--L46
\item[] Buckley M R and Randall L 2010 arXiv:1009.0270
\item[] Burrows A, Ram R S, Bernath P, Sharp C M and Milsom J A 2002
  \apj {\bf 577} 986--992
\item[] Burrows A, Dulick M, Bauschlicher C W, Bernath P F, Ram R S,
  Sharp C M and Milsom J A 2005 \apj {\bf 624} 988--1002
\item[] Busquet M \etal 2007 \hedp {\bf 3} 8--11
\item[] Busso M, Gallino R and Wasserburg G J 1999 \araa {\bf 37}
  239--309
\item[] Byun D-Y, Koo B-C, Tatematsu K and Sunada K 2006 \apj {\bf
  637} 283--295
\item[] Calder A \etal 2002 \apj {\bf 143} 201--229
\item[] Calvet N and Gullbring E 1998 \apj {\bf 509} 802--818
\item[] Calzavara A J and Matzner C D 2004 \mnras {\bf 351} 694--706
\item[] Cami J, Bernard-Salas J, Peeters E and Malek S E 2010 \sci
  {\bf 329} 1180--1182
\item[] Canto J, Tenorio-Tagle G and Rozyczka M 1988 \aanda {\bf 192}
  287--294
\item[] Carena M, Nardini G, Quiros M and Wagner C E M 2009 \npb {\bf
  812} 243--263
\item[] Cartledge S I B, Lauroesch J T, Meyer D M and Sofia U J 2006
  \apj {\bf 641} 327--346
\item[] Cartledge S I B, Meyer D M and Lauroesch J T 2003 \apj {\bf
  597} 408--413
\item[] Castelli F, Gratton R G and Kurucz R L 1997 \aanda {\bf 318}
  841--869
\item[] Castelli F and Kurucz R L 2004 \aanda {\bf 419} 725--733
\item[] CDMS II Collaboration 2010 \sci {\bf 327} 1619--1621
\item[] Cernicharo J, Gu\'{e}lin M, Ag\'{u}ndez M, Kawaguchi K,
  McCarthy M and Thaddeus P 2007 \aanda {\bf 467} L37--L40
\item[] Chafa A \etal 2007 \prc {\bf 75} 035810
\item[] Chakraborty N, Fields B D and Olive K A 2010 \prd {\bf 83} 
  063006
\item[] Chakravorty S, Kembhavi A K, Elvis M, Ferland G and Badnell N
  R 2008 \mnras {\bf 384} L24--L28
\item[] Chakravorty S, Kembhavi A K, Elvis M, Ferland G and Badnell N
  R 2009 \mnras {\bf 393} 83--98
\item[] Charbonneau D \etal 2005 \apj {\bf 626} 523--529
\item[] Charbonneau D, Knutson H A, Barman T, Allen L E, Mayor M,
  Megeath S T, Queloz D and Udry S 2008 \apj {\bf 686} 1341--1348
\item[] Charbonnel C, Tosi M, Primas and Chiappini C 2010 {\it IAU
  Symposium No. 268: Light Elements in the Universe, Versoix,
  Switzerland} (Cambridge: Cambridge University Press)
\item[] Chastaing D, Le Picard S D, Sims I R and Smith I W M 2001
  \aanda {\bf 365} 241--247
\item[] Chen D, Gao H and Kwong V H 2003 \pra {\bf 68} 052703
\item[] Chen G-X, Smith R K, Kirby K, Brickhouse N S and Wargelin B J
  2006 \pra {\bf 74} 042709
\item[] Chen G-X 2008 \mnras {\bf 386} L62--L66
\item[] Chen J-W and Savage M J 1999 \prc {\bf 60} 065205
\item[] Chevalier R A 1992 \nat {\bf 355} 691--696
\item[] Chevalier R A and Fransson C 2008 \apj {\bf 683} L135--L138
\item[] Chevalier R A and Imamura J N 1982 \apj {\bf 261} 543--549
\item[] Chiar J E, Tielens A G G M 2006 \apj {\bf 637} 774–85
\item[] Chipps K A \etal 2009 \prl {\bf 102} 152502
\item[] Chowdhury P K, Merer A J, Rixon S J Bernath P F and Ram R S
  2005 \pccp {\bf 8} 822--826
\item[] Christensen-Dalsgaard J, di Mauro M P, Houdek G and Pijpers F
  2009 \aanda {\bf 494} 205--208
\item[] Chugai N N and Chevalier R A 2006 \apj {\bf 641} 1051--1059
\item[] Ciardi A, Lebedev S V, Frank A, Suzuki-Vidal, F, Hall G N,
  Bland S N, Harvey-Thompson A, Blackman E G and Camenzind M 2009 \apj
  {\bf 691} L147--L150
\item[] Cirigliano V, Profumo S and Ramsey-Musolf M J 2006 \jhep {\bf
  0607} 002
\item[] Cirigliano V, Li Y, Profumo S and Ramsey-Musolf M J 2010,
  \jhep {\bf 1001}, 002
\item Clemett S J, Sandford S A, Nakamura-Messenger K, Horz F and
  McKay D S 2010 \mps {\bf 45}, 701--722
\item[] Coc A and Vangioni E 2010 \jpcs {\bf 202} 012001
\item[] Cody G D \etal 2008 \mps {\bf 43}, 353--365
\item[] Cohen D, MacFarlane J, Bailey J and Liedahl D 2003 \rsi {\bf
  74} 1962--1965
\item[] Cohen D H, Leutenegger M A, Wollman E E, Zsarg\'o J, Hillier D
  J, Townsend R H D and Owocki S P 2010 \mnras {\bf 405} 2391--2405
\item[] Commins E D, Ross S B, DeMille D and Regan B C 1994 \pra {\bf
  50}, 2960--2977
\item[] Confortola F \etal 2007 \prc {\bf 75} 065803
\item[] Cordiner M and Millar T J 2009 \apj {\bf 697} 68--78
\item[] Costantini H, Formicola A, Imbriani G, Junker M, Rolfs C and
  Strieder F 2009 \rpp {\bf 72} 086301
\item[] Cothran C D, Brown M R, Grat T, Schaffer M J and Marklin G
  2009 \prl {\bf 103} 215002
\item[] Cowan J J and Sneden C 2006 \nat {\bf 440} 1151--1156
\item[] Cowan J J and Thielemann F-K 2004 \phystoday {\bf 57(10)}
  47--53
\item[] Cranmer S R, Panasyuk A V and Kohl J L 2008 \apj {\bf 678}
  1480--1497
\item[] Cravens T E 1997 \grl {\bf 24} 105--108
\item[] Cravens T E, Robertson I P and Snowden S L 2001 \jgr {\bf 106}
  24883--24892
\item[] Crespo Lopez-Urrutia J R, Neger T and Jager H 1994 \jpd {\bf
  27}, 994--998
\item[] Crovisier J, Leech K, Bockelee-Morvan D, Brooke T Y, Hanner M
  S, Altieri B, Keller H U, Lellouch E 1997. \sci {\bf 275} 1904–7
\item[] Crovisier J \etal 2004 \aanda {\bf 418} 1141--1157
\item[] Curry J J, den Hartog E A and Lawler J E 1997 \josab {\bf 14}
  2788--2799
\item[] Dalgarno, A 1976 {\it Atomic Processes and Applications} ed P
  G Burke (Amsterdam: North-Holland) 109
\item[] Davoudiasl H, Morrissey D E, Sigurdson K and Tulin S 2010 \prl
  {\bf 105} 211304
\item[] De Vries M S, Reihs K, Wendt H R, Golden W G, Hunziker H E,
  Fleming R, Peterson E and Chang S 1993 \gca {\bf 57} 933--938
\item[] Deamer D, Dworkin J P, Sandford S A, Bernstein M P and
  Allamandola L J 2002 \astrobio {\bf 2} 371--381
\item[] Decrock P \etal 1991 \prl {\bf 67} 808-811
\item[] Delbar Th \etal 1993 \prc {\bf 48} 3088--3096
\item[] DeMeo F E, Barucci M A, Merlin F, Guilbert-Lepoutre A,
  Alvarez-Candal A, Delsanti A, Fornasier S and de Bergh C 2010 \aanda
  {\bf 521} A35
\item[] Deming D, Seager S, Richardson L J and Harrington, J 2005 \nat
  {\bf 434} 740--743
\item[] Demorest P B, Pennucci T, Ransom S M, Roberts M S E and
  Hessels J W T 2010 \nat {\bf 467} 1081--1083
\item[] den Hartog E A, Curry J J, Wickliffe M E and Lawler J E 1998
  \solphys {\bf 178} 239--244
\item[] den Hartog E A, Wiese L M and Lawler J E 1999 \josab {\bf 16}
  2278--2284
\item[] den Hartog E A, Fedchak J A and Lawler J E 2001 \josab {\bf
  18} 861--865
\item[] den Hartog E A, Wickliffe M E and Lawler J E 2002 \apjs {\bf
  141} 255--265
\item[] den Hartog E A, Lawler J E, Sneden C and Cowan J J 2003 \apjs
  {\bf 148} 543--566
\item[] den Hartog E A, Herd M T, Lawler J E, Sneden C, Cowan J J and
  Beers T C 2005 \apj {\bf 619} 639--655
\item[] den Hartog E A, Lawler J E, Sneden C and Cowan J J 2006 \apjs
  {\bf 167} 292--314
\item[] den Hartog E A, Lawler J E, Sobeck J S, Sneden C and Cowan J J
  2011 \apjs {\bf 194} 35
\item[] Dennerl K 2010 \ssr {\bf 157} 57--91
\item[] d'Enterria D, Engel R, Pierog T, Ostapchenko S and Werner K
  2011 \app {\bf 35} 98--113
\item[] di Leva A \etal 2009 \prl {\bf 102} 232502
\item[] Diemand J, Kuhlen M and Madau P 2007 \apj {\bf 657} 262--270
\item[] Dine M and Kusenko A 2003 \rmp {\bf 76} 1--30
\item[] Dobler G, Finkbeiner D P, Cholis I, Slatyer T and Weiner N
  2010 \apj {\bf 717} 825--842
\item[] Doron R and Behar E 2002 \apj {\bf 574} 518--526
\item[] Dorschner J and Henning T 1995 \aarev {\bf 6} 271--333
\item[] Dos Santos S F, Kokoouline V and Greene C H 2007 \jcp {\bf
  127} 124309
\item[] Doss F W 2011 {\it Ph.D. thesis} University of Michigan
\item[] Doss F W, Drake R P and Kuranz C C 2010 \hedp {\bf 6} 157--161
\item[] Doss F W, Robey H F, Drake R P and Kuranz C C 2009 \physplas
  {\bf 16} 112705
\item[] Draine B T 2003 \araa {\bf 41} 241-–289
\item[] Drake R P 1999 \jgr {\bf 104} 14505--14515
\item[] Drake J J, Ratzlaff P W, Laming J M and Raymond J 2009 \apj
  {\bf 703} 1224--1229
\item[] Duffy L D and van Bibber K 2009 \njp {\bf 11} 105008
\item[] Duffy A R, Schaye J, Kay S T, Dalla Vecchia C, Battye R A and
  Booth C M 2010 \mnras {\bf 405} 2161--2178
\item[] Dworkin L P, Deamer D W, Sandford S A and Allamandola L J 2001
  \pnasusa {\bf 98} 815--819
\item[] Edens A D, Ditmire T, Hansen J F, Edwards M J, Adams R G,
  Rambo P K, Ruggles L, Smith I C and Porter J L 2005 \prl {\bf 95}
  244503
\item[] Edwards M J, MacKinnon A J, Zweiback J, Shigemori K, Ryutov D
  D, Rubenchik A M, Keitlty K A, Liang E, Remington B A and Ditmire T
  2001 \prl {\bf 87} 085004
\item[] Eggert J, Brygoo S, Loubeyre P, McWilliams R S, Celliers P M,
  Hicks D G, Boehly T R, Jeanloz R and Collins G W 2008 \prl {\bf 100}
  124503
\item[] Elsila J E, Dworkin J P, Bernstein M P, Martin M P and
  Sandford S A 2007 \apj {\bf 660} 911--918
\item[] Elsila J E, Glavin D P and Dworkin J P 2009 \mps {\bf 44},
  1323--1330
\item[] Emery J P, Cruikshank D P and van Cleve J 2006 \ica {\bf 182}
  496-–512
\item[] Epp S W \etal 2007 \prl {\bf 98} 183001
\item[] Fabian J 1996 \prb {\bf 53} 13864--13870
\item[] Falgarone E \etal 2010 \aanda {\bf 521} L15
\item[] Falize E, Bouquet S and Michaut C 2009a \apss {\bf 322} 107--111
\item[] Falize E, Michaut C, Cavet C, Bouquet S, Koenig M,
  Loupias B, Ravasio A and Gregory C G 2009b \apss {\bf 322} 71--75
\item[] Falize E \etal 2011 \apss {\bf 726} 41--49
\item[] Farley D R, Estabrook K G, Glendinning S G, Glenzer S H,
  Remington B A, Shigemori K, Stone J M, Wallace R J, Zimmerman G B
  and Harte J A 1999 \prl {\bf 83} 1982--1985
\item[] Fedchak J A, den Hartog E A, Lawler J E, Palmeri P, Quinet P
  and Bi\'{e}mont E 2000 \apj {\bf 542} 1109--1118
\item[] Federman S R, Knauth D C and Lambert D L 2004 \apj {\bf 603}
  L105--L108
\item[] Ferland G F, Korista K T, Verner D A, Ferguson J W, Kindgon J
  B, and Verner E M 1998 \pasp {\bf 110} 761--778
\item[] Fesen R A, Zastrow J A, Hammell M C, Shull J M and Silvia D W
  2011 \apj {\bf 736} 109
\item[] Field G B, Goldsmith D W and Habing H J 1969 \apj {\bf 155}
  L149
\item[] Fiksel G, Almagri A F, Chapman B E, Mirnov V V, Ren Y, Sarff J
  S and Terry P W 2009 \prl {\bf 103} 145002
\item[] Fink U 2009, \ica {\bf 201} 311--334
\item[] Fitzpatrick A L, Hooper D and Zurek K M 2010 \prd {\bf 81}
  115005
\item[] Flynn \etal 2006 \sci {\bf 314} 1731--1735
\item[] Foord M E \etal 2004 \prl {\bf 93} 055002
\item[] Foord M E \etal 2006 \jqsrt {\bf 99} 712--729
\item[] Formicola A \etal 2004 \plb {\bf 591} 61--68
\item[] Fortney J J and Nettelmann N 2010, \ssr {\bf 152} 423--447
\item[] Fortney J J, Glenzer S H, Koenig M, Militzer B, Saumon D and
  Valencia D 2009 \physplas {\bf 16} 041003
\item[] Foster J M \etal 2005 \apj {\bf 634} L77--L80
\item[] Fox C, Iliadis C, Champagne A E, Fitzgerald R P, Longland R,
  Newton J, Pollanen J and Runkle R 2005 \prc {\bf 71} 055801
\item[] Fox P J and Poppitz E 2009 \prd {\bf 79} 083528
\item[] Fragile P C, Murray S D, Anninos P and van Breugel W 2004 \apj
  {\bf 604} 74--87
\item[] Frum C I, Engleman R Jr, Hedderich H G, Bernath P F, Lamb L D
  and Huffman D R 1991 \cpl {\bf 176} 504--508
\item[] Fujioka S \etal 2009 \natphy {\bf 5} 821--825
\item[] Fukugita M and Yanagida T 1986 \plb {\bf 174} 45--47
\item[] Fullerton A W, Massa D L and Prinja R K 2006 \apj {\bf 637}
  1025--1039
\item[] Fussen D and Kubach C 1986 \jpb {\bf 19} L31--L34
\item[] Gailitis A \etal 2000 \prl {\bf 84} 4365--4368
\item[] Gallo L C, Boller Th, Brandt W N, Fabian A C and Vaughan S 2004 
  \aanda {\bf 417} 29--38
\item[] Gao H and Kwong V H 2003 \pra {\it 68} 052704
\item[] Garc\'{i}a-Hern\'{a}ndez D A, Manchado A, Garc\'{i}a-Lario P,
  Stanghellini L, Villaver E, Shaw R A, Szczerba R and
  Perea-Calder\'{o}n J V 2010 \apj {\bf 724} L39--L43
\item[] Garg U \etal 2007 \npa {\bf 788} 36--43
\item[] Garrod R T and Herbst E 2006 \aanda {\bf 457} 927--936
\item[] Geppert W D and Larsson M 2008 \molphys {\bf 106}
  2199--2226
\item[] Gerbier G 2010 arXiv:1012.2260
\item[] Gibb E L, Whittet D C B, Boogert A C A and Tielens A 2004
  \apjs {\bf 151} 35--73
\item[] Gillaspy J D, Lin T, Tedesco L, Tan J N, Pomeroy J M, Laming J
  M, Brickhouse N, Chen G-X and Silver E 2011 \apj {\bf 728} 132--143 
\item[] Gillett F C, Forrest W J and Merrill K M 1973 \apj {\bf 183}
  87--93
\item[] Giudice G F, Notari A, Raidal M and Strumia A 2004 \npb {\bf
  685} 89--149
\item[] Glosik J, Korolov I, Plasil R, Novotny O, Kotrik T, Hlavenka
  P, Varju J, Mikhailov I A, Kokoouline V and Greene C H 2008 \jpb
  {\bf 41} 191001
\item[] Glos\'{\i}k J, Pla\v{s}il R, Korolov I, Kotr\'{\i}k T,
  Novotn\'y O, Hlavenka P, Dohnal P, Varju J, Kokoouline V and Greene
  C H 2009 \pra {\bf 79} 052707
\item[] Glover S C, Savin D W and Jappsen A-K 2006 \apj {\bf 640}
  553--568
\item[] Glover S C O and Abel T 2008 \mnras {\bf 388} 1627--1651
\item[] Goodman M W and Witten E 1985 \prd {\bf 31} 3059--3063
\item[] Gorham P W, Saltzberg D, Field R C, Guillian E,
  Milin\v{c}i\'{c} R, Mio\v{c}inovi\v{c} P, Walz D and Williams D 2005
  \prd {\bf 72} 023002
\item[] Greenwood J B, Williams I D, Smith S J and Chutjian A 2000
  \pra {\bf 63} 062707
\item[] Greisen K 1966 \prl {\bf 16} 748--750
\item[] Griffith W C, Swallows M D, Loftus T H, Romalis M V, Heckel B
  R, Fortson E N 2009 \prl {\bf 102} 101601
\item[] Grillmair C J, Burrows A, Charbonneau D, Armus L, Stauffer J,
  Meadows V, van Cleve J, von Braun K and Levine D 2008 \nat {\bf 456}
  767--769
\item[] Grosdidier Y, Moffat A F J, Joncas G and Acker A 1998 \apj {\bf 506} 
  L127--L131
\item[] Grosskopf M J, Drake R P, Kuranz C C, Miles A R, Hansen J F,
  Plewa T, Hearn N, Arnett D and Wheeler J C 2009 \apss {\bf 322}
  57--63
\item[] Grun J, Stamper J, Manka C, Resnick J, Burris R, Crawford J
  and Ripin B H 1991 \prl {\bf 66} 2738--2741
\item[] Grundy W M and Schmitt B 1998 \jgr {\bf 103} 25809--25822
\item[] Grundy W M, Young L A, Spencer J R, Johnson R E, Young E F and
  Buie M W 2006 \ica {\bf 184} 543--555
\item[] Grupen C 2005 {\it Astroparticle physics} (Berlin: Springer)
\item[] Gu M F 2004 \apjs {\bf 153} 389--393
\item[] Gu M F, Holczer T, Behar E and Kahn S M 2006 \apj {\bf 641}
  1227--1232
\item[] Gudipati M S 2004 \jpca {\bf 108} 4412--4419
\item[] Gudipati M S and Allamandola L J 2004 \apj {\bf 615}
  L177--L180
\item[] Gudipati M S and Allamandola L J 2006 \jpca {\bf 110}
  9020--9024
\item[] Gudipati M S \etal 2009 arXiv:0910.0442
\item[] Gunther H M, Lewandowska N, Hundertmark M P G, Steinle H, 
  Schmitt J, Buckley D, Crawford S, O'Donoghue D and Vaisanen P 2010 \aanda
  {\bf 518} A54
\item[] Guo B and Li Z H 2007 \chinpl {\bf 24} 65--68
\item[] Gupta H \etal 2010 \aanda {\bf 521} L47
\item[] Guzman J and Plewa T 2009 \nonlin {\bf 22} 2775--2797
\item[] Gyurky G \etal 2007 \prc {\bf 75} 035805
\item[] Halfen D T and Ziurys L M 2004 \apj {\bf 611} L65--L68
\item[] Halfen D T, Sun M, Clouthier D J and Ziurys L M 2009 \jcp {\bf
  130} 014305
\item[] Hammache F \etal 1998 \prl {\bf 80} 928
\item[] Hammache F \etal 2001 \prl {\bf 86} 3985
\item[] Hanner M S and Zolensky M E 2010 in {\it Astromineralogy,
  Lecture Notes in Physics Vol. 815} ed T Henning (Berlin:
  Springer-Verlag) 275--315
\item[] Hansen J F, Edwards M J, Froula D H, Edens A D, Gregori G and
  Ditmire T 2006 \physplas {\bf 13} 112101
\item[] Harada N, Herbst E and Wakelam V 2010, \apj {\bf 721}
  1570--1578
\item[] Harrison J J, Brown J M, Halfen D T and Ziurys L M 2006 \apj
  {\bf 637} 1143--1147
\item[] Hartigan P and Morse J 2007 \apj {\bf 660} 426--440
\item[] Hartigan P, Foster J M, Wilde B H, Coker R F, Rosen P A,
  Hansen J F, Blue B E, Williams R J R, Carver R and Frank A 2009 \apj
  {\bf 705} 1073--1094
\item[] Hartigan P, Frank A, Foster J M, Wilde B H, Douglas M, Rosen P
  A, Coker R F, Blue B E and Hansen J F 2011 \apj {\bf 736} 29
\item[] Hauschildt P, Warmbier R, Schneider R and Barman T 2009 \aanda
  {\bf 504} 225--229
\item[] Haxton W C and Serenelli A M 2008 \apj {\bf 687} 678--691
\item[] Heger A, Langanke K, Martinez-Pinedo G and Woosley S E 2001
  \prl {\bf 86} 1678--1681
\item[] Heger A, Kolbe E, Haxton W C, Langanke K, Mart\'{\i}nez-Pinedo
  G and Woosley S E 2005 \plb {\bf 606} 258--264
\item[] Heil M, K\"{a}ppeler F, Uberseder E, Gallino R, Bisterzo S and
  Pignatari, M 2008 \prc {\bf 78} 025802
\item[] Henning T 2010 \araa {\bf 48} 21-–46
\item[] Henning T and Salama F 1998 \sci {\bf 282} 2204--2210
\item[] Henry R B C, Cowan J J and Sobeck J S 2010 \apj {\bf 709},
  715--724
\item[] Herbst E and Klemperer W 1973 \apj {\bf 185} 505--534
\item[] Herbst E 1981 \nat {\bf 289} 656--657
\item[] Herbst E and van Dishoeck E F 2009 \araa {\bf 47} 427--480
\item[] Hernanz M, Jos\'e J, Coc A, G\'omez-Gomar J and Isern J 1999 
\apj {\bf 526} L97--L100
\item[] Heros C d l 2010 arXiv:1012.0184
\item[] Hersant F, Wakelam V, Dutrey A, Guilloteau S and Herbst E 2009
  \aanda {\bf 493} L49--L52
\item[] Hessels J W T, Ransom S M, Stairs I H, Freire P C C, Kaspi V M
  and Camilo F 2006 \sci {\bf 311} 1901--1904
\item[] Hicks D G, Boehly T R, Celliers P M, Eggert J H, Moon S J,
  Meyerhofer D D and Collins G W 2009, \prb {\bf 79} 014112
\item[] Hohenberger M \etal 2010 \prl {\bf 105} 205003
\item[] Hooper D, Finkbeiner D P and Dobler G 2007 \prd {\bf 76}
  083012
\item[] Hora J L, Latter W B, Smith H A and Marengo M 2006 \apj {\bf
  652} 426--441
\item[] Horn A, Mollendal H, Sekiguchi O, Uggerud E, Roberts H, Herbst
  E, Viggiano A A and Fridgen T D 2004, \apj {\bf 611} 605--614
\item[] H\"orz \etal 2006 \sci {\bf 314} 1716--1719
\item[] Hovde D C and Saykally R J 1987 \jcp {\bf 87} 4332--4338
\item[] Huet P and Nelson A E 1996 \prd {\bf 53} 4578--4597
\item[] Hwang U, Flanagan K A and Petre R 2005 \apj {\bf 635} 355--364
\item[] IAU Commission 14 {\tt http://iacs.cua.edu/IAUC14/}
\item[] Iglesias-Groth S, Manchado A, Garc\'{\i}a-Hern\'andez D A,
  Gonz\'alez-Hern\'andez J I and Lambert D L 2008 \apj {\bf 685}
  L55--L58
\item[] Iglesias-Groth S, Manchado A, Rebolo R, Gonz\'alez-Hern\'andez
  J I, Garc\'{\i}a-Hern\'andez D A and Lambert D L 2010 \mnras {\bf
    407} 2157--2165; Erratum {\bf 409} 880--880
\item[] Imbriani G \etal 2005 \epja {\bf 25} 455--466
\item[] Indriolo N, Geballe T R, Oka T and McCall B J 2007 \apj {\bf
  671} 1736--1747
\item[] Indriolo N, Fields B and McCall B J 2009 \apj {\bf 694}
  257--267
\item[] Innes D E, Giddings J R and Falle S A E G 1987 \mnras {\bf
  226} 67--93
\item[] Ivarsson S, Litz\'{e}n U and Wahlgren G M 2001 \physcr {\bf
  64} 455--461
\item[] Ivarsson S, Andersen J, Nordstr\"{o}m B, Dai X, Johansson S,
  Lundberg H, Nilsson H, Hill V, Lundqvist M and Wyart J F 2003 \aanda
  {\bf 409} 1141--1149
\item[] Izotov Y I and Thuan T X 2010 \apj {\bf 710} L67--L71
\item[] Jager C, Huisken F, Mutschke H, Henning Th, Poppitz W and
  Voicu I 2007 \car {\bf 47} 2981--2994
\item[] Jewitt D 1999 \areps {\bf 27} 287--312
\item[] Jewitt D C and Luu J 2004 \nat {\bf 432} 731--733
\item[] Ji H, Burin M, Schartman E and Goodman J 2006 \nat {\bf 444},
  343--346
\item[] Juhasz A, Henning Th, Bouwman J, Dullemond C P, Pascucci I,
  Apai D 2009 \apj {\bf 695} 1024–41
\item[] Julien K and Knobloch E 2010 \ptrsa {\bf 368} 1607--1633
\item[] Junghans A R, Mohrmann E C, Snover K A, Steiger T D,
  Adelberger E G, Casandjian J M, Swanson H E, Buchmann L, Park S H
  and Zyuzin A 2002 \prl {\bf 88} 041101
\item[] Junghans A R \etal 2003 \prc {\bf 68} 065803
\item[] Junghans A R, Snover K A, Mohrmann E C, Adelberger E G and
  Buchmann L 2010 \prc {\bf 81} 012801
\item[] K{\"a}ppeler F, Beer H and Wisshak K 1989 \rpp {\bf 52}
  945--1013
\item[] K\"appeler F, Gallino R, Bisterzo S and Aoki W 2011 \rmp
  {\bf 83} 157–-193
\item[] Kallman T R 2010 \ssr {\bf 157} 177--191
\item[] Kallman T R and Palmeri P 2007 \rmp {\bf 79} 79--133
\item[] Kaltenegger L and Sasselov D 2010 \apj {\bf 708} 1162--1167
\item[] Kalvans J and Shmeld I 2010 \aanda {\bf 521} A37
\item[] Kang Y G \etal 2000 \pSPIE {\bf 3886} 489--495
\item[] Kaspi S \etal 2002 \apj {\bf 574} 643--662
\item[] Kaspi S, Netzer H, Chelouche D, George I M, Nandra K and 
  Turner T J 2004 \apj {\bf 611} 68--80
\item[] Kastner J H, Huenemoerder D P, Schulz N S, Canizares C R and
  Weintraub D A 2002 \apj {\bf 567} 434--440
\item[] Kawaguchi K, Kasai Y, Ishikawa S and Kaifu N 1995 \pasj {\bf
  47} 853--876
\item[] Keller \etal 2006 \sci {\bf 314} 1728--1731
\item[] Kelley M S and Wooden D H 2009 \pss {\bf 57} 1133-–1145
\item[] Kharchenko V and Dalgarno A 2000 \jgr {\bf 105} 18351--18360
\item[] Kifonidis K, Plewa T, Janka H-T and Muller E 2000 \apj
  {\bf 531} L123--L126
\item[] Kifonidis K, Plewa T, Janka H-T and Muller E 2003 \aanda
  {\bf 408} 621--649
\item[] Kifonidis K, Plewa T, Sheck L, Janka H-T and Muller E 2006
  \aanda {\bf 457} 963--986
\item[] Kim H, Wyrowski F, Menten K M, Decin L 2010 \aanda {\bf 516}
  A68
\item[] Kimoto P A and Chernoff D F 1997 \apj {\bf 485} 274--284
\item[] Kirkpatrick J D 2005 \araa {\bf 43} 195--245
\item[] Klein R I, McKee C F and Colella P 1994 \apj {\bf 420},
  213--236
\item[] Klein R I, Budil K S, Perry T S and Bach D R 2003 \apj {\bf
  583} 245--259
\item[] Klypin A, Trujillo-Gomez S and Primack J 2011 \apj {\bf 740}
  102
\item[] Kobayashi K \etal 2005 \prd {\bf 72} 052007
\item[] Koenig M \etal 2006 \physplas {\bf 13} 056504
\item[] Kohl J L, Noci G, Cranmer S R and Raymond J L 2006 \aarev {\bf
  13} 31--157
\item[] Koldoba A V, Ustyugova G V, Romanova M M and Lovelace R V E 2008
  \mnras {\bf 388} 357--366
\item[] Kotera K and Olinto A V 2011 \araa {\bf 49} 119--153
\item[] Koutroumpa D, Lallement R, Kharchenko V, Dalgarno A, Pepino R,
  Izmodenov V and Quemerais E 2006 \aanda {\bf 460} 289--300
\item[] Knutson H A, Charbonneau D, Allen L E, Fortney J J, Agol E,
  Cowan N B, Showman A P, Cooper C S and Megeath, S T 2007 \nat {\bf
    447} 183--186
\item[] Knutson, H A, Charbonneau, D, Allen, L E, Burrows, A, and 
    Megeath, S T 2008 \apj {\bf 673} 526--531 
\item[] Konacki M, Torres G, Jha S and Sasselov D D 2003 \nat {\bf
  421} 507--509
\item[] Konacki M, Torres G, Sasselov D D, Pietrzy\'{n}ski G, Udalski
  A, Jha S, Ruiz M T, Gieren W and Minniti D 2004 \apj {\bf 609}
  L37--L40
\item[] Konigl A 1991 \apj {\bf 370} L39--L43
\item[] Kotr\'{\i}k T, Dohnal P, Korolov I, Pla\v{s}il R, Rou\v{c}ka
  \v{S}, Glos\'{\i}k J, Greene C H and Kokoouline V 2010 \jcp {\bf
    133} 034305
\item[] Kraemer S B, Ferland G J and Gabel J R 2004 \apj {\bf 604}
  556--561
\item[] Kr\"{a}tschmer W, Lamb L D, Fostiropoulos K and Huffman D R
  1990 \nat {\bf 347} 354--358
\item[] Krasnopolsky V A, Mumma M J, Abbott M, Flynn B C, Meech K J,
  Yeomans D K, Feldman P D and Cosmovici C B 1997 \sci {\bf 277}
  1488--1491
\item[] Krasnopolsky V A and Mumma M J 2001 \apj {\bf 549} 634--634
\item[] Kratz K-L, Pfeiffer B, Thielemann F-K and Walters W B 2000
  \hyperf {\bf 129} 185-221
\item[] Krause O, Tanaka M, Usuda T, Hattori T, Goto M, Birkmann S and
  Nomoto K 2008, \nat {\bf 456} 617--619
\item[] Kravchencko I \etal 2006 \prd {\bf 73} 082002
\item[] Kreckel H \etal 2005 \prl {\bf 95} 263201
\item[] Kreckel H \etal 2010 \pra {\bf 82} 042715
\item[] Kreckel H, Bruhns H, \v{C}\'{\i}\v{z}ek M, Glover S C O,
  Miller K A, Urbain X and Savin D W 2010 \sci {\bf 329} 69--71
\item[] Kre{\l}owski J, Beletsky Y, Galazutdinov G A, Ko{\l}os R,
  Gronowski M and LoCurto G 2010 \apj {\bf 714} L64--L67
\item[] Krolik J, McKee C M and Tarter C B 1981 \apj {\bf 249}
  422--442
\item[] Krongold Y, Nicastro F, Brickhouse N S, Elvis M, Liedahl D A 
  and Mathur S 2003 \apj {\bf 597} 832--850
\item[] Krongold Y, Nicastro F, Elvis M, Brickhouse N S, Mathur S and
  Zezas A 2005 \apj {\bf 620} 165--182
\item[] Kroto H W, Heath J R, Obrien S C, Curl R F and Smalley, R E
  1985 \nat {\bf 318} 162--163
\item[] Kuhlen M, Weiner N, Diemand J, Madau P, Moore B, Potter D,
  Stadel J and Zemp M 2010 \jcap {\bf 1002} 030
\item[] Kuranz C C 2009 \etal \apj {\bf 696} 749--759
\item[] Kuranz C C, Drake R P, Grosskopf M J, Fryxell B, Budde A,
  Hansen J F, Miles A R, Plewa T, Hearn N C and Knauer J P 2010
  \physplas {\bf 17} 052709
\item[] Kurucz R L and Bell B 1995 {\it Kurucz CD-ROM 23: Atomic Line
  List} (Cambridge, MA: Smithsonian Astrophysical Observatory)
\item[] Kuzmin V A, Rubakov V A and Shaposhnikov M E 1985 \plb {\bf
  155} 36--42
\item[] Laming J M 2004 \pre {\bf 70} 057402
\item[] Laming J M \etal 2000 \apj {\bf 545} L161--L164
\item[] Landi E and Cranmer S R 2009 \apj {\bf 691} 794--805
\item[] Langanke K and Martinez-Pinedo G 2003 \rmp {\bf 75} 819--862
\item[] Lattanzi V, Walters A, Drouin B J, Pearson J C 2007 \apj {\bf
  662} 771--778
\item[] Lawler J E, Bonvallet G and Sneden C 2001a \apj {\bf 556}
  452--460
\item[] Lawler J E, Wickliffe M E, Cowley C R and Sneden C 2001b \apjs
  {\bf 137} 341--349
\item[] Lawler J E, Wickliffe M E, den Hartog E A and Sneden C 2001c
  \apj {\bf 563} 1075--1088
\item[] Lawler J E, Sneden C and Cowan J J 2004 \apj {\bf 604}
  850--860
\item[] Lawler J E, den Hartog E A, Sneden C and Cowan J J 2006 \apjs
  {\bf 162} 227--260
\item[] Lawler J E, den Hartog E A, Labby Z E, Sneden C, Cowan J J and
  Ivans I I 2007 \apjs {\bf 169} 120--136
\item[] Lawler J E, den Hartog E A, Sneden C and Cowan J J 2008a, \cjp
  {\bf 86} 1033--1038
\item[] Lawler J E, Sneden C, Cowan J J, Wyart J-F, Ivans I I, Sobeck
  J S, Stockett M H and den Hartog E A 2008b, \apj {\bf 178} 71--88
\item[] Lawler J E, Sneden C, Cowan J J, Ivans I I and den Hartog E A
  2009 \apjs {\bf 182} 51--79
\item[] Lebedev S V \etal 2002 \apj {\bf 564} 113--119
\item[] Lebedev S V, Ampleford D, Ciardi A, Bland S N, Chittenden J P,
  Haines M G, Frank A, Blackman E G and Cunningham A 2004 \apj {\bf
    616} 988--997
\item[] Lebedev S V \etal 2005 \mnras {\bf 361} 97--108
\item[] Lee C, Cirigliano V and Ramsey-Musolf M J 2005 \prd {\bf 71}
  075010
\item[] Leith C E 1990 \pfa {\bf 2} 297--299
\item[] Lemut A \etal 2006 \plb {\bf 634} 483--487
\item[] Leonard D S, Karwowski H J, Brune C R, Fisher B M and Ludwig E
  J 2006 \prc {\bf 73} 045801
\item[] Lestinsky M \etal 2009 \apj {\bf 698} 648--659
\item[] Li Y, Profumo S and Ramsey-Musolf M 2008 \prd {\bf 78} 075009
\item[] Li Y, Profumo S and Ramsey-Musolf M 2009 \plb {\bf 673}
  95--100
\item[] Li Y, Profumo S and Ramsey-Musolf M 2010 \jhep {\bf 1008} 062
\item[] Li Z H \etal 2006 \prc {\bf 74} 035801
\item[] Lin C-H, Antia H M and Basu S 2007 \apj {\bf 668} 603--610
\item[] Lin C-J 2011 arXv:1101.0261 
\item[] Linden T and Profumo S 2010 \apj {\bf 714} L228--L232
\item[] Lis D C \etal 2010 \aanda {\bf 521} L9
\item[] Lisse C M \etal 1996 \sci {\bf 274} 205--209
\item[] Lisse C M, Christian D J, Dennerl K, Meech K J, Petre R,
  Weaver H A and Wolk S J 2001 \sci {\bf 292} 1343--1348
\item[] Liszt H S, Lucas R and Pety J 2006 \aanda {\bf 448} 253--259
\item[] Loch S D, Pindzola M S, Ballance C P and Griffin D C 2006 \jpb
  {\bf 39} 85--104
\item[] Lodders K 2003 \apj {\bf 591} 1220--1247
\item[] Lodders K and Fegley B Jr 1999 in {\it IAU Symposium 191:
  Asymptotic Giant Branch Stars, Montpellier, France} ed T le Bertre,
  A L\'ebre and C Waelkens (Cambridge: Cambridge University Press)
  279--290
\item[] Lopez R E and Turner M S 1999 \prd {\bf 59} 103502
\item[] Luki\'c D \etal 2007 \apj {\bf 664} 1244--1252
\item[] Lundberg H, Johansson S, Nilsson H and Zhang Z 2001 \aanda
  {\bf 372} L50--L52
\item[] Maercker M, Schoier F L, Olofsson H, Bergman P, Ramstedt S
  2008 \aanda {\bf 479} 779--791
\item[] Maier J P, Walker G A H, Bohlender D A, Mazzotti F J,
  Raghunandan R, Fulura J, Garkusha I and Nagy A 2011 \apj {\bf 726}
  41
\item[] Malloci G, Joblin C and Mulas G 2007 \cp {\bf 332} 353--359
\item[] Maloney P R, Hollenbach D J and Tielens A G G M 1996 \apj {\bf
  466} 561--584
\item[] Mancini R C, Bailey J E, Hawley J F, Kallman T, Witthoeft M,
  Rose S J and Takabe H 2009 \physplas {\bf 16} 041001
\item[] Marta M \etal 2008 \prc {\bf 78} 022802
\item[] Martin M C, Koller D and Mihaly L 1993 \prb {\bf 47}
  14607--14610
\item[] Mastrapa R M E and Brown R H 2006 \ica {\bf 183} 207--214
\item[] Matsumoto C, Leighly K M and Marshall H L 2004 \apj {\bf 603}
  456--462
\item[] Matthews C N and Minard R D 2006 \far {\bf 133} 393--401
\item[] Maxted P F L \etal 2010 \aj {\bf 140} 2007--2012
\item[] McCall B J \etal 2003 \nat {\bf 422} 500--502
\item[] McCarthy M C, Gottlieb C A Gupta H and Thaddeus P 2006 \apj
  {\bf 652} L141--L144
\item[] McClendon J H 1999 \esr {\bf 47} 71--93
\item[] McKeegan \etal 2006 \sci {\bf 314} 1724--1728
\item[] McKernan B, Yaqoob T and Reynolds C S 2007 \mnras {\bf 379} 
  1359--1372
\item[] McWilliam, A. 1997, \araa {\bf 35} 503--556
\item[] Medvedev M V and Loeb A 1999 \apj {\bf 526} 697--706
\item[] Menten K M, Wyrowski F, Belloche A, G\"{u}sten R, Dedes L and
  M\"{u}ller H S P 2011 \aanda {\bf 525} A77
\item[] Milam S N, Halfen D T, Tenenbaum E D, Apponi A J, Woolf N J
  and Ziurys L M 2008 \apj {\bf 684} 618--625
\item[] Militzer B 2009 \prb {\bf 79} 155105
\item[] Miles A R, Braun D G, Edwards M J, Robey H F, Drake R P and
  Leibrandt D R 2004 \physplas {\bf 11} 3631--3645.
\item[] Miller-Ricci E, Seager S and Sasselov D 2009 \apj {\bf 690}
  1056--1067
\item[] Miyake S, Stancil P C, Sadeghpour H R, Dalgarno A, McLaughlin
  B M and Forrey R C 2010 \apj {\bf 709} L168--L171
\item[] M{\"o}ller P, Pfeiffer B and Kratz K-L 2003 \prc {\bf 67}
  055802
\item[] Molster F J and Waters L B F M 2003 in {\it Astromineralogy,
  Lecture Notes in Physics Vol. 609} ed T Henning (Berlin:
  Springer-Verlag) 121--170
\item[] Moore A S, Gumbrell E T, Lazarus J, Hohenberger M, Robinson J S, 
  Smith R A, Plant T J A, Symes D R and Dunne M 2008 \prl {\bf 100}
  055001
\item[] Monchaux R \etal 2007 \prl {\bf 98} 044502
\item[] Morton D C 2000 \apjs {\bf 130} 403--436
\item[] Morton D C 2003 \apjs {\bf 149} 205--238
\item[] Moseley J, Aberth W and Peterson J R 1970 \prl {\bf 24}
  435--439
\item[] Muller E, Fryxell B and Arnett D 1991 \aanda {\bf 251}
  505--514
\item[] M\"{u}ller H S P, Schl\"{o}der F, Stutzki J and Winnewisser G
  2005 \jmst {\bf 742} 215--227
\item[] Mu\~{n}oz Caro G M, Meierhenrich U J, Schutte W A, Barbier B,
  Arcones Segovia A, Rosenbauer H, Thiemann W H-P, Brack A and
  Greenberg J M 2002 \nat {\bf 416} 403--406
\item[] Murphy A St J \etal 2009 \prc {\bf 79} 058801
\item[] Murphy G C, Dieckmann M E, Bret A and Drury L O C 2010 \aanda
  {\bf 524} A84
\item[] M\"{u}rtz P, Zink L R, Evenson K M and Brown J M 1998 \jcp
  {\bf 109} 9744--9752
\item[] Nemes L, Ram R S, Bernath P F, Tinker F A, Zumwalt M C, Lamb L
  D and Huffman D R 1994 \cpl {\bf 218} 295--303
\item[] Netzer H \etal 2003 \apj {\bf 599} 933--948
\item[] Netzer H 2004 \apj {\bf 604} 551--555
\item[] Neufeld D A \etal 2010 \aanda {\bf 521} L10
\item[] Nilsson H, Zhang Z G, Lundberg H, Johansson S and Nordstrom B
  2002a \aanda {\bf 382} 368--377
\item[] Nilsson H, Ivarsson S, Johansson S and Lundberg H 2002b \aanda
  {\bf 381} 1090--1093
\item[] Nilsson H, Ljung G, Lundberg H and Nielsen K E 2006, \aanda {\bf
  445} 1165--1168
\item[] Nishikawa K I, Niemiec J, Hardee P E, Medvedev M, Sol H,
  Mizuno Y Zhang B, Pohl M, Oka M and Hartmann D H 2009 \apj {\bf 698}
  L10--L13
\item[] Nishimura S \etal 2011 \prl {\bf 106} 052502
\item[] Nishino H \etal 2009 \prl {\bf 102} 141801
\item[] Nitz D, Kunau A E, Wilson K L and Lentz L R 1999 \apjs {\bf
  122} 557--561
\item[] Nollett K M and Burles S 2000 \prd {\bf 61} 123505
\item[] Nuevo M, Milam S N, Sandford S A, Elsila J E and Dworkin J P
  2009 \astrobio {\bf 9} 683--695
\item[] \"{O}berg K I, van Dishoeck E F and Linnartz H 2009a \aanda
  {\bf 496} 281--293
\item[] \"{O}berg K I, Garrod R T, van Dishoeck E F and Linnartz H
  2009b \aanda {\bf 504} 891--913 
\item[] Olive K A 1999 arXiv:astro-ph/9901231
\item[] Oliver P and Hibbert A 2010 \jpb {\bf 43} 074013
\item[] Osterbrock D E and Ferland G J 2006 {\it Astrophysics of
  Gaseous Nebulae and Active Galactic Nuclei, Second Edition} (Mill
  Valley: University Science Press) (AGN3)
\item[] Otranto S and Olson R E 2011 \pra {\bf 83} 032710
\item[] Palmeri P, Quinet P, Wyart J-F and Bi\'{e}mont E 2000 \physcr
  {\bf 61} 323--334
\item[] Palomaries C 2009 \pos {\bf EPS-HEP2009} 275
\item[] Pascoli G and Polleux A 2000 \aanda {\bf 359} 799--810
\item[] Patel H H and Ramsey-Musolf M J 2011 arXiv:1101.4665
\item[] Patel N A \etal 2011 \apjs {\bf 193} 17
\item[] Persson C M \etal 2010 \aanda {\bf 521} L45
\item[] Pessah M E 2010 \apj {\bf 716} 1012--1027
\item[] Pessah M E and Goodman J 2009 \apj {\bf 698} L72--L76
\item[] Petrignani A \etal 2009 \jpcs {\bf 192} 012022
\item[] Pettini M, Zych B J, Murphy M T, Lewis A and Steidel C S 2008
  \mnras {\bf 391} 1499--1510
\item[] Pfeiffer B, Kratz K-L, Thielemann F-K and Walters W B 2001
  \npa {\bf 693} 282--324
\item[] Piekarewicz J 2010 \jpg {\bf 37} 064038
\item[] Pierre Auger Collaboration 2010 \prb {\bf 685} 239--246
\item[] Pietroni M 1993 \npb {\bf 402} 27--45
\item[] Pilaftsis A 2005 \prl {\bf 95} 081602
\item[] Pilaftsis A 2009 \jpcs {\bf 171} 012017
\item[] Pilaftsis A and Underwood T E J 2005 \prd {\bf 72} 113001
\item[] Piran T 1999 \pr {\bf 314} 575--667
\item[] Pitman K M, Dijkstra C, Hofmeister A M and Speck A K 2010
  \mnras {\bf 406} 460--481
\item[] Polehampton E T, Menten K M, van der Tak F F S, White G J 2010
  \aanda {\bf 510} A80
\item[] Poludnenko A Y, Dannenberg K K, Drake R P, Frank A, Knauer J,
  Meyerhofer D D, Furnish M, Asay J R and Mitran S 2004 \apj {\bf 604}
  213--221
\item[] Porter R L, Ferland G J and MacAdam K B 2007 \apj {\bf 657}
  327--337
\item[] Porter R L, Ferland G J, MacAdam K B and Storey P 2009 \mnras
  {\bf 393} L36--L40
\item[] Porter S B, Desch S J and Cook J C 2010 \ica {\bf 208}
  492--498
\item[] Pospelov M, Ritz A and Voloshin M B 2008 \plb {\bf 662} 53--61
\item[] Pounds K A, Reeves J N, O'Brian P T, Page K A, Turner M J L
  and Nayakshin S 2001 \apj {\bf 559} 181--186
\item[] Pounds K A, Reeves J N, King R A and Page K L 2004 \mnras {\bf
  350} 10--20
\item[] Prager S C, Rosner R, Ji H T and Cattaneo F 2010 {\it Research
  Opportunities in Plasma Astrophysics, Princeton Plasma Physics
  Laboratory} \hfill \\
  {\tt http://www.pppl.gov/conferences/2010/WOPA/index.html}
\item[] Profumo S 2008 arXiv:0812.4457
\item[] Profumo S, Ramsey-Musolf M J and Shaughnessy G 2007\jhep {\bf
  0708} 010
\item[] Pudritz R E, Ouyed R, Fendt C and Brandenburg, A 2007 in {\it
  Protostars and Planets V} ed B Reipurth, D Jewitt and K Keil
  (Tucson: University of Arizona Press) 277--294
\item[] Pulliam R L, Savage C, Ag\'undez M, Cernicharo J, Gu\'elin M
  and Ziurys L M 2010 \apj {\bf 725} L181--L184
\item[] Puget J L and L\'eger A 1989 \araa {\bf 27} 161--198
\item[] Qian Y-Z and Wasserburg G J \pr {\bf 442} 237--268
\item[] Quinet P, Palmeri P, Biemont E, McCurdy M M, Rieger G,
  Pinnington E H Wickliffe M E and Lawler J E 1999 \mnras {\bf 307}
  934--940
\item[] Raiteri C M, Gallino R, Busso M, Neuberger D and K\"{a}ppeler
  F 1993 \apj {\bf 542} 400--403
\item[] Rapaport J and Sugarbaker E 1994 \arnps {\bf 44} 109--153
\item[] Raymond J C, Wallerstein G and Balick B 1991 \apj {\bf 383} 
  226--232
\item[] Redman M P, Viti S, Cau P and Williams D A 2003 \mnras {\bf
  345} 1291--1296
\item[] Reighard A B \etal 2006 \physplas {\bf 13} 082901
\item[] Reipurth B and Bally J 2001 \araa {\bf 39} 403--455
\item[] Remijan A J, Hollis J M, Lovas F J Cordiner M A, Millar T J,
  Markwick-Kemper A J and Jewell, P R 2007 \apj {\bf 664} L47--L50
\item[] Remington B A, Drake R P and Ryutov D D 2006 \rmp {\bf 78}
  755--807
\item[] Ren Y, Yamada M, Ji H, Dorfman S, Gerhardt S P and Kulsrud R M
  2008 \physplas {\bf 15} 082113
\item[] Ren Y, Almagri A F, Fiksel G, Prager S C, Sarff J F and Terry
  P W 2009 \prl {\bf 103} 145002
\item[] Rest A \etal 2008 \apj {\bf 680} 1137--1148
\item[] Reynolds S P, Borkowski K J, Hwang U, Hughes, J P, Badenes C,
  Laming J M and Blondin J M 2007 \apj {\bf 668} L135--L138
\item[] Ricketts C, Contreras C, Walker R and Salama F 2011 \ijms,
  {\bf 300} 26--30
\item[] Richardson J E, Melosh H J, Lisse C M amd Carcich B 2007 \ica
  {\bf 190} 357--390
\item[] Ritchey A M, Federman S R, Sheffer Y and Lambert D L 2011
  \apj, {\bf 728} 70
\item[] Roberts H, Herbst E and Millar T J 2004 \aanda {\bf 424}
  905--917
\item[] Robey H F, Perry T S, Klein R I, Kane J O, Greenough J A and
  Boehly T R 2002 \prl {\bf 89} 085001
\item[] Rogers F J and Iglesias C A 1994 \sci {\bf 263} 50--55
\item[] Rosen P A, Foster J M, Wilde B H, Hartigan P, Blue B E, Hansen
  J F, Sorce C, Williams R J R, Coker R and Frank A 2009 \apss {\bf
    322} 101--105
\item[] Rosner R and Hammer D A 2010 {\it Basic Research Needs for
  High Energy Density Laboratory Physics} (Washington: U.S. Department
  of Energy) \hfill \\
  {\tt  http://www.science.energy.gov/fes/news-and-resources/workshop-reports/}
\item[] Rothman L S \etal 2005 \jqsrt {\bf 96} 139--204
\item[] Rubbia A 2011 March 16 talk presented at 2011 Neutrino
  Telescopes Conference \hfill \\
  {\tt http://agenda.infn.it/conferenceOtherViews.py?view=standard\&confId=3101}
\item[] Runkle R C, Champagne A E, Angulo C, Fox C, Iliadis C,
  Longland R and Pollanen J 2005 \prl {\bf 94} 082503
\item[] Ryan S G, Beers T C, Olive K A, Fields B D and Norris J E 2000
  \apj   {\bf 530} L57--L60
\item[] Ryu D and Vishniac E T 1991 \apj {\bf 368} 411--425
\item[] Ryu D and Vishniac E T 1991 \apj {\bf 368} 411--425
\item[] Ryutov D D, Drake R P, Kane J, Liang E, Remington B A and
  Wood-Vasey M 1999 \apj {\bf 518} 821--832
\item[] Sakai N, Shiino T, Hirota T, Sakai T and Yamamoto S 2010 \apj
  {\bf 718} L49--L52
\item[] Sako M \etal 2001 \aanda {\bf 365} L168--L173
\item[] Sako M \etal 2003 \apj {\bf 596} 114--128
\item[] Salama F 1998, in {\it Solar System Ices} ed B Schmitt, C de
  Bergh and M Festou (Dordrecht: Kluwer Academic Publishers) 259--280
\item[] Salama F 1999, in {\it Solid Interstellar Matter: The ISO
  Revolution, Les Houches, France} ed L d’Hendecourt, C Joblin and A
  Jones (Berlin: Springer-Verlag) 65--87
\item[] Salama F 2008 in {\it IAU Symposium 251: Organic Matter in
  Space} ed S Kwok and S A Sanford (Cambridge: Cambridge University
  Press) 357--366
\item[] Salama F, Galazutdinov G, Krelowski J, Biennier L, Beletsky Y
  and Song I 2011 \apj {\bf 728} 154--162
\item[] Saltzberg D, Gorham P, Walz D, Field C, Iverson R, Odian A,
  Resch G, Schoessow P and Williams D 2001 \prl {\bf 86} 2802--2805
\item[] Sandford \etal 2006 \sci {\bf 314} 1720--1724
\item[] Sargent B A \etal 2009 \apj {\bf 690} 1193-–1207
\item[] Sarre P J 1980 \jcppcb {\bf 77} 769--771
\item[] Sarre P J 2006 \jmsp {\bf 238} 1--10
\item[] Sasselov D D 2003 \apj {\bf 596} 1327--1331
\item[] Saumon D and Guillot T 2004 \apj {\bf 609} 1170--1180
\item[] Savage C, Apponi A J and Ziurys L M 2004 \apj {\it 608}
  L73--L76
\item[] Savin D W \etal 2011 arXiv:1103.1341
\item[] Schartman E, Ji H, Burin M and Goodman J 2011 arXiv:1102.3725
\item[] Schectman R M, Cheng S, Curtis L J, Federman S R, Fritts M C
  and Irving R E 2000 \apj {\bf 419} 207--223
\item[] Schilke P \etal 2010 \aanda {\bf 521} L11
\item[] Schippers S 2009 \jpcs {\bf 163} 012001
\item[] Schippers S, Lestinsky M, M\"uller A, Savin D W, Schmidt E W
  and Wolf A 2010 \iramp {\bf 1} 109--120
\item[] Schlemmer S, Asvany O, Hugo E and Gerlich D 2006 in {\it
  Astrochemistry: Recent Successes and Current Challenges} ed D C Lis,
  G A Blake and E Herbst (Cambridge: Cambridge University Press)
  p~125\
\item[] Schmidt E W \etal 2006 \apj {\bf 641} L157--L160
\item[] Schmidt E W \etal 2008 \aanda {\bf 492} 265--275
\item[] Schoier F L, Maercker M, Justtanout K, Olofsson H, Black J H,
  Decin L, de Koter A, Waters R 2011 \aanda {\bf 530} A83.
\item[] Seager S, Richardson L J, Hanse B M S, Menou K, Cho Y Y-K and
  Deming D 2005 \apj {\bf 632} 1122--1131
\item[] Sellgren K 1984 \apj {\bf 277} 623--633
\item[] Sellgren K, Werner M W, Ingalls J G, Smith J D T, Carleton T M
  and Joblin C 2010 \apj {\bf 722} L54--L57
\item[] Shelton J and Zurek K M 2010 \prd {\bf 82} 123512
\item[] Silk J 1999 in {\it The Third Stromlo Symposium: The Glactic
  Halo, Canberra, Australia, ASP Conference Series Volime 165} ed B K
  Gibson, R S Axelrod and M E Putnam (San Francisco: Astronomical
  Society of the Pacific) 27--33
\item[] Silva L O, Fonseca R A, Tonge J W, Dawson J M, Mori W B and
  Medvedev M V 2003 \apj {\bf 596} L121--L124
\item[] Simon M C \etal 2010 \prl {\bf 105} 183001
\item[] Sims I R 2006 in {\it Astrochemistry: Recent Successes and
  Current Challenges} ed D C Lis, G A Blake and E Herbst (Cambridge:
  Cambridge University Press) 97
\item[] Singh B S, Hass M, Nir-El Y and Haquin G 2004 \prl {\bf 93} 262503
\item[] Sisan D R, Mujica N, Tillotson W A, Huang Y M, Dorland W, Hassam A B, 
Antonsen T M and Lathrop D P 2004 \prl {\bf 93} 114502
\item[] Smith N and Morse J A 2004 \apj {\bf 605} 854--863
\item[] Smith R K, Chen G-X, Kirby K P and Brickhouse N S 2009 \apj
  {\bf 700} 679--683 (erratum 2009 {\bf 701} 2034)
\item[] Sneden C, Cowan J J and Gallino R 2008 \araa {\bf 46} 241--288
\item[] Sneden C, Lawler J E, Cowan J J, Ivans I I and den Hartog E A
  2009 \apjs {\bf 182} 80--96
\item[] Snow T P and McCall B J 2006 \araa {\bf 44} 367--414
\item[] Sobeck J S, Lawler J E and Sneden C 2007 \apj {\bf 667}
  1267--1282
\item[] Soderberg A M \etal 2008 \nat {\bf 453} 469--474
\item[] Sofia U J, Meyer D M and Cardelli J A 1999 \apj {\bf 522}
  L137--L140
\item[] Sogoshi N, Kato Y, Wakabayashi T, Momose T, Tam S, DeRose M E
  and Fajardo M E 2000 \jpca {\bf 104} 3733--3742
\item[] Spence E J, Nornberg M D, Jacobson C M, Parada C A, Taylor N,
  Kendrick R D and Forest C B 2007 \prl {\bf 98} 164503
\item[] Spitkovsky A 2008 \apj {\bf 673} L39--L42
\item[] Springer P T \etal 1997 \jqsrt {\bf 58} 927--935
\item[] Steenbrugge K C, Kaastra J S, de Vries C P and Edelson R 2003 
  \aanda {\bf 402} 477--486
\item[] Steenbrugge K C, Kaastra J S, Sako M, Branduardi-Raymont G,
  Behar E, Paerels F B S, Blustin A J and Kahn S M 2005 \aanda {\bf
    432} 453--462
\item[] Stefani F, Gundrum T, Gerbeth G, Rudiger G, Schultz M, Szklarski J 
and Hollerbach R 2006 \prl {\bf 97} 184502
\item[] Steigman G 2011 \pos {\bf NIC XI} 001
\item[] Strieder F \etal 2001 \npa {\bf 696} 219
\item[] Sun X, Intrator T P, Dorf L, Sears J, Fumo I and Lapenta G
  2010 \prl {\bf 105} 255001
\item[] Springel V \etal 2005 \nat {\bf 435} 629--636
\item[] Stenrup M, Larson {\AA} and Elander N 2009 \pra {\bf 79}
  012713
\item[] Storey P J and Hummer D G 1995 \mnras {\bf 272} 41--48
\item[] Su M, Slatyer T R and Finkbeiner D P 2010 \apj {\bf 724}
  1044--1082
\item[] Symes D R \etal 2010 \hedp {\bf 6} 274--279
\item[] Tangri V, Terry P W and Fiksel G 2008 \physplas {\bf 15}
  112501
\item[] Tenenbaum E D and Ziurys L M 2008 \apj {\bf 680} L121--L124
\item[] Tenenbaum E D and Ziurys L M 2010 \apj {\bf 712} L93--L97
\item[] Tenenbaum E D, Woolf N J and Ziurys L M 2007 \apj {\bf 666}
  L29--L32
\item[] Tenenbaum E D, Milam S N, Woolf N J and Ziurys L M 2009 \apj
  {\bf 704} L108--L112
\item[] Tenenbaum E D, Dodd J L, Milam S N, Woolf N J and Ziurys L M
  2010a \apj {\bf 720} L102--L107
\item[] Tenenbaum E D, Dodd J L, Milam S N, Woolf N J and Ziurys L M
  2010b \apjs {\bf 190} 348--417
\item[] Terzieva R and Herbst E 2000 \ijms {\bf 201} 135--142
\item[] Thaddeus P, Gottlieb C A, Gupta H, Br\"unken S, McCarthy M C,
  Ag\'undez M, Gu\'elin M and Cernicharo J 2008 \apj {\bf 677}
  1132--1139
\item[] The L-S, El Eid M F and Meyer B S 2007 \apj {\bf 655}
  1058--1078
\item[] Tielens A 2005 {\it The Physics and Chemistry of the
  Interstellar Medium} (Cambridge, U.K.: Cambridge University Press)
\item[] Tikhonchuk V T \etal 2008 \ppcf {\bf 50} 124017
\item[] Tinetti G \etal 2007 \nat {\bf 448} 169--171
\item[] Tom B A \etal 2009 \jcp {\bf 130} 031101
\item[] Tornow W, Czakon N G, Howell C R, Hutcheson A, Kelley J H,
  Litvinenko V N, Mikhailov S F, Pinayev I V, Weisel G J, and Witala H
  2003 \plb {\bf 574} 8--13
\item[] Torres G \etal 2011 \apj {\bf 727} 24
\item[] Trotta R, Feroz F, Hobson M P, Roszkowski L and Ruiz de Austri
  R 2008 \jhep {\bf 0812} 024
\item[] Trujillo C A, Brown M E, Barkume K M, Schaller E L and
  Rabinowitz D L 2007 \apj {\bf 655} 1172--1178
\item[] Tucker-Smith D and Weiner N 2001 \prd {\bf 64} 043502
\item[] van Boekel R, Min M, Waters L B F M, de Koter A, Dominik C,
  van den Ancker M E, Bouwman J 2005 \aanda {\bf 437} 189–208
\item[] van der Holst B, Toth G, Sokolov I V, Powell K G, Holloway J
  P, Myra E S, Stout Q, Adams M L, Morel J E and Drake R P 2011 \apjs
  {\bf 194} 23
\item[] van Veelen B, Langer N, Vink J, Garcia-Segura G and
  van Marle A J 2009 \aanda {\bf 503} 495--503
\item[] Vastel C, Caselli P, Ceccarelli C, Phillips T, Wiedner M C,
  Peng R, Houde M and Dominik C 2006 \apj {\bf 645} 1198--1211
\item[] Vernazza J E, Avrett E H and Loeser R 1976 \apjs {\bf 30}
  1--60
\item[] Vishniac E T 1983 \apj {\bf 274} 152--167
\item[] Walker K M, Federman S R, Knauth D C and Lambert D L 2009 \apj
  {\bf 706} 614--622
\item[] Walsh C, Harada N, Herbst E and Millar T J 2009 \apj {\bf 700}
  752--761
\item[] Wang F L, Fujioka S, Nishimura H, Kato D, Li Y T, Zhao G, Zhang J
  and Takabe H 2008 \physplas {\bf 15} 073108
\item[] Wang L, Howell D A, H\"oflich P and Wheeler J C 2001 \apj {\bf
  550} 1030--1035
\item[] Wang X K, Lin X W, Mesleh M, Jarrold M F, Dravid V P,
  Ketterson J B and Chang R P H 1995 \jmr {\bf 10} 1977--1938
\item[] Watson W D 1973 \apj {\bf 183} L17--L20
\item[] Waxman E 2006 \ppcf {\bf 48} B137--B151
\item[] Westmoquette M S, Exter K M, Smith L J, and Gallagher J S 2007
  \mnras {\bf 381} 894--912
\item[] Whittet D C B 1997 \oleb {\bf 27} 249--262
\item[] Whittet D C B 2003 {\it Dust in the Galactic Environment (Second 
Edition)} (Bristol: Institute of Physics)
\item[] Wickliffe M E and Lawler J E 1997 \josab{\bf 14} 737--753
\item[] Wickliffe M E, Lawler J E and Nave G 2000 \jqsrt {\bf 66}
  363--404
\item[] Wiescher M, Gorres J, Uberseder E, Imbriani G and Pignatari M
  2010 \arnps {\bf 60} 381--404
\item[] Wood K and Raymond J C 2000 \apj {\bf 540} 563--571
\item[] Xu H L, Svanberg S, Quinet P, Garnir H P and Bi\'{e}mont E
  2003 \jpb {\bf 36} 4773--4787
\item[] Yamada M 2007 \physplas {\bf 14} 058102
\item[] Yamada M, Kulsrud R and Ji H 2010 \rmp {\bf 82} 603--664
\item[] Yirak K, Frank A, Cunningham A and Mitran S 2008 \apj {\bf
  672} 996--1005
\item[] Yirak K, Frank A and Cunningham A 2011 \apj in press
  arXiv:1101.6020
\item[] Young K, Cox P, Huggins P J, Forveille T and Bachiller R 1999
  \apj {\bf 522} 387--396
\item[] Zack L N, Ziegler N and Ziurys L M 2011 \apj submitted
\item[] Zatsepin G T and Kuz'min V A 1966 \JETPL {\bf 4} 78--80
\item[] Zheng W J, Jewitt D and Kaiser R I 2009 \jpca {\bf 113} 
11174--11181
\item[] Ziurys L M 2006 \pnasusa {\bf 103} 12274--12279
\item[] Zolensky \etal 2006 \sci {\bf 314} 1735--1739
\item[] Zweibel E G and Yamada M 2009 \araa {\bf 47} 292--332
\end{harvard}

\end{document}